\documentclass[
 reprint,
superscriptaddress,
nofootinbib,
 amsmath,amssymb,
 aps,
 prd,
floatfix,
]{revtex4-1}

\usepackage{graphicx,color}% Include figure files
\graphicspath{{./figures/}}
\usepackage{dcolumn}% Align table columns on decimal point
\usepackage{bm}% bold math
\usepackage{hyperref}% add hypertext capabilities
\usepackage{physics}
\usepackage{soul}
\usepackage{cancel}
\usepackage{mathtools}
\usepackage{cprotect}
\usepackage[normalem]{ulem}
\usepackage{comment}
\newcommand{\red}[1]{{\color{black} #1}}

\usepackage{tabularx}
\usepackage{adjustbox}

\newcommand\Tstrut{\rule{0pt}{2.6ex}}         % = `top' strut
\usepackage[multidot]{grffile}

\usepackage[caption=false]{subfig}

\usepackage[dvipsnames]{xcolor}
\definecolor{C0}{RGB}{31, 119, 180}
\definecolor{C1}{RGB}{255, 127, 14}
\definecolor{C2}{RGB}{44, 160, 44}
\definecolor{C3}{RGB}{214, 39, 40}
\definecolor{C4}{RGB}{148, 103, 189}
\definecolor{C5}{RGB}{140, 86, 75}

\begin{document}

\title{
Postmerger multimessenger analysis of binary neutron stars: Effect of the magnetic field strength and topology}

\author{Jamie Bamber}
\affiliation{%
Department of Physics, University of Illinois Urbana-Champaign, Urbana, IL 61801, USA
}
\author{Antonios Tsokaros}
\affiliation{%
Department of Physics, University of Illinois Urbana-Champaign, Urbana, IL 61801, USA
}%
\affiliation{National Center for Supercomputing Applications, University of Illinois Urbana-Champaign, Urbana, IL 61801, USA}
\affiliation{Research Center for Astronomy and Applied Mathematics, Academy of Athens, Athens 11527, Greece}
\author{Milton Ruiz}%
\affiliation{%
 Departament d’Astronomia i Astrof\'{i}sica, Universitat de Val\`{e}ncia, C/ Dr Moliner 50, 46100, Burjassot (Val\`{e}ncia), Spain
}
\author{Stuart L. Shapiro}
\affiliation{%
Department of Physics, University of Illinois Urbana-Champaign, Urbana, IL 61801, USA
}%
\affiliation{
Department of Astronomy, University of Illinois Urbana-Champaign, Urbana, IL 61801, USA
}
\affiliation{National Center for Supercomputing Applications, University of Illinois Urbana-Champaign, Urbana, IL 61801, USA}

\date{\today}% It is always \today, today,

\begin{abstract} 
The oscillation modes of neutron star (NS) merger remnants, as encoded by the kHz postmerger gravitational wave (GW) signal, hold great potential for constraining the as-yet undetermined equation of state (EOS) of dense nuclear matter. Previous works have used numerical relativity simulations to derive quasi-universal relations for the key oscillation frequencies, but most of them omit the effects of a magnetic field. We conduct full general-relativistic magnetohydrodynamics simulations of NSNS mergers with two different masses and two different EOSs (SLy and ALF2) with three different initial magnetic field topologies (poloidal and toroidal only, confined to the interior, and ``pulsar-like'': dipolar poloidal extending from the interior to the exterior), with four different magnetic field strengths with maximum values ranging from from $5.5\times 10^{15}G$ to $2.2\times 10^{17}G$ at the time of insertion. 
We find that magnetic braking and magnetic effective turbulent viscosity drives the merger remnants towards uniform rotation and increases their overall angular momentum loss. As a result, the $f_2$ frequency of the dominant postmerger GW
mode shifts upwards over time. The overall shift is up to $\sim 200$ Hz for the strongest magnetic field we consider and $\sim 50$Hz for the median case and is therefore detectable in principle by future GW observatories, which should include the magnetic field in their analyses. We also explore the impact of the magnetic field on the postmerger electromagnetic emission, and demonstrate that an extremely large magnetic field, or alternatively a significant shear viscosity mechanism, can cause a supramassive NS remnant to collapse to a BH in less than 100ms and lead to jet formation, although we do not expect the conditions for such an outcome to be realistic. 

\end{abstract}

\maketitle

\section{Introduction}

Gravitational wave (GW) and electromagnetic (EM) observations of neutron star-neutron star (NSNS) mergers hold enormous potential for constraining their as-yet undetermined nuclear equation of state (EOS), as well as unresolved astrophysical questions such as the mechanism powering short $\gamma$-ray bursts. Advanced LIGO and Virgo have already identified two NSNS merger GW events, GW170817 \cite{LIGOScientific:2017vwq} and GW190425 \cite{LIGOScientific:2020aai}, with GW170817 accompanied by a coincident short $\gamma$-ray burst \cite{LIGOScientific:2017zic,Savchenko:2017ffs} and kilonova \cite{Cowperthwaite:2017dyu,Smartt:2017fuw,Kasliwal:2017ngb} observed across the EM spectrum. %\cite{Lamb:2018qfn,Coulter:2017wya,Troja:2020pzf,Mooley:2018clx,Hajela:2019mjy,DAvanzo:2018zyz,Resmi:2018wuc,Alexander:2017aly,DAvanzo:2018zyz,Resmi:2018wuc}. 
Finite-size effects on the inspiral GW signal from these events have already been used to place constraints on the NS tidal deformability \cite{LIGOScientific:2017vwq,LIGOScientific:2018hze}, which, along with constraints on the NS maximum mass and mass-radius relation %\cite{NANOGrav:2019jur,Riley:2021pdl,Fonseca:2016tux,Antoniadis:2013pzd,Pang:2021jta,Miller:2019cac,Riley:2019yda,Tsokaros:2020hli,Ruiz:2017due,Rezzolla:2017aly,Riley:2019yda,Musolino:2024sju}, 
\cite{NANOGrav:2019jur,Riley:2021pdl,Fonseca:2016tux,Antoniadis:2013pzd,Miller:2019cac,Riley:2019yda,Riley:2019yda}, has been used to constrain the range of viable EOS models. %(see \cite{Ozel:2016oaf,Miller:2019nzo,Burgio:2021vgk,Lattimer:2021emm,Koehn:2024set} and the references therein for recent reviews). 
We expect to see many more NSNS merger GW events in the near future \cite{KAGRA:2013rdx} with corresponding improvements in our knowledge of the EOS %\cite{HernandezVivanco:2019vvk,Lackey:2014fwa,Bauswein:2017vtn,DelPozzo:2013ala}.
\cite{HernandezVivanco:2019vvk,Lackey:2014fwa}.

The fate of the postmerger remnant depends on its mass, angular momentum and the nuclear EOS \cite{Shibata:2006nm}. If the total mass is above some dynamically determined threshold mass, $M_{\textup{thresh}}$, the remnant will undergo prompt collapse to a black hole (BH) within a few ms \cite{Bauswein2013}.  %\cite{Koppel:2019pys,Kolsch:2021lub}. 
If the mass is below $M_{\textup{thresh}}$ the merger will form a hot, differentially rotating NS remnant. For masses above the supramassive limit, $M_{\textup{sup}}$, the maximum mass for a stable, cold, rigidly rotating star \cite{Cook92b}, the remnant is termed a hypermassive neutron star (HMNS) \cite{Baumgarte:1999cq}. It is only metastable, and collapses to a BH after several orbital periods (see e.g.~\cite{Duez:2004nf,Ruiz:2017due}) as viscosity, magnetic winding, and the emission of GWs drive it towards uniform rotation. For masses below $M_{\textup{sup}}$ the remnant survives for much longer timescales as a uniformly rotating star. If the mass is above $M_{\textup{TOV}}$, the maximum mass for a cold, spherical, non-rotating star, but below $M_{\textup{sup}}$, the remnant is termed a supramassive neutron star (SMNS) \cite{Cook92b}. A SMNS eventually spins down and collapses to a BH \cite{Friedman:1988er} with lifetime that depends on its magnetic field, its spin, and how close the star is to the turning point, but is typically of the order $\sim 10^3(B/10^{15}G)^{-2}$s where $B$ is the strength of the magnetic field \cite{Shapiro:1983,Thompson:2004wi,Lander:2018und,Beniamini:2021tpy}. 
%\MR{This paragraph feels excessively detailed and long}

A particularly rich source of information about the nuclear EOS is the kHz postmerger GW signal produced by a hot, non-axisymmetric NS remnant, with the frequency of the dominant quadrupole mode located between 1500 - 4000Hz 
\cite{Zhuge94}.
While LIGO-Virgo was unable to detect this postmerger signal in GW170817 and GW190425 \cite{LIGOScientific:2017fdd,Grace:2024xty}, expected improvements in sensitivity \cite{KAGRA:2013rdx}, third-generation detectors 
%\cite{LIGOScientific:2016wof,Dhani:2023ijt,Reitze:2019iox,Punturo:2010zz}, 
\cite{LIGOScientific:2016wof,Reitze:2019iox,Punturo:2010zz}, 
and dedicated high frequency observatories 
%\cite{Ackley:2020atn,Martynov:2019gvu,Page:2020zbr,Sarin:2021qqo} 
\cite{Ackley:2020atn,Page:2020zbr} 
should allow robust detections of the postmerger GW component. Numerical relativity (NR) simulations of NSNS mergers over the past two decades have provided predictions of the postmerger GW emission, along with descriptions of the dynamics, remnant evolution, and EM luminosity. 

While the postmerger regime involves many complex physical processes including thermal effects, turbulence, possible phase transitions and magnetic instabilities, attempts have been made to describe the signal in terms of characteristic frequencies corresponding to peaks in its power spectral density, related to oscillation modes of the NS remnant. By measuring these frequencies across a range of NR merger simulations using various NS masses and equations of state, it has been shown that one can construct \textit{quasi-universal} relations (although see \cite{Raithel:2022orm} for possible violations) between the frequencies and NS stellar properties like the compactness, chirp mass and tidal deformability (see e.g. \cite{Shibata:2005ss,Shibata:2006nm,Stergioulas:2011gd,Takami:2014tva,Bauswein:2014qla,Rezzolla:2016nxn,Vretinaris:2019spn}). These relations have been used to inform and construct waveform models \cite{Clark:2015zxa,Bose:2017jvk,Breschi:2019srl,Wijngaarden:2022sah} and can in principle be used with a strong postmerger GW observation to place tight constraints on the EOS.

However, these relations are typically constructed from datasets of NR simulations using relativistic hydrodynamics only, with no magnetic fields. The surface magnetic field of NSs in binary systems is typically $10^8 - 10^{12.2}$G as inferred from observations of binary pulsars \cite{Tauris:2017omb,Lorimer:2008se}.  
During and following the merger several mechanisms work to amplify the magnetic field to magnetar strengths of $\gtrsim 10^{15}$G \cite{Shapiro:2000zh,Duez:2006qe,Rosswog:2006ue,Obergaulinger:2010gf,Oechslin:2006uk,Zrake:2013mra}. The effect of strong magnetic fields on NS oscillation modes has been studied for isolated axisymmetric magnetars. Flores et al. (2020) \cite{Flores:2019lvq}, using perturbation theory, found that magnetic field strengths of up to $3 \times 10^{18}$G alter the fundamental oscillation frequencies by less than 50Hz. Leung et al. (2022) \cite{Leung:2022mvm} examined the pulsation modes of non-rotating stars with strong toroidal magnetic fields with general relativistic magnetohydrodynamics (GRMHD) simulations, also finding that the oscillation modes are largely insensitive to the magnetic field for $E_{\textup{mag}}/\vert W \vert \lesssim 10^{-2}$, where $E_{\textup{mag}}$ and $W$ are the electromagnetic energy and gravitational binding energy respectively. 
These studies use simplistic models of magnetized NSs. On the other hand, our previous works \cite{Ruiz:2021qmm,Bamber:2024kfb} suggested that the inclusion of magnetic fields in GRMHD simulations of NSNS binary mergers can result in a shift of the frequency of the dominant quadrupolar mode of the merger remnant by up to several hundred Hz, an imprint of the magnetic field that may be detectable by third generation observatories \cite{Ruiz:2021qmm}. 

In this work we extend our previous analyses by conducting merger simulations with a range of different initial magnetic field strengths and magnetic field topologies, inserted at 1ms prior to merger, for two different binary ADM masses and two different equations of state. We find that in most cases the presence of the magnetic field causes the NS remnant to lose more angular momentum due to the effective turbulent magnetic viscosity and magnetic braking, resulting in an increase in remnant compactness and an associated time-dependent shift in the dominant quadrupolar $f_2$ frequency of the postmerger GW signal and a decrease in the lifetime of HMNS remnants, consistent with the results of \cite{Ruiz:2021qmm}. The maximum increase in frequency, as obtained from the whole postmerger signal, is $\sim 200$Hz for the strongest magnetic field considered and $\sim 50$Hz for the median case. 

We also find that very strong magnetic fields (initial maximum strength at insertion of $2.2\times 10^{17}$G, the strongest field we explored) can cause SMNS remnants to lose sufficient angular momentum that they collapse to BHs, despite being well below the supramassive limit. That efficient angular momentum loss could drive SMNS remnants to collapse is a possibility previously raised by Ma et al. (2018) \cite{Ma:2017yva}, but not observed in hydrodynamics-only simulations \cite{Radice:2018xqa}. Our results present the first GRMHD simulations of the collapse of SMNS merger remnants with binary masses $< M_{\textup{sup}}$ on $\sim 10$ms timescales. We show that these systems subsequently form magnetically driven outflows consistent with incipient jets, powered by accreting BHs, similar to those seen in our previous simulations with HMNS remnants \cite{Ruiz:2016rai,Ruiz:2021qmm}. However, it should be noted that the magnetic fields required are unrealistically large, and produce EM luminosities of order $\sim 10^{54}\textup{erg}\;\textup{s}^{-1}$, corresponding to isotropic equivalent $\gamma-$ray luminosities of $L_{\gamma,\textup{iso}} \sim 10^{54}-10^{56}\textup{erg}\;\textup{s}^{-1}$, outside the observed range for sGRBs. Therefore, the constraints on the NS maximum mass based on the assumption that the GW170817 merger must have produced a remnant with mass above or close to $M_{\textup{sup}}$ in order to produce a BH powered jet (e.g. \cite{Ruiz:2017due}) remain valid, unless a more realistic source of efficient angular momentum loss can be established. 

The remainder of the paper is organized as follows. Sec.~\ref{sec:methods} presents a brief summary of our numerical methods, referring the reader to our previous works~\cite{Etienne:2010ui,Ruiz:2020via,Ruiz:2019ezy,Ruiz:2017inq} for further details. A detailed description of our initial data, including the prescription for the initial magnetic fields, along with the simulation grid structure is given in Sec.~\ref{subsec:Initialdata} and \ref{subsec:Grid}, respectively. A partial set of diagnostics used to analyse our numerical results is presented in Sec.~\ref{subsec:diagnostic}, and Sec.~\ref{subsec:M_processes} gives a short summary of the important mechanisms associated with the magnetic field that we expect to occur in the postmerger remnant. Sec.~\ref{sec:jet_theory} presents a brief discussion of aspects of the theory of ideal MHD jets, which will be used in Sec.~\ref{sec:EM_jets}.
We present our results in Sec.~\ref{Sec:Results}. Finally, we summarize our results and conclude in Sec.~\ref{sec:Conclusions}. Throughout the paper we adopt geometrized units ($G = c = 1$) except where stated explicitly. Greek indices denote all four spacetime dimensions, while Latin indices imply spatial parts only.

\section{Methods}
\label{sec:methods}

We conduct our GRMHD simulations using the in-house and well-tested \verb|Illinois GRMHD| code \cite{Etienne:2010ui,Etienne:2011re,Farris:2012ux} using the same methods as in our previous works (see~e.g.~\cite{Bamber:2024kfb,Ruiz:2020via,Ruiz:2019ezy,Ruiz:2017inq}). For the spacetime evolution we use the Baumgarte–Shapiro–Shibata–Nakamura (BSSN) formulation of the Einstein equations \cite{Shibata:1995we,Baumgarte:1998te} with the moving-puncture gauge condition (Eqs. (2)-(4) in \cite{Etienne:2007jg}). The damping parameter $\eta$ in the shift evolution equation is set to $\eta \approx 2.0/M$, where $M$ is the total Arnowitt–Deser–Misner (ADM) mass of the system. At the boundaries we apply outgoing-wave (Sommerfeld) boundary conditions to all the BSSN variables. We use fourth-order centered stencils for spatial derivatives, except for shift advection terms where fourth-order upwind stencils are used, along with fifth-order Kreiss-Oliger dissipation \cite{Baker:2006yw}. Time integration is performed using the Method of Lines with a fourth order Runge-Kutta integration scheme with a Courant-Friedrichs-Lewy (CFL) factor set to 0.45. For the matter evolution we evolve the equations of ideal MHD in conservative form using a high-resolution shock capturing method (see Eqs. (27)-(29) in \cite{Etienne:2010ui}) which employs a modified piecewise parabolic reconstruction scheme (PPM+) \cite{Colella:1982ee,Duez:2005sf} and the Harten, Lax, and van Leer (HLL) approximate Riemann solver \cite{Harten1983}. We evolve the magnetic field by integrating the magnetic induction equation using a vector potential (see Eqs. (19)-(20) in \cite{Etienne:2010ui}) and the generalized Lorentz gauge in \cite{Farris:2012ux,Etienne:2011re} with a damping factor of $1.5/\dd t_{\rm max}$ where $\dd t_{\rm max}$ is the time step on the coarsest level. As described in \cite{Ruiz:2021qmm} Eqs (1)-(3) we include a thermal component in the EOS to account for shock heating, writing the total pressure as $P = P_{\textup{th}} + P_{\textup{cold}}$ where $P_{\textup{cold}}$ is the cold component (the piecewise polytropic representation of the EOS), and $P_{\textup{th}}$ is a thermal component given by 
\begin{equation}
    {P_{\textup{th}} = (\Gamma_{\textup{th}}-1) \rho_0(\epsilon - \epsilon_{\textup{cold}}).}
\end{equation}
Here $\epsilon$ is the specific internal energy, $\epsilon_{\textup{cold}}$ is the zero-temperature component of $\epsilon$, and $\Gamma_{\textup{th}} = 5/3$ appropriate for ideal nonrelativistic baryons \cite{Bauswein:2010dn,Paschalidis:2011ez}. 

Following our previous works \cite{Paschalidis:2014qra,Ruiz:2016rai,Ruiz:2019ezy,Ruiz:2021qmm,Bamber:2024kfb}, to reliably evolve the exterior magnetic fields within the assumptions of ideal MHD we initially add a low-density artificial atmosphere exterior to the NSs in regions where the magnetic field dominates over the fluid pressure gradient. The density of this artificial atmosphere is chosen such that at $t=0$ the plasma parameter $\beta$ satisfies $\beta = P_{\textup{gas}}/P_{\textup{magnetic}} = 0.01$ everywhere exterior to the stars, with an additional density floor of $\rho_0^{\textup{min}} = 10^{-10}\rho_0^{\textup{max}}$, where $\rho^{\textup{max}}_0$ is the maximum value of the initial rest-mass density of the system. Further implementation details can be found in \cite{Ruiz:2018wah} Sec. II. B and \cite{Bamber:2024kfb}. The artificial atmosphere increases the total rest mass of the system by $\lesssim 2\%$, and was shown previously to have a negligible effect on the dynamical evolution \cite{Paschalidis:2014qra}. Our evolution code has been extensively tested (see e.g. \cite{Duez:2005sf,Etienne:2010ui,Etienne:2011re,Etienne:2015cea}). However, as a refresher we present a simple convergence test for one of the cases examined in this work in Appendix \ref{sec:conv} and verify we obtain the expected convergence behavior.

\subsection{Initial data}
\label{subsec:Initialdata}

\begin{figure}
    \centering
    \includegraphics[width=\linewidth]{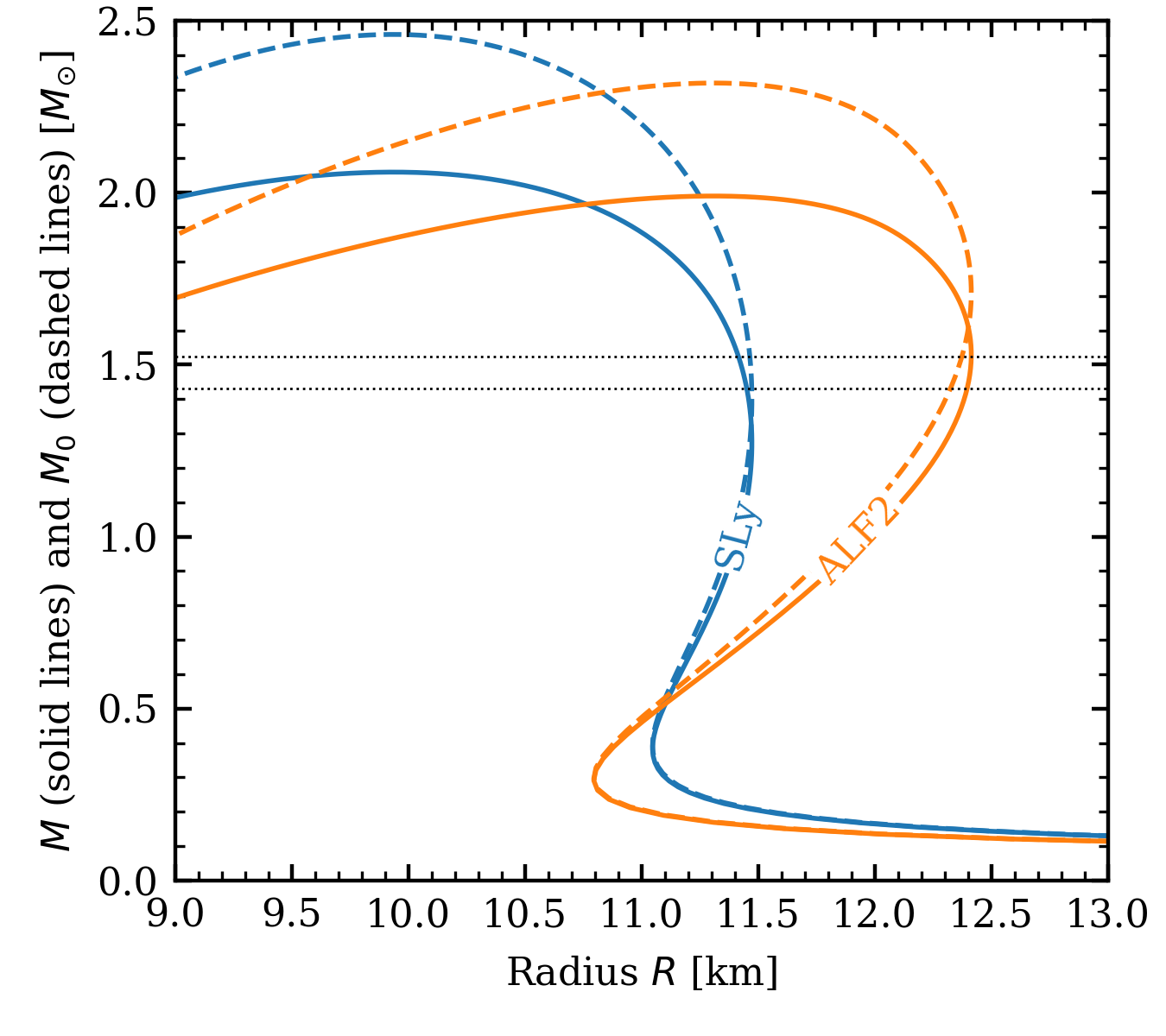}
    \caption{Mass vs. radius curves for the SLy and ALF2 equations of state for isolated, cold, spherical neutron stars. The ADM mass and rest mass curves are shown as solid and dashed lines respectively. The two black horizontal dotted lines show the two different initial NS rest masses used in our simulations.}
    \label{fig:EOS_M_vs_radius}
\end{figure}

Our initial data consists of a binary of two equal mass, irrotational, NSs that start in a quasiequilibrium circular orbit and then inspiral and merge. The data is generated using the Compact Object CALculator (\texttt{COCAL}) code (see e.g.~\cite{Tsokaros:2015fea,Tsokaros:2018dqs}) with two different equations of state (EOS): the hadronic Skyrme-Lyon (SLy) \cite{Douchin:2001sv} and the nuclear matter-quark matter hybrid ALF2 \cite{Alford:2004pf}, both modeled using a piecewise polytropic representation \cite{Read:2008iy}. The critical mass limits $M_{\textup{TOV}}, M_{\textup{sup}}$ and $M_{\textup{thresh}}$ for these EOSs are shown in Table \ref{tab:mass_lims}. SLy and ALF2 are both  relatively soft, and broadly compatible with current observational constraints \cite{Koehn:2024set,Lattimer:2021emm,Miller:2019nzo} using data from pulsar timing \cite{NANOGrav:2019jur,Fonseca:2016tux,Antoniadis:2013pzd}, NICER observations \cite{Pang:2021jta,Miller:2019cac,Riley:2019yda,Riley:2021pdl}, and GW measurements \cite{Tsokaros:2020hli,Ruiz:2017due,LIGOScientific:2018hze} (see the discussion in \cite{Bamber:2024kfb} Sec. I. A. for more detail). Mass vs. radius curves for the two EOSs used this work are shown in Fig. \ref{fig:EOS_M_vs_radius}. 

\def\arraystretch{1.0}%
\vspace{0.5cm}
\begin{table}
%\begin{ruledtabular}
\begin{tabular}{l|l|cc}
        & critical  & $\;$rest mass $M_0$$\;$$\;$  &  $\;$$\;$ADM mass $M$ \\
    EOS & mass & [$M_{\odot}$] & [$M_{\odot}$] \\[.05cm] 
    \hline
    & $M_{\textup{TOV}}$ & 2.46 & 2.06 \Tstrut \\
    SLy & $M_{\textup{sup}}$ & 2.96 & 2.49 \\
    & $M_{\textup{thresh}}$ & $\dots$ & $\sim 2.76$ \\
    \hline
    & $M_{\textup{TOV}}$ & 2.32 & 1.99 \Tstrut \\
    ALF2 & $M_{\textup{sup}}$ & 2.89 & 2.48 \\
    & $M_{\textup{thresh}}$ & $\dots$ & $\sim 2.96$ \\
\end{tabular}
%\end{ruledtabular}
\caption{Critical mass limits for the SLy and ALF2 EOSs \cite{Read:2008iy,Tsokaros:2020hli,Kolsch:2021lub,Koppel:2019pys}. The estimate for the gravitational $M_{\textup{thresh}}$ is from a series of GRHD merger simulations of initially irrotational neutron stars (see \cite{Kolsch:2021lub} for details).}
\label{tab:mass_lims}
\end{table}

\begin{table}
\begin{ruledtabular}
\begin{tabular}{c|ccccccc}
Case & EOS & $M$ & $M_0/2$ & $\bar{M}$ & $R_x$ & ${\Omega}M$ & $\Lambda$ \\
\colrule
SLy\_2.57 & SLy & 2.57 & 1.43 & 1.30 & 9.25 & 0.029 & 498 \Tstrut \\
ALF2\_2.57 & ALF2 & 2.57 & 1.43 & 1.29 & 9.25 & 0.029 & 918 \\
ALF2\_2.70 & ALF2 & 2.70 & 1.52 & 1.36 & 8.67 & 0.034 & 688
\end{tabular}
\end{ruledtabular}
\caption{Summary of the initial base properties of the hydrodynamics-only NSNS cases. We list the name of the case, the gravitational (ADM) mass of the binary system $M$, the rest mass of each star $M_0/2$, the gravitational mass of each star at infinite separation $\bar{M}$ (all masses given in units of $M_{\odot})$, the equatorial coordinate radius of each star measured along the axis of the binary $R_x$ (in km), the dimensionless quantity $\Omega M$ where $\Omega$ is the orbital angular velocity of the binary, and the dimensionless tidal deformability at infinite separation $\Lambda$. The initial coordinate separation between the centers of the stars is set to $3.98 R_x$. For each of the hydro-only cases we also conducted six simulations with different initial magnetic fields inserted 1 ms before merger.}
\label{tab:initial_NS}
\end{table} 

\subsubsection{Magnetic fields}
\label{sec:magn_fields}

For each EOS and binary mass we ran one hydrodynamics-only simulation, along with magnetized cases. The properties of the hydrodynamics-only simulations are shown in Table \ref{tab:initial_NS}. We use an initial NSNS separation of $3.98 R_x$ where $R_x$ is the radius of the stars along the axis of the binary which gives $3-4$ orbits of inspiral before the stars merge. 

\begin{figure*}
\begin{tabular}{ccc}
  \includegraphics[width=0.33\textwidth]{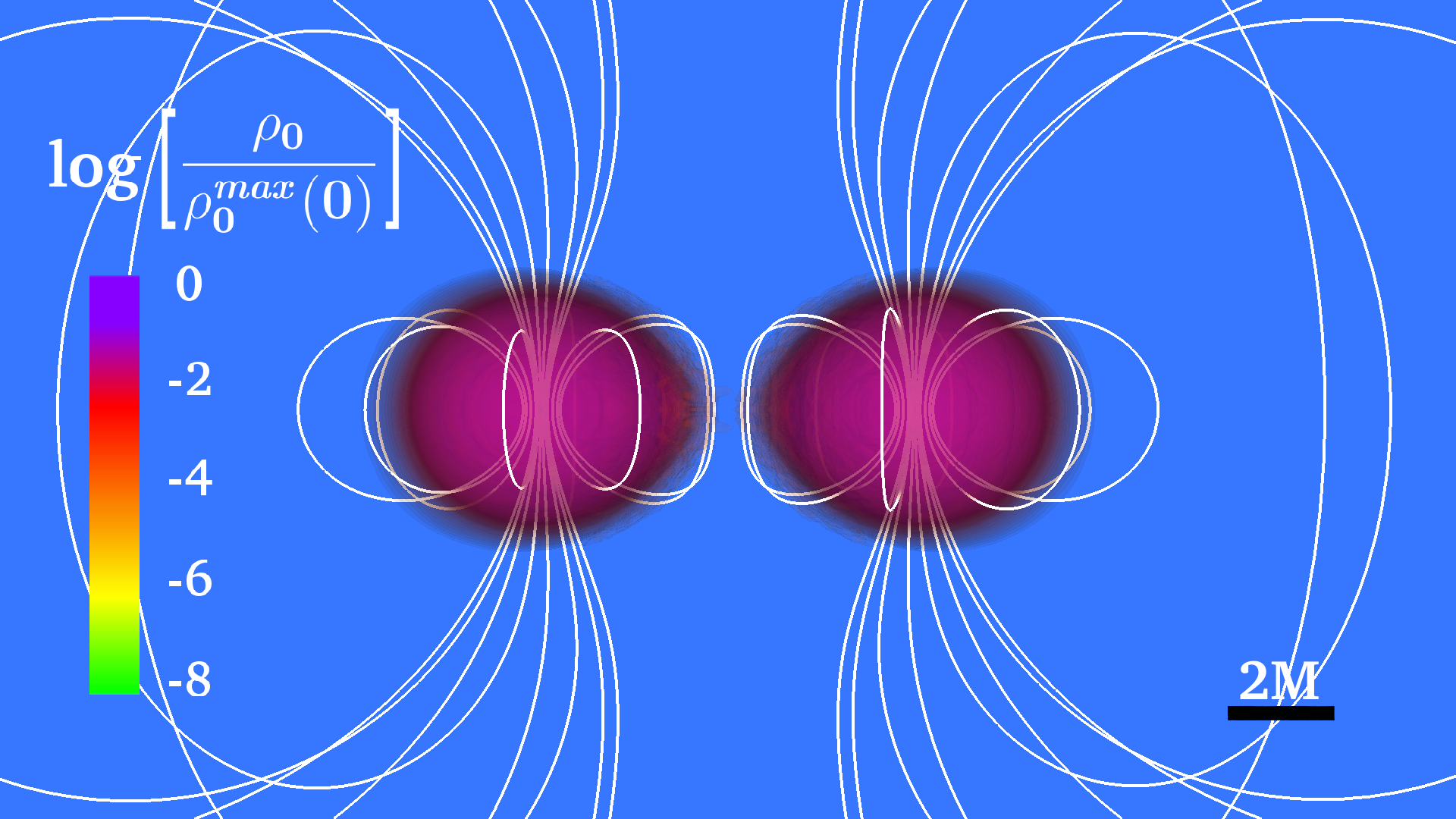} &   \includegraphics[width=0.33\textwidth]{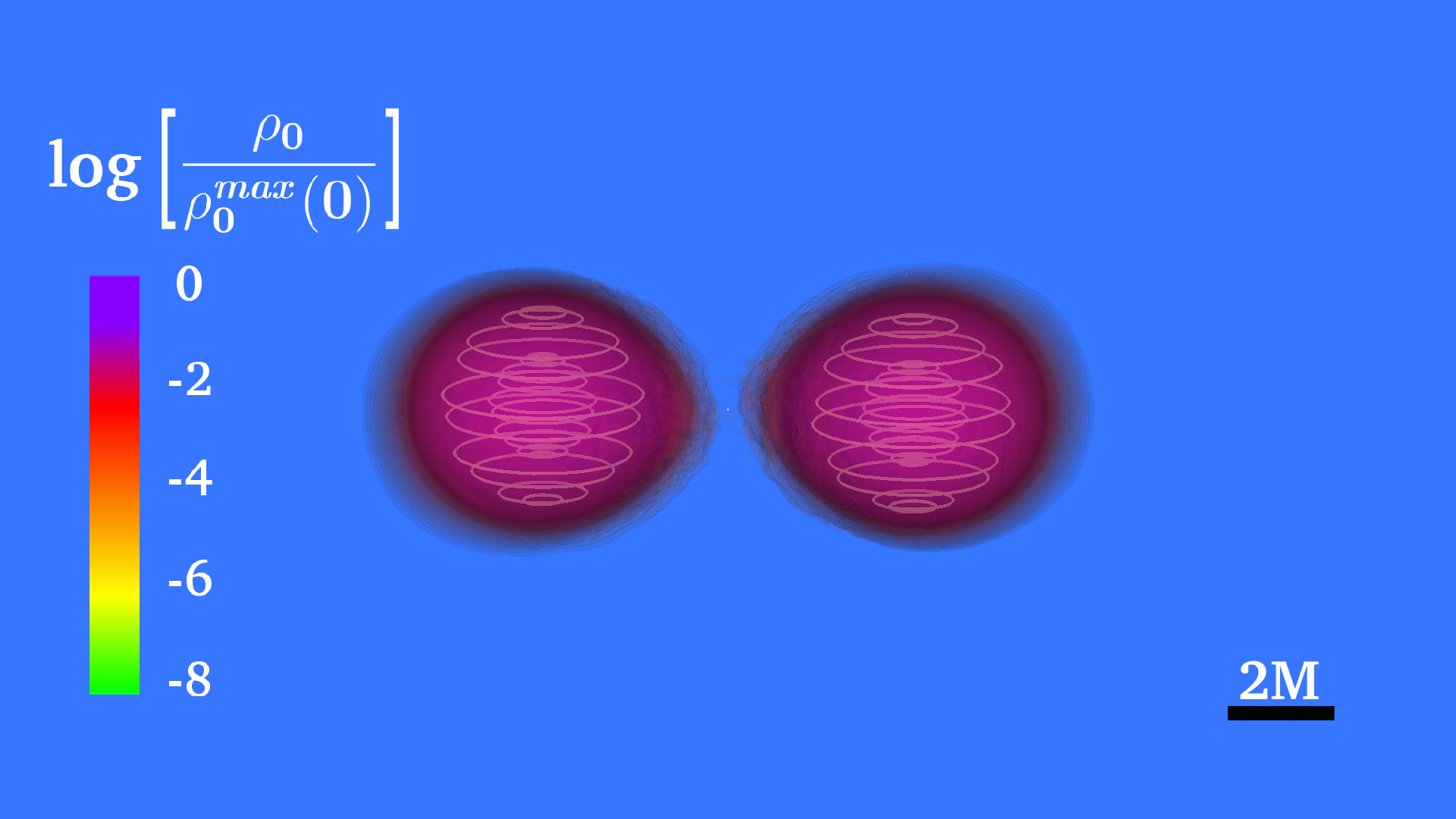} &
 \includegraphics[width=0.33\textwidth]{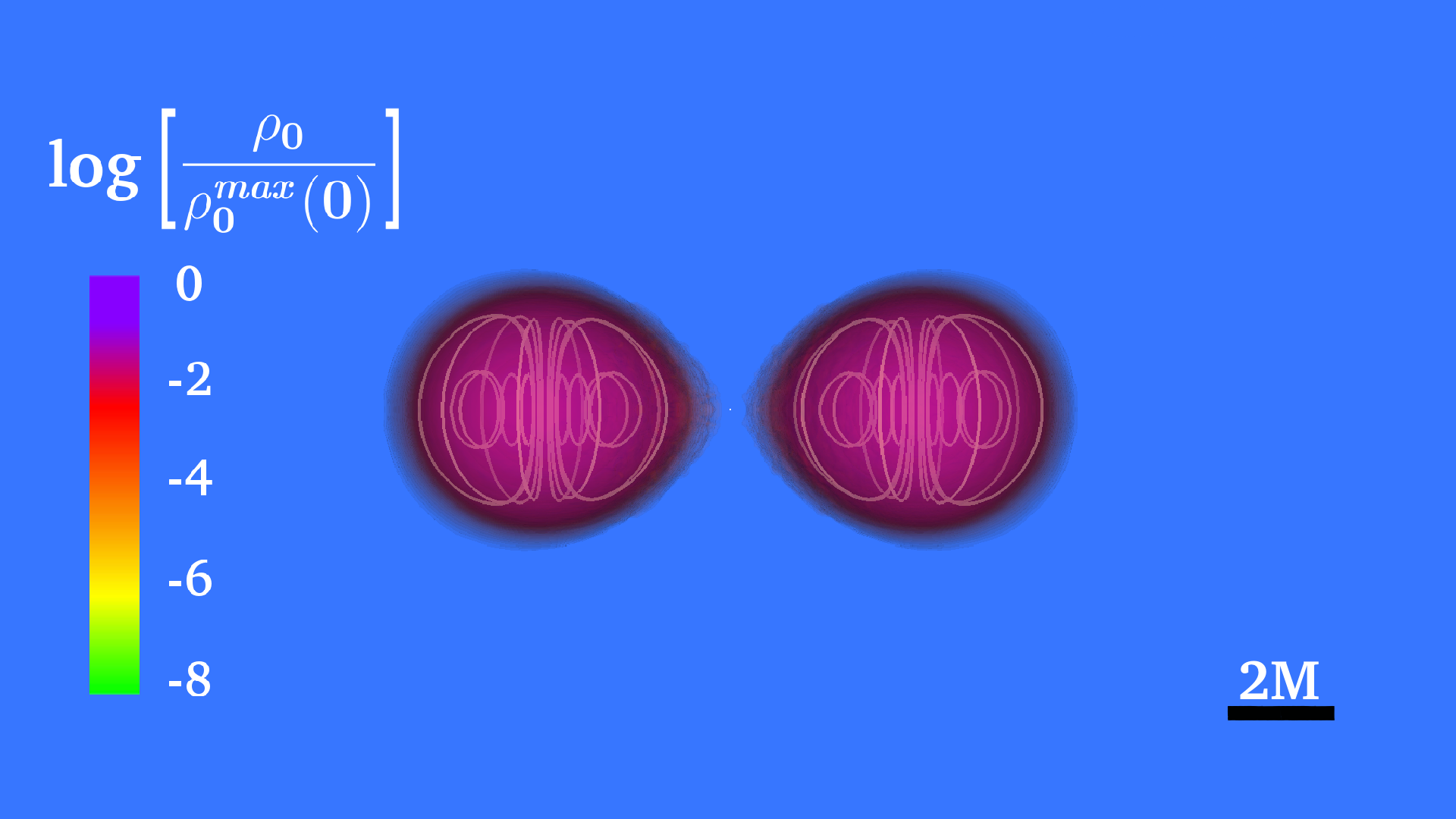} 
\end{tabular}
\caption{3D snapshots of the magnetic field lines and density isosurfaces just after magnetic field insertion for the three different magnetic field topologies: pulsar-like (left), interior only toroidal (middle) and interior only poloidal (right).
}
\label{fig:initial_topologies}
\end{figure*}

We insert the magnetic field at 1ms (approximately one orbit) before the stars merge. The short time between magnetic field insertion and merger gives us better control over the magnetic field topology in the merger regime, as we avoid magnetic winding during the inspiral phase. As the true premerger magnetic field structure inside neutron stars remains uncertain \cite{Braithwaite:2005md,Bilous:2019knh,Tsokaros:2021pkh} we test three different idealized topologies, as shown in Fig. \ref{fig:initial_topologies}. These are: 
\begin{itemize}
    \item \textbf{Pulsar-like}: a purely dipolar poloidal field that extends from the interior to the exterior of the stars, as used in \cite{Farris:2012ux,Paschalidis:2013jsa,Ruiz:2017inq,Bamber:2024kfb}, generated by the magnetic vector potential 
    \begin{equation}
        A_{\phi} = \frac{15\pi \varpi^2 I_0 r^2_0}{23(r^2_0 + r^2)^{3/2}}\left[1 + \frac{15r^2_0(r^2_0+\varpi^2)}{8(r^2_0+r^2)^2}\right],
    \end{equation}
    which approximately corresponds to that generated by a current loop inside the NS with radius $r_0$ and current $I_0$, with $\varpi^2 = (x - x_{\rm NS})^2 + (y - y_{\rm NS})^2$ and $r^2 = \varpi^2 + (z - z_{\rm NS})^2$ where $\{x_{\rm NS},y_{\rm NS},z_{\rm NS}\}$ are the coordinates of the center of the star, defined as the position of maximum rest-mass density. The maximum magnetic field, as measured by normal observers, is located at the center of the star and given by 
    \begin{equation}
        \vert B \vert_{\textup{max}} = \frac{15}{4}\frac{I_0\pi}{r_0} \Psi^{-6}.
    \end{equation}
    where $\Psi$ is the conformal factor.
    We set $r_0 = 2R_x/3$. The specific values for $I_0$ required to achieve the desired magnetic field strengths for the different cases are shown in Table \ref{tab:I0_Ab} in Appendix \ref{app:I0}.

    \item \textbf{Interior-only toroidal}: a purely toroidal field confined to the interior of the stars, using the vector potential defined in \cite{Hayashi:2022cdq} Eq. (2), given by
    \begin{equation}
    \begin{split}
        A_{\varpi} =& \;\varpi ((z-z_{\rm NS})^2 - R_x^2) A(r), \\
        A_{z} =& - (z-z_{\rm NS})(\varpi^2 - R_x^2), \\
        A_{\phi} =& \;0,
    \end{split}
    \end{equation}
    with $A(r) = A_b(1 + \cos(\pi r/(0.95R_x)))$ for $r < 0.95R_x$, $A(r)=0$ otherwise, where $A_b$ is a constant coefficient. One can show that for this potential 
    \begin{align}
        B^{\phi} =& \;A_b R_x^2 \sin(2\theta) f(\Tilde{r}) \Psi^{-6}, \\
        f(\Tilde{r}) =& \;\Tilde{r}^2\left(2 + 2\cos(\pi\Tilde{r}/a_0) + (1-\tfrac{1}{2}\Tilde{r}^2)\frac{\sin(\pi\Tilde{r}/a_0)}{\pi\Tilde{r}/a_0}\right),
    \end{align}
    where $\Tilde{r} = r/R_x$, $a_0 = 0.95$ and $\theta$ is the polar angle. The maximum value is 
    \begin{equation}
        \vert B \vert_{\mathrm{max}} \approx 3.45 A_b R_x^2 \Psi^{-6},
    \end{equation}
    for $\theta = \pi/4$ and $r \approx 0.56 R_x$.

    \item \textbf{Interior-only poloidal}: a purely dipolar poloidal field confined to the interior of the stars, as used in our previous works, generated by the vector potential defined in \cite{Etienne:2011ea} Eq. (12) given by 
    \begin{equation}
    A_{\phi} = A_b \varpi^2 \textup{max}((P- P_{\textup{cut}})/P_{\textup{max}},0)^{2},
    \end{equation}
    where $P_{\textup{cut}} = 0.01 P_{\textup{max}}$ is the cut-off pressure defining the star's surface, with a maximum value of the magnetic field $\vert B \vert_{\textup{max}} \approx 2 A_b \Psi^{-6}$ at the center of the star. For both interior-only topologies the specific values of $A_b$ required to achieve the desired magnetic field strengths are shown in Table \ref{tab:I0_Ab} in Appendix \ref{app:I0}.
    
\end{itemize}

While pure poloidal and toroidal magnetic fields are known to be unstable on Alfv\'{e}n timescales \cite{Ciolfi:2011xa,Lasky:2012ju}, these timescales are $\gtrsim 1$ms for the cases we consider here, even for the strongest magnetic field used. For each EOS and binary ADM mass we conducted four simulations with pulsar-like magnetic fields with initial maximum field strength at insertion of $\vert B \vert_{\textup{max},\textup{insert}} = 5.5\times 10^{15}\textup{G},\;1.4\times 10^{16}\textup{G},\;5.5\times 10^{16}\textup{G}$ and $2.2\times 10^{17}\textup{G}$, along with one interior-only toroidal case and one interior-only poloidal case both with initial $\vert B \vert_{\textup{max},\textup{insert}} = 5.5\times 10^{16}\textup{G}$, the canonical field strength used in our previous study \cite{Bamber:2024kfb}. Including the hydrodynamics-only $\vert B \vert_{\textup{max},\textup{insert}} = 0$ case, this makes seven different magnetic fields in total. Note that the ratio of total electromagnetic energy $E_{\textup{mag}}$ to gravitational binding energy $W$ is always $< 10^{-3}$ for all cases, so the magnetic field insertion has an insignificant effect on the initial stellar structure \cite{Leung:2022mvm}. 

The maximum magnetic field strength as measured by observers comoving with the fluid, $\vert B \vert^{\textup{co}}_{\textup{max},\textup{insert}}$ is related to $\vert B \vert_{\textup{max},\textup{insert}}$ as $\vert B \vert^{\textup{co}}_{\textup{max},\textup{insert}} \approx 1.5 \vert B \vert_{\textup{max},\textup{insert}}$. The volume-averaged magnetic field inside the stars is roughly an order of magnitude lower than the maximum: $\langle \vert B \vert \rangle \sim 0.1 \vert B \vert_{\textup{max}}$ for the pulsar-like topology and $\langle \vert B \vert \rangle \sim (0.4-0.5) \vert B \vert_{\textup{max}}$ for the interior-only topologies. For the pulsar-like fields the initial magnetic field strength at the neutron star poles is $B_{\textup{pole}} \approx 0.01\vert B \vert_{\textup{max},\textup{insert}}$ and $\approx 0.03\vert B \vert_{\textup{max},\textup{insert}}$ for the stars with the ALF2 and SLy EOS respectively, corresponding to values between $10^{13.8}$ to $10^{15.8}$G. This is significantly larger than the expected surface magnetic field strengths inferred for pulsars observed in binary systems \cite{Tauris:2017omb,Lorimer:2008se}. However, as discussed in our previous works \cite{Bamber:2024kfb,Sun:2022vri,Ruiz:2021qmm}, after the stars collide, magnetic instabilities, initially driven by the Kelvin-Helmholtz Instability (KHI) at the shearing surface can significantly boost the magnetic field. Special relativistic and approximate-GR simulations have shown that the field can be amplified to magnetar levels of $\gtrsim 10^{15}$G as the stretching and folding of the magnetic field lines converts kinetic energy to magnetic \cite{Price:2006fi,Rosswog:2006ue,Obergaulinger:2010gf,Oechslin:2006uk,Zrake:2013mra}. The magnetorotational instability (MRI), magnetic winding, an $\alpha \Omega$ dynamo \cite{Kiuchi:2023obe} and potentially the Rayleigh-Taylor instability \cite{Skoutnev:2021chg} also work to boost the magnetic field. Very high resolution GRMHD simulations have also shown similar results, as in $\Delta x_{\textup{min}} = 17.5$m simulation of Kiuchi et al. (2015) \cite{Kiuchi:2015sga} where an initial field strength of $10^{13}$G is amplified to values of $\gtrsim 10^{16}$G in the core of the remnant, with Kiuchi et al. (2024) \cite{Kiuchi:2023obe} reporting the subsequent generation of a large-scale poloidal field due to a combination the MRI and an $\alpha\Omega$ dynamo mechanism. 

Unfortunately, the electromagnetic energy growth rate due to the KHI in the linear regime is inversely proportional to the minimum resolvable wavelength, and thus to the numerical resolution, at least down to a resolution $\Delta x_{\textup{min}} \sim 12.5$m \cite{Price:2006fi,Kiuchi:2023obe} (see Section \ref{subsec:M_processes}). Conducting GRMHD simulations at ultra-high $\Delta x_{\textup{min}} < 10$m resolutions in order to achieve asymptotic convergence (such that the magnetic field doesn't change on increasing the resolution further) is computationally prohibitive. This leads to two different approaches to try to model the magnetic fields in neutron star mergers. First, one can take the simpler approach used our past works \cite{Bamber:2024kfb,Ruiz:2016rai,Ruiz:2021qmm} and in this work
and compensate for the missing amplification by starting with an artificially strong magnetic field, as is common practice in the field ~\cite{Ciolfi:2019fie,Musolino:2024sju}. 

The second, alternative, approach is to use Large-Eddy-Simulations (LESs) and subgrid-scale models as discussed in \cite{Giacomazzo:2014qba,Aguilera-Miret:2020dhz,Palenzuela:2021gdo,Aguilera-Miret:2021fre,Aguilera-Miret:2023qih,Most:2023sme,Izquierdo:2024rbb,Aguilera-Miret:2024cor}. Proponents of such schemes (e.g. \cite{Aguilera-Miret:2024cor,Aguilera-Miret:2023qih}) have strongly argued that using initial artificially strong large-scale magnetic fields is unrealistic, as the initial structure of the large scale field may bias the results. However, this assessment was done by comparison to the results of higher resolution LESs \cite{Aguilera-Miret:2024cor}. The problem is that such simulations themselves depend on the details of the subgrid model used, with no guarantee that they necessarily reflect the behavior of a real merger or a standard simulation with sufficiently high resolution to achieve asymptotic convergence. 

By using a range of magnetic field strengths in the regime where the MRI is still resolved (as the MRI wavelength is proportional to the poloidal magnetic field strength $\vert b^P \vert$) we can explore the ``best case" scenario for a signature of the magnetic field on the GW waveform, which can be extrapolated to smaller field strengths if necessary. 

\subsection{Grid structure.}

\label{subsec:Grid}
The grid structure used here is identical to that of \cite{Bamber:2024kfb}. We use the ``moving-box" approach, with two sets of nested grids centered on each star. There are nine refinement levels of nested grids differing in size and resolution by a factor of two, plus the coarsest level which covers the whole simulation box. The simulation box is a half-cube (using equatorial symmetry across the $xy$ plane) of spatial extent $L_0/2 \simeq 5748\textup{km} \simeq 3891 M_{\odot}$, where $L_0$ is the total width, and grid spacing 
$\Delta x_0 \simeq 46\textup{km} \simeq 31 M_{\odot}$, so that each subsequent level has half-width 
$L_n/2 \simeq 5748/2^n \textup{km}$
and grid spacing  
$\Delta x_n \simeq 46/2^n \textup{km}$
for $n = 1,2 \dots 9$. The maximum resolution is 
$\Delta x_9 = \Delta x_{\textup{min}} \simeq 90\textup{m}$. The number of grid points covering the equatorial diameter of the NS, denoted $N_{\rm NS}$, is then between $193$ and $224$ for the most and least compact cases, respectively. We use the same grid topology for all the cases. When two grid boxes overlap they are replaced by a combined box centered on the center of mass of the system. 

\subsection{Diagnostics}
\label{subsec:diagnostic}

\begin{figure}
    \centering
    \includegraphics[width=\linewidth]{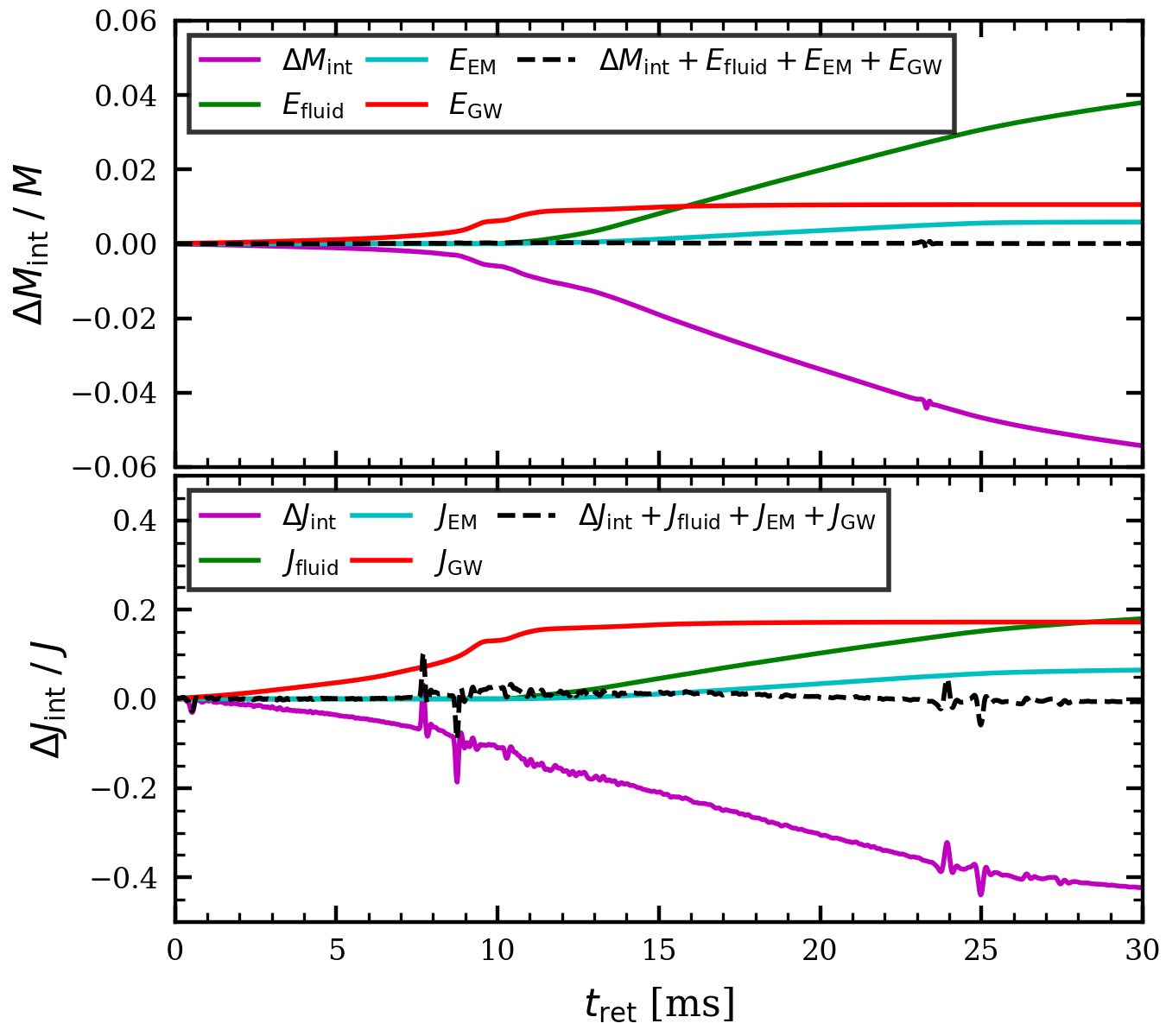}
    \caption{Conservation of the mass and angular momentum integrals $M_{\textup{int}}$ and $J_{\textup{int}}$ for the ALF2 $M=2.57M_{\odot}$, $\vert B \vert_{\textup{max},\textup{insert}} = 2.2\times 10^{17}$ case (corresponding results are found for the other cases) vs, retarded time $t_{\textup{ret}} := t - r_*$, where $r_*$ is the extraction radius in tortoise coordinates $r_* = r + 2M\ln(r/(2M)-1)$. The magenta lines denote the change in the interior $M$ and $J$ surface integrals, Eqs. (37) and (39) in \cite{Etienne:2007jg}, evaluated on spherical surfaces at $r_{\textup{ext}}\simeq 273M$. The cyan line denotes the total EM energy (upper plot) and $z$ component of the total EM angular momentum (lower plot) that has passed across that surface. The green and red lines do the same for the fluid and GWs respectively. The black dashed line gives the change in the total energy in the top panel, and total angular momentum in the bottom panel, correcting for the lost energy and angular momentum. The black dashed lines in both plots remain close to zero, showing that both total mass and total angular momentum are well conserved.}
    \label{fig:MJ_conservation}
\end{figure}

We extract the GW signal by computing the Weyl scalar $\Psi_4$ using the \texttt{Psikadelia} thorn, then decompose it into $s = -2$ spin-weighted spherical harmonic modes extracted over spherical surfaces at seven different extraction radii between $120M_{\odot}$ and $840M_{\odot}$. We then convert these values to $h_{+/\times}$ strain polarizations using the relation
\begin{equation}
    \Psi_4 = \ddot{h}_{+} - i \ddot{h}_{\times},
\end{equation}
valid for outgoing waves far from the source in the tranverse-traceless gauge. We also compute the energy and angular momentum flux radiated away in GWs (for further details see \cite{Ruiz:2007yx}). The GW luminosity can be obtained from the $\Psi_4$ Weyl scalar as 
\begin{equation}
    L_{\textup{GW}} = \lim_{r\rightarrow\infty}\frac{r^2}{16\pi}\int \left\vert\int^t_{-\infty}\Psi_4 \dd t'\right\vert^2 \dd \Omega, 
\end{equation}
which we approximate via a surface integral at a finite radius in the wave zone. We also calculate the fluid thermal and kinetic energy luminosity \cite{Gottlieb:2023est}
\begin{equation}
L_{\textup{fluid}} = \int \sqrt{-g}\,(-T^{i\;(\textup{fluid})}_t - \rho_0 u^i)\,\dd S_i\,,
\label{eq:fel}
\end{equation}
(where in Eq. \eqref{eq:fel} we subtract the contribution from the rest-mass-energy flux), and the EM Poynting luminosity 
\begin{equation}
L_{\textup{EM}} = - \int \sqrt{-g}\,T^{i\;(\textup{EM})}_t \dd S_i\ , 
\label{eq:pl}
\end{equation}
over ten spherical radii between $\simeq 57$km and $\simeq 4320$km, or approximately $15M$ and $1200M$, as well as the corresponding fluxes over the outer surface of the simulation domain (note that we do not calculate thermal EM emission from the gaseous debris). The conservation of total mass and angular momentum can be confirmed by computing $M_{\textup{int}}$ and $J_{\textup{int}}$ (defined via Eqs. (37) and (39) in \cite{Etienne:2007jg} with an integral over a spherical surface of finite radius) which correspond to the ADM mass and the $z$ component of the ADM angular momentum respectively when evaluated at spatial infinity ($r=\infty$), and comparing the change in these integrals with the loss of energy and angular momentum via fluid outflow, EM and GW emission. Consistent with \cite{Tsokaros:2019anx,Ruiz:2019ezy,Bamber:2024kfb} we find that mass is conserved to $< 1\%$ and angular momentum is conserved to $< 2\%$, as shown in Fig. \ref{fig:MJ_conservation}. 

For cases where the remnant collapses to a BH we use the \verb|AHFinderDirect| thorn \cite{Thornburg:2003sf} to track the apparent horizon (AH) and estimate the BH mass and dimensionless spin using the isolated horizon formalism of \cite{Dreyer:2002mx}. The total electromagnetic energy measured by a normal observer is 
\begin{equation}
  E_{\rm mag} = \int n^{\mu} n^{\nu}T^{\textup{EM}}_{\mu\nu} \dd V = \int \frac{1}{8\pi}\gamma_{ij}(E^iE^j + B^i B^j) \dd V.
\end{equation}
We can decompose this into toroidal and poloidal components by writing the total EM energy in terms of the magnetic field only using the ideal MHD condition
\begin{equation}
  E_{\rm mag} = \int n^{\mu} n^{\nu}T^{\textup{EM}}_{\mu\nu} \dd V = \int \frac{1}{8\pi}q_{ij}B^i B^j \dd V,
\end{equation}
where 
\begin{align}
    q_{ij} =& \gamma_{ij}(1 + w_k w^k) - w_i w_j, \\
    w^i =& (v^i + \beta^i)/\alpha.
\end{align}
We can then split $B^i$ into 
\begin{equation}
    B^i = B_{\textup{tor}} \hat{e}^i_{\textup{tor}} + B_{\textup{pol}} \hat{e}^i_{\textup{pol}}, 
\end{equation}
where $\hat{e}^i_{\textup{tor}},\hat{e}^i_{\textup{pol}}$ are orthonormal 3-vectors such that $q_{ij}\hat{e}^i_{\textup{tor}}\hat{e}^j_{\textup{tor}} = q_{ij}\hat{e}^i_{\textup{pol}}\hat{e}^j_{\textup{pol}} = 1$, $q_{ij}\hat{e}^i_{\textup{tor}}\hat{e}^j_{\textup{pol}} = 0$ and $\hat{e}^i_{\textup{tor}} \propto (-y,x,0)$ in Cartesian coordinates. Then 
\begin{align}
  E^{\rm tor}_{\rm mag} =& \int \frac{1}{8\pi}B^2_{\textup{tor}} \dd V, \\
  E^{\rm pol}_{\rm mag} =& \int \frac{1}{8\pi}B^2_{\textup{pol}} \dd V, \\
  E_{\rm mag} =& \;E^{\rm tor}_{\rm mag} + E^{\rm pol}_{\rm mag}. 
\end{align}

We monitor the outflow of matter by computing the unbound rest mass outside a radius $r_0$ respectively as 
\begin{equation}
\begin{split}
    M_{\textup{esc}} =& \int_{r>r_0} \rho_{*} \Theta(-u_t-1)\Theta(v^r)\dd x^3 \\
    &+ \int^t_{t'=0}\int_{\delta\mathcal{D}}\rho_{*} \Theta(-u_t-1)\Theta(v^r)v^i\dd S_i \dd t',
    \label{eq:esc_mass}
\end{split}
\end{equation}
where $\rho_* = \sqrt{-g}\rho_0 u^t$, $\dd S_i$ is the surface element on the sphere, and the Heaviside functions $\Theta$ ensure we only include material with a positive specific energy $E = -u_t - 1$ (i.e. unbound material) and a positive radial velocity. Note that we also add in the contribution from the rest mass leaving the boundary of the simulation domain, $\delta \mathcal{D}$, although this contribution remains negligible by the end of the simulation. Here $\rho_0$ is the rest-mass density, $g$ the determinant of the 4-metric, $u^{\mu}$ the four-velocity of the fluid and $v^r$ the radial component of the three-velocity. We measure at radii $r_0 = 30M,50M,70M,100M$ and confirm that the difference between them is less than $\lesssim 4.5\%$.

We also measure the rest mass and angular momentum of the NSs/NS remnant, defined via
\begin{align}
  M_0 =& \int_{\rho_0/\rho^{\textup{max}}_{0,t=0}>10^{-3}} \rho_{*} \dd x^3, \label{eq:M0_def} \\
  J =& \int_{\rho_0/\rho^{\textup{max}}_{0,t=0}>10^{-3}} \rho_{*} h u_{\phi}\dd x^3, \label{eq:J_def}
\end{align}
where $h$ is the specific enthalpy and $u_{\phi}$ the azimuthal component of the covariant 4-velocity and the integral is taken over the volume where the rest-mass density is greater than $10^{-3}\rho^{\textup{max}}_{0,t=0}$ where $\rho^{\textup{max}}_{0,t=0}$ is the maximum rest-mass density at the start of the simulation. 

To estimate the effective viscosity produced by the magnetic field in the binary remnant we calculate the Shakura–Sunyaev viscosity parameter \cite{Shakura:1973boa,Shafee:2008mm} 
\begin{equation}
\alpha_{\textup{SS}} \sim \frac{\textup{magnetic stress}}{\textup{pressure}} =
\frac{T^{\textup{EM}}_{\hat{r}\hat{\phi}}}{P}
\label{eq:alphass}  
\end{equation}
where $T^{EM}_{\hat{r}\hat{\phi}} = e_{\hat{r}}^{\ \mu}e_{\hat{\phi}}^{\ \nu}T_{\mu\nu}^{\textup{EM}}$ is the orthonormal $r\phi$ component of the electromagnetic stress-energy tensor in the local comoving frame and $e_{\hat{i}}^{\ \mu}$ the corresponding basis of local tetrads (see Eq. (26) in \cite{Penna:2010hu}).

\subsection{NSNS postmerger oscillation modes}

The general understanding of the postmerger GW signal, at least from hydrodynamics-only NSNS merger simulations, is as follows. There is a strong consensus that the dominant peak in the postmerger GW power spectrum corresponds to the frequency of the fundamental quadrupolar $l=m=2$ mode in the remnant \cite{Stergioulas:2011gd,Bauswein:2015vxa,Takami:2014tva,Rezzolla:2016nxn,Vretinaris:2019spn,Soultanis:2021oia}, denoted $f_{\textup{peak}}$ or $f_2$, generically excited in all mergers that do not produce prompt collapse to a BH \cite{Vretinaris:2019spn,Shibata:2005ss}. In linear perturbation theory this mode corresponds to the $f$-mode, the fundamental mode of the pressure mode sequence \cite{Vretinaris:2019spn,Kruger:2019zuz}. There is less agreement on the nature of the sub-dominant peaks. In \cite{Stergioulas:2011gd} two side peaks were identified, denoted $f_{2-0}$, $f_{2+0}$ in \cite{Vretinaris:2019spn}, as arising from quasi-linear combinations of the fundamental quadrupolar mode $f_2$ and the fundamental quasi-radial mode $f_0$ such that $f_{2\pm0} \sim f_2 \pm f_0$, forming a triplet with $f_2$. In addition, they identify an additional peak, with a frequency denoted $f_{\textup{spiral}}$ between $f_{2-0}$ and $f_2$, arising from spiral tidal bulges in the outer parts of the remnant that rotate more slowly than the inner core \cite{Bauswein:2015yca,Vretinaris:2019spn,Soultanis:2021oia}. In contrast Refs. \cite{Takami:2014tva,Takami:2014zpa,Rezzolla:2016nxn} identify a triplet of peaks, denoted $f_1,f_2,f_3$ where $f_2$ is again the dominant fundamental quadrupolar mode, and $f_2 \sim (f_1 + f_3)/2$. However, they characterise $f_1$ as distinct from the $f_{2-0}$ mode arising from a quasi-linear combination of the quadrupolar and quasi-radial modes \cite{Rezzolla:2016nxn}. In this work we follow the notation of $f_{2-0},f_{\textup{spiral}},f_2,f_{2+0}$ used in \cite{Vretinaris:2019spn}. 

For a cold non-rotating star it has been long established that the $f$-mode frequency scales approximately with the square root mean density $\sim \bar{\rho}^{1/2} \sim \sqrt{M_{\mathrm{star}}/R^3}$ \cite{Andersson:1997rn,Chakravarti:2019sdc} (see \cite{Kruger:2019zuz} for analysis of the quasinormal mode spectrum of uniformly rotating neutron stars). 
%Yip et al. (2024) \cite{Yip:2024kky} studied the fundamental mode frequencies of differentially rotating stars with 2D axisymmetric simulations, finding that both $f_0$ and $f_2$ scale almost linearly with stellar compactness $M_{\mathrm{star}}/R$. 

For a very close GW event at a polar distance of 20Mpc, Advanced LIGO has been estimated to be able to measure $f_2$ to within uncertainties of $50$Hz to at most $160$Hz \cite{Bauswein:2011tp}, while third generation detectors have been estimated to be able to constrain $f_2$ to an average of $\delta f_2 \sim 138$Hz for sources at the detector horizon \cite{Clark:2015zxa}. Stacking multiple events may further reduce this to $\delta f_2 \sim 4 - 20$Hz \cite{Yang:2017xlf}. 

The characteristic frequencies can shift with time, as cooling of the remnant, the loss of mass, and the redistribution of angular momentum leads to a change in the stellar structure. Rezzolla and Takami \cite{Rezzolla:2016nxn} used short-time Fourier transforms to identify an initial short-lived transient regime, up till $\sim 3$ms postmerger, where the dominant quadrupolar mode has a different frequency $f_{2,i}$, before it evolves to the quasi-stationary regime with frequency $f_2$. Soultanis, Bauswein \& Stergioulas \cite{Soultanis:2021oia} used a continuous wavelet decomposition, also identifying an evolution of the dominant $f_2$ mode from an transient regime before $\sim 6$ms post merger to a subsequent quasi-stationary regime. 

\subsection{Magnetic processes}
\label{subsec:M_processes}

As described in e.g. \cite{Bamber:2024kfb} Section I E the presence of the magnetic field produces several important effects in the evolution of the NS remnant. In a highly conducting plasma where magnetic field lines are ``frozen-in", winding, stretching and folding of the field lines due to shear and compression, e.g. through differential rotation or turbulent motion, can significantly amplify the field as the rotational kinetic energy of the remnant is converted to electromagnetic energy~\cite{Duez:2006qe,Brandenburg:2004jv,Shapiro:2000zh} while producing a redistribution of angular momentum and additional effective viscosity.

\subsubsection{Kelvin-Helmholtz and Rayleigh-Taylor instabilities}

Immediately postmerger the strongest mechanism for magnetic amplification \cite{Kiuchi:2015sga,Kiuchi:2017zzg} is the Kelvin-Helmholtz instability (KHI) \cite{Chandrasekhar:1961} occuring at the shear interface that develops for a non-synchronized binary \cite{Rasio:1999ku} when the stars come into contact. A simple Newtonian linear stability analysis suggests small perturbations should grow as $\propto \exp(\sigma_{\textup{KH}}(t - t_{\textup{merge}}))$ \cite{Chandrasekhar:1961,Schekochihin:2005rd} where $\sigma_{\textup{KH}}$ is inversely proportional to the wavelength. For a shear layer of finite width $d$, unstable modes have wavelength $\lambda \gtrsim d$. However, in a numerical simulation the minimum wavelength and shear layer thickness are both limited by the resolution with $\lambda,d \gtrsim \Delta x_{\textup{min}}$ where $\Delta x_{\textup{min}}$ is the resolution, leading Price and Rosswog \cite{Price:2006fi} to suggest and Kiuchi et al. \cite{Kiuchi:2023obe} to find numerically that the KHI growth rate is inversely proportional to $\Delta x_{\textup{min}}$ down to $\Delta  x_{\textup{min}} \sim 12.5$m. The total electromagnetic energy also grows exponentially, $\propto \exp(2\gamma_{\textup{KH}}(t-t_{\textup{merge}}))$, with some characteristic growth rate $\gamma_{\textup{KH}}$ which can in principle be very different from $\sigma_{\textup{KH}}$ \cite{Rincon:2019coh}. Growth terminates either when the magnetic field saturates, with near equipartition between magnetic and turbulent kinetic energy \cite{Schekochihin:2005rd,Rincon:2019coh,Skoutnev:2021chg}, or when the shear layer is destroyed \cite{Kiuchi:2015sga,Kiuchi:2023obe}. The Rayleigh-Taylor instability (RTI), which occurs when the density gradient between two fluid layers is misaligned with the local gravitational field, has also been proposed as a source of turbulence and therefore magnetic field amplification in the outer regions of the remnant \cite{Skoutnev:2021chg}, and which may complement the KHI-induced amplification, which is strongest in the core \cite{Palenzuela:2021gdo}. 
\begin{figure}
    \centering
    \begin{tabular}{c}
    \includegraphics[width=\linewidth]{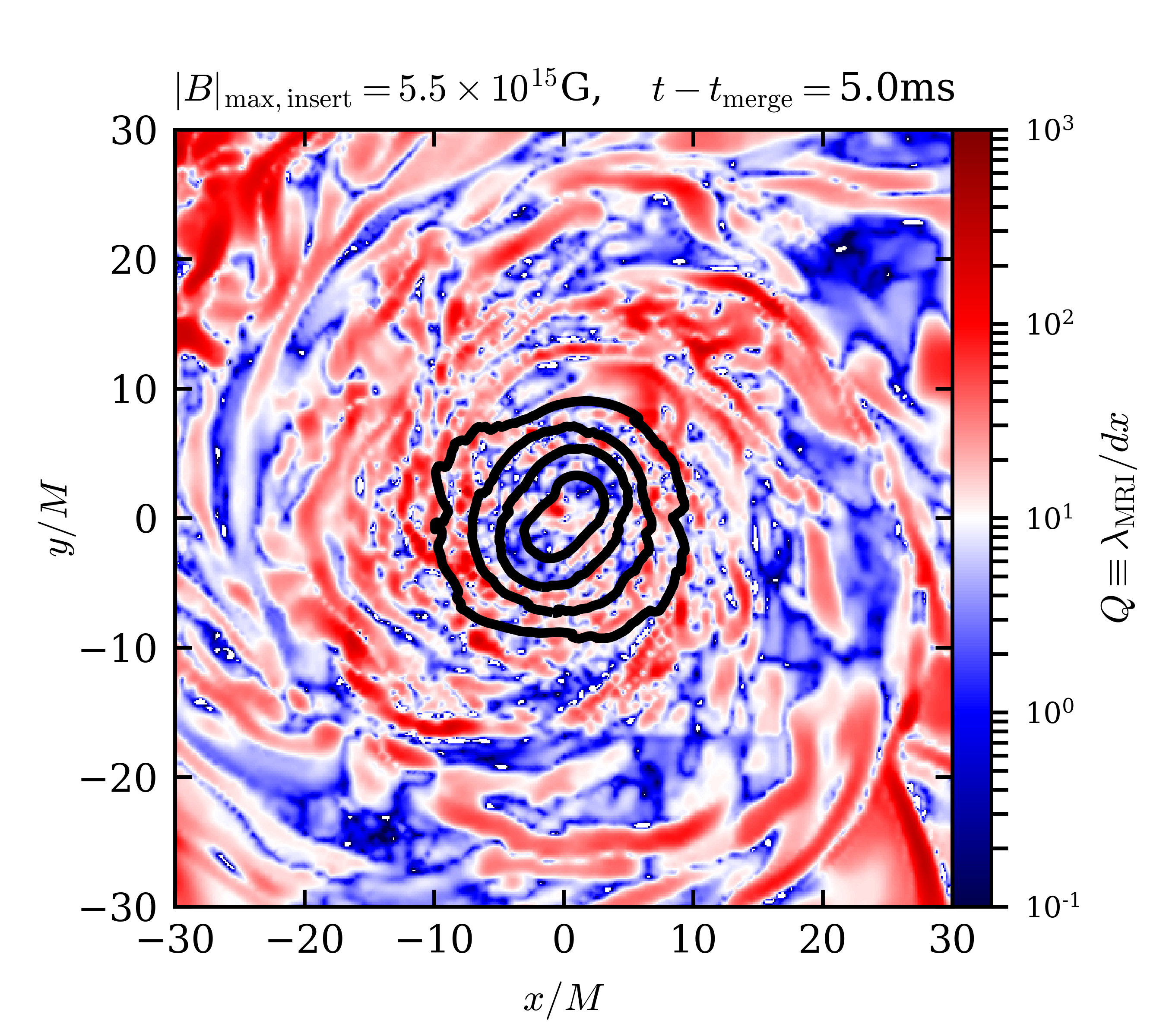} \\
    \includegraphics[width=\linewidth]{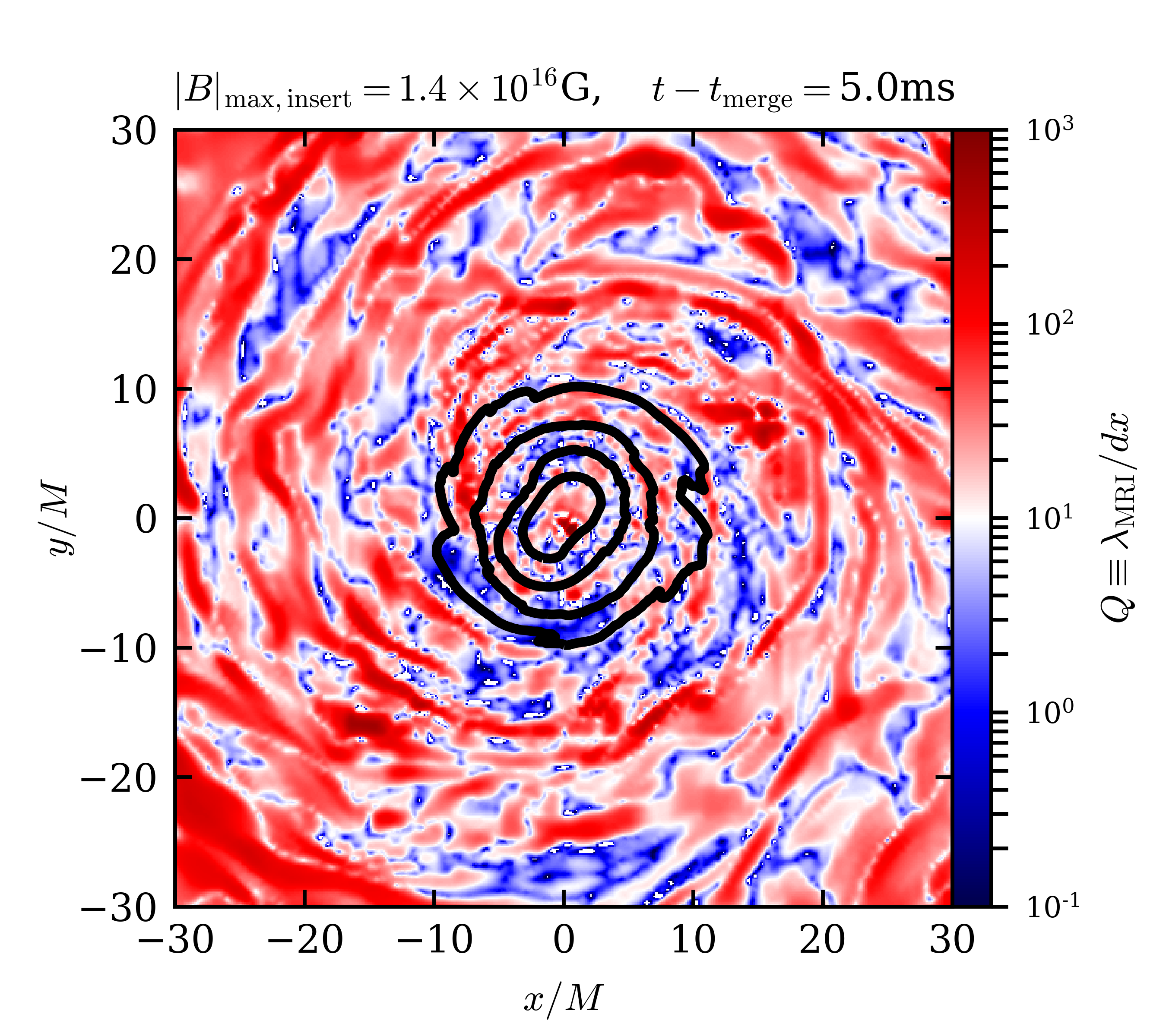}
    \end{tabular}
    \caption{Quality factor $Q \equiv \lambda_{\textup{MRI}}/dx$ on the equatorial plane for the SLy, $M=2.57M_{\odot}$ cases with the smallest initial magnetic fields (the equivalent plots for the ALF2 cases with the same magnetic fields look very similar). The thick black lines show the rest-mass density contours $\rho_0/\rho^{\textup{max}}_{0,t=0} = 0.01,0.1,0.5,1.0,2.0.$}
    \label{fig:MRI_xy}
\end{figure}

\subsubsection{Magnetorotational instability}

At around $\sim 5$ms after the merger, when the KHI amplification terminates~\cite{Kiuchi:2014hja}, other mechanisms, such as the MRI and magnetic winding and braking, gradually take over. The MRI \cite{Balbus:1991ay,Balbus:1998ja} occurs in any magnetized rotating astrophysical fluid whenever the angular velocity $\Omega$ decreases with radius $\partial_{\varpi}\Omega < 0$, where $\varpi$ is the cylindrical radius. Again, initial exponential growth transitions to a non-linear regime and then saturates, generating turbulence and boosting the magnetic field via a dynamo mechanism \cite{Duez:2006qe,Shibata:2006hr,Siegel:2013nrw,Kiuchi:2023obe} while transporting angular momentum from the inner to outer layers of the binary remnant, which induces the formation of a central core surrounded by a Keplerian disk. 
The MRI growth rate as well as its fastest-growing wavelength are \cite{Balbus:1998ja,Shibata:2006hr,Etienne:2012te}
\begin{align}
    \sigma_{\textup{MRI}} =& \;\tfrac{1}{2} \pdv{\Omega}{\ln \varpi} \ , \label{eq:MRIsigma} \\
    \lambda_{\textup{MRI}} \approx& \frac{2\pi v_A}{\Omega} \approx  \frac{2\pi\sqrt{|b^P|^2/(b^2 + \rho_0 h)}}{\Omega} \ ,
    \label{eq:MRIlambda}
\end{align}
where $v_A$ is the Alfv\'{e}n speed, $|b^P|=\sqrt{b^\mu b_\mu - (b_\mu (e_{\hat{\phi}})^\mu)^2}$, $b^{\mu} = B^{\mu}_{\textup{co}}/\sqrt{4\pi}$ where $B^{\mu}_{\textup{co}}$ is the magnetic field in the reference frame comoving with the fluid and
$(e_{\hat{\phi}})^\mu$ is the toroidal orthonormal vector in the frame comoving with the fluid.
For a Keplerian distribution, $\Omega \propto \varpi^{-3/2}$, Eqs. (\ref{eq:MRIsigma}), (\ref{eq:MRIlambda}) 
give
\begin{align}
    \sigma_{\textup{MRI}} \sim& \;\tfrac{3}{4}\Omega = 
         1.0 \;\textup{ms}^{-1} \; \left(\tfrac{10^{3}\textup{rad s}^{-1}}{\Omega}\right), \\
    \lambda_{\textup{MRI}} \sim& \;2\textup{km}\; 
                   \left(\tfrac{10^{3}\textup{rad s}^{-1}}{\Omega}\right)
                   \left(\tfrac{b^P}{10^{15}\textup{G}}\right)
                   \left(\tfrac{\rho_0}{10^{15}\textup{g\,cm}^{-3}}\right)^{-1/2}.
\end{align}
To monitor whether we can resolve the MRI we calculate the MRI-quality factor $Q_{\textup{MRI}} := \lambda_{\textup{MRI}}/\Delta x$, where $\Delta x$ is the local grid spacing; thus $Q_{\textup{MRI}}$ which measures the number of grid points per wavelength of the faster growing MRI mode \cite{Ruiz:2021qmm}. Previous works \cite{Sano:2003bf,Shiokawa:2011ih} suggest we need $Q_{\textup{MRI}} \gtrsim 10$ to properly capture the instability, and the local height of the remnant must be $>\lambda_{\textup{MRI}}$ for the instability to be active. In Fig. \ref{fig:MRI_xy} we show the quality factor in the $xy$ plane at $5$ms after merger for the SLy $M=2.57M_{\odot}$, $\vert B \vert_{\textup{max},\textup{insert}}=5.5\times 10^{15}$G (top) and $\vert B \vert_{\textup{max},\textup{insert}}=1.4\times 10^{16}$G (bottom). In the top panel we can see that only about half the remnant and its disk are red, indicating $Q>10$, so the MRI is only well resolved in about half the region of interest. 
For this reason $\vert B \vert_{\textup{max},\textup{insert}}=5.5\times 10^{15}$G is the smallest initial magnetic field we examine, since to resolve the MRI at all with weaker magnetic fields we would need much higher resolution which would be prohibitively costly. The bottom panel shows the next highest initial magnetic field of $\vert B \vert_{\textup{max},\textup{insert}}=1.4\times 10^{16}$G. Here most of the plot is red, indicating that for this case we can resolve the MRI across most of the remnant and disk.

\begin{figure}
    \centering
    \includegraphics[width=\linewidth]{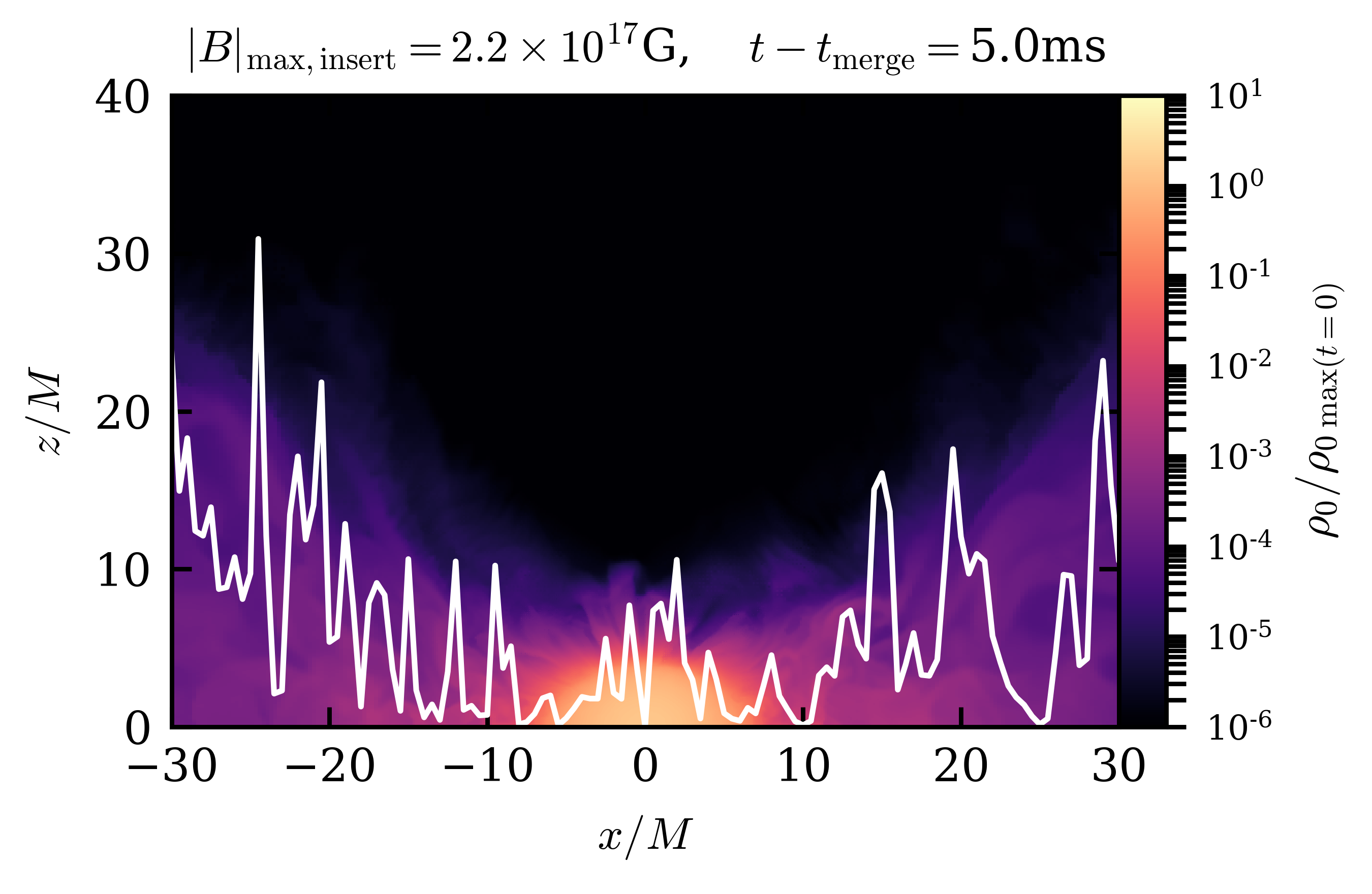}
    \caption{Rest-mass density normalised to its initial maximum value on a log scale along with $\lambda_{\textup{MRI}}$ (white line) on the meridional plane for the SLy, $M=2.57M_{\odot}$ with the largest initial magnetic field. The height of the line above the $x$ axis indicates the value of $\lambda_{\textup{MRI}}$ along the $x$-axis.}
    \label{fig:MRI_xz}
\end{figure}

In Fig. \ref{fig:MRI_xz} we show the value of $\lambda_{\textup{MRI}}$ along the $x$ axis together with the rest-mass density on the meridional $xz$ plane for the SLy case with the strongest initial magnetic field, and therefore the largest MRI wavelength. We see that even for this case there are a number of regions where $\lambda_{\textup{MRI}}$ is smaller than the height of the remnant and its surrounding disk, indicating that the MRI should be active. The ALF2 cases show a very similar behavior. 

\subsubsection{Magnetic winding and braking} 

This is not an instability but rather a secular consequence of the differential rotation and magnetic induction equation (Eq. (14) and (15) in \cite{Etienne:2010ui}) \cite{Baumgarte:1999cq,Spruit:1999cc,Shapiro:2000zh,Cook:2003ku,Duez:2004nf,Duez:2006qe}. Assuming axisymmetry, a magnetic field small enough that it has negligible back-reaction on the fluid, and quasiequilibrium conditions such that the fluid velocities are solely axial and slowly varying with time, we obtain (see Eqs. (2)-(7) in \cite{Duez:2006qe})
\begin{equation}
    \partial_t(\tilde{B}^{\phi}) \approx \tilde{B}^{i}\partial_i\Omega, \label{eq:induction}
\end{equation}
for $i \in (\varpi,z)$ and where $\tilde{B}^{i} = \sqrt{\gamma}B^i$ and $\gamma$ is the determinant of the spatial metric. If the poloidal field is approximately constant, or only slowly changing, we can integrate \eqref{eq:induction} to find the toroidal field $B^T \equiv \varpi B^{\phi}$ grows linearly with time as \cite{Shibata:2006hr}
\begin{align}
    B^T \approx& \;t\varpi\tilde{B}^{i}(t=0)\partial_i\Omega(t=0) \sim \tfrac{3}{2}t\Omega(0,\varpi,z)\vert B^{\varpi}\vert, \\
    \sim& 10^{15}\textup{G} \left(\tfrac{t}{100\; \textup{ms}}\right)\left(\tfrac{\Omega}{10^3 \textup{rad s}^{-1}}\right)\left(\tfrac{\vert B^{\varpi}\vert}{10^{13}\textup{G}}\right),
    \label{eq:BT_winding}
\end{align}
assuming a Keplerian angular velocity profile. As the magnetic fields lines are wound up and the toroidal field increases this creates magnetic tension that acts to resist the differential rotation via magnetic braking \cite{Shapiro:2000zh}, changing the velocity profile towards $\Omega = const.$ inside the star on the Alfv\'{e}n timescale 
\begin{equation}
    t_{A} \sim \frac{R}{v_A} \sim 10\;\textup{ms} \left(\tfrac{\vert B^{\varpi}\vert}{10^{13}\textup{G}}\right)^{-1}\left(\tfrac{R}{10\;\textup{km}}\right)\left(\tfrac{\rho_0}{10^{15}\textup{g\, \rm{cm}}^{-3}}\right)^{1/2}.
\end{equation}

\subsection{Ideal MHD jets}
\label{sec:jet_theory}

While jet formation and sGRBs are not the primary focus of this work, it is useful for the discussion in section \ref{sec:EM_jets} to highlight some features of jet theory. As we discussed in \cite{Bamber:2024kfb} section I. D., a key challenge is estimating the asymptotic Lorentz factor $\Gamma_{\infty}$ at large distances where the $\gamma-$ray emission is produced given our simulations are limited in spatial extent, and we expect the velocity of the outflow to increase with distance. Semi-analytic works (see e.g. \cite{Vlahakis:2003si,Beskin:2005uq,Komissarov:2010zm,Lyubarsky:2010yd}) on jets in ideal MHD, adopting special relativity, steady-state and axisymmetry, find that integrating the equations of motion produces several quantities that are conserved along the poloidal component of the field lines $\boldsymbol{B}_p$, which is parallel to the poloidal component of the fluid velocity $\boldsymbol{v}_p$. In particular, the ratio of total energy flux to rest-mass flux along a bundle of field lines, given by 
\begin{align}
    \mu :=& \frac{\textup{energy flux}}{\textup{rest-mass flux}}= \;\frac{\Gamma^2 \rho_0 h v_p + \tfrac{1}{4\pi}( \boldsymbol{E} \times \boldsymbol{B})\cdot \hat{\boldsymbol{v}}_p}{\Gamma \rho_0 v_p},%, \\
    \nonumber\\
    =& \;\Gamma + \Gamma(h - 1) + \sigma \Gamma,
    \label{eq:mu}
\end{align}
is conserved. In \eqref{eq:mu} $\hat{\boldsymbol{v}}_p$ is the unit vector in the direction of $\boldsymbol{v}_p$ and in the second line we have decomposed it into the contributions from the specific kinetic $+$ rest-mass energy $\Gamma$, the thermal enthalpy contribution $\Gamma(h-1)$, and the EM Poynting flux contribution $\sigma \Gamma = \tfrac{1}{4\pi}\vert \boldsymbol{E} \times \boldsymbol{B}_{\phi}\vert /(\Gamma \rho_0 v_p)$, where $\sigma$ is the magnetization parameter \cite{Narayan:2010bi,Lyubarsky:2010yd}. The gravitational energy is typically neglected, assuming a flat Minkowski metric. At the base of the jet the flow is sub-relativistic with $\Gamma \approx 1$, with the energy flow dominated by the EM Poynting flux with $\sigma \gg 1$ and $\mu \sim \sigma$ (for $\mu \gg 1$ the enthaply component is a sub-dominant contribution). As the fluid is accelerated upwards along the magnetically-dominated low density funnel, electromagnetic energy is converted to kinetic energy, so $\sigma$ decreases as $\Gamma$ increases in such a way as to keep $\mu$ constant. If the acceleration were perfectly efficient the final asymptotic Lorentz factor would be $\Gamma_{\infty} \approx \mu$, while a self-similar model \cite{Vlahakis:2003si} predicts the final state is rough equipartition between kinetic and electromagnetic energy with $\Gamma_{\infty} \sim \mu/2$ and cylindrical flow parallel to the jet axis. For strongly poloidal flow and tightly coiled magnetic fields we can approximate the magnetization at the base of the jet as \cite{Bamber:2024kfb}
\begin{equation}
    \sigma \approx \frac{b^2}{\rho_0},
\end{equation}
and, as $\Gamma \sim h \sim 1$ in this region, approximate $\Gamma_{\infty} \sim \sigma/2 \sim b^2/(2\rho_0)$ for $b^2/(2\rho_0) \gg 1$. Here $b^2/(2\rho_0)$ is the ratio of the magnetic energy density $\rho_{B}$ to the rest-mass density $\rho_0$, $b^{\mu} = B^{\mu}_{\textup{co}}/\sqrt{4\pi}$ where $B^{\mu}_{\textup{co}}$ is the comoving magnetic field, and $b^2 = b_{\mu}b^{\mu}$.

\begin{figure*}
    \centering
    \includegraphics[width=\linewidth]{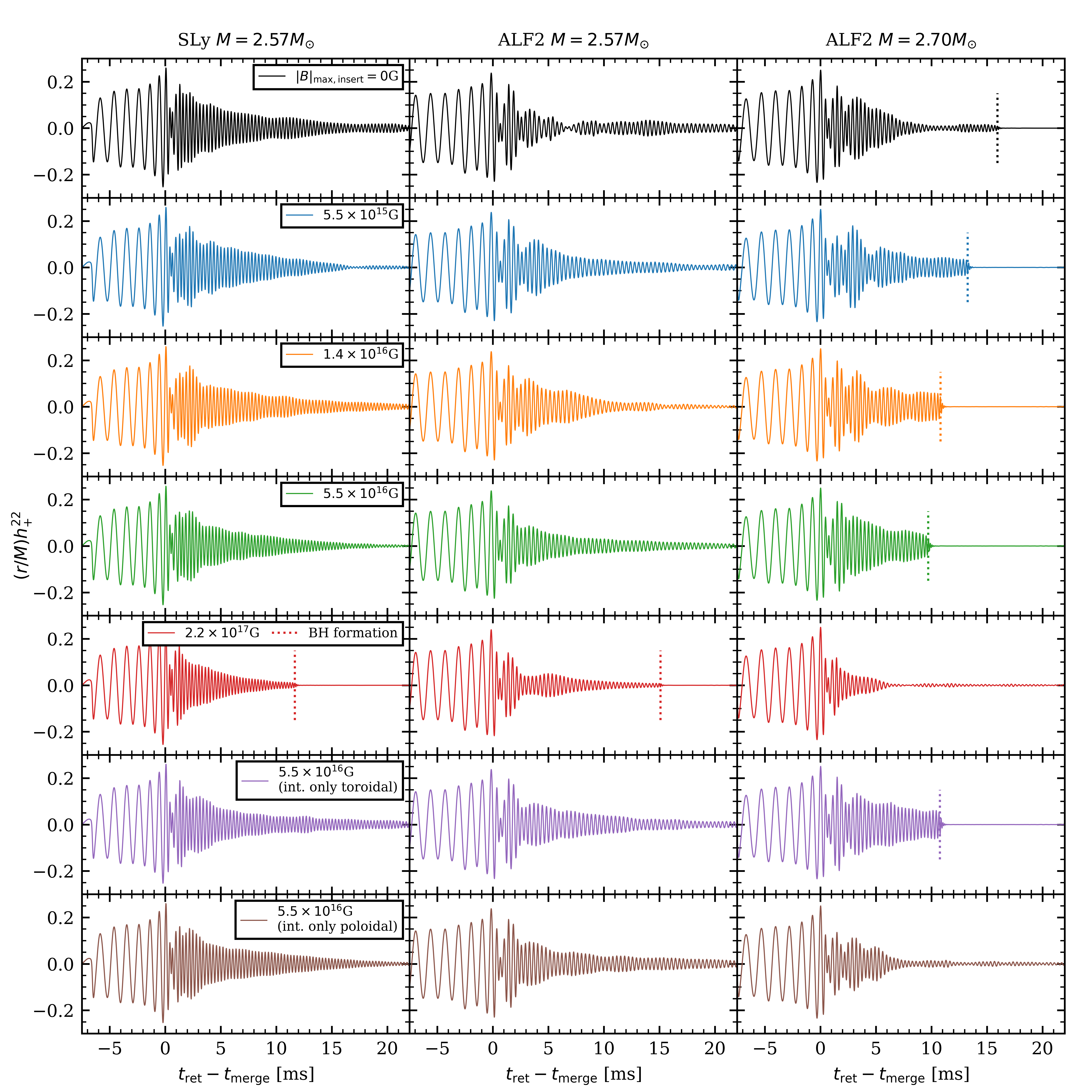}
    \caption{Gravitational wave strain $h^{22}_{+}$ (the 22 is the dominant mode) as a function of retarded time $t_{\mathrm{ret}}$ measured from the peak of the GW signal at $t_{\textup{merge}}$, extracted at coordinate radius $r_{\textup{ext}} = 683\textup{km} \approx 180M$. BH formation is denoted by a vertical dotted line.}
    \label{fig:GW_combined}
\end{figure*}

\section{Results}
\label{Sec:Results}

\subsection{Gravitational waves}

First, we shall discuss the GW waveforms, the only observable for NSNS mergers without an identified electromagnetic counterpart.

\subsubsection{Time domain}

The gravitational waveforms corresponding to the dominant $l$=$m$=2 mode and the $h_{+}$ polarization are shown in Fig. \ref{fig:GW_combined}. The different rows and colors indicate a different magnetic field strength or magnetic field topology, while the different columns refer to the different EOS and mass combinations. The waveforms are aligned with respect to the merger time $t_{\textup{merge}}$ defined as the time of maximum GW amplitude $\vert h \vert := (\vert h_+ \vert^2 + \vert h_{\times} \vert^2)^{1/2}$ and BH formation (if it occurs before the end of the simulation) is indicated by a vertical dotted line. One can see that after the magnetic field insertion at $t - t_{\textup{merge}} = -1$ms the GW signal is largely unaffected by the presence of the magnetic field until a few ms after merger, after which clear differences in the waveforms become apparent. Rows one to five correspond to initially pulsar-like magnetic fields of different strengths. 

\begin{figure}
    \centering
    \includegraphics[width=\linewidth]{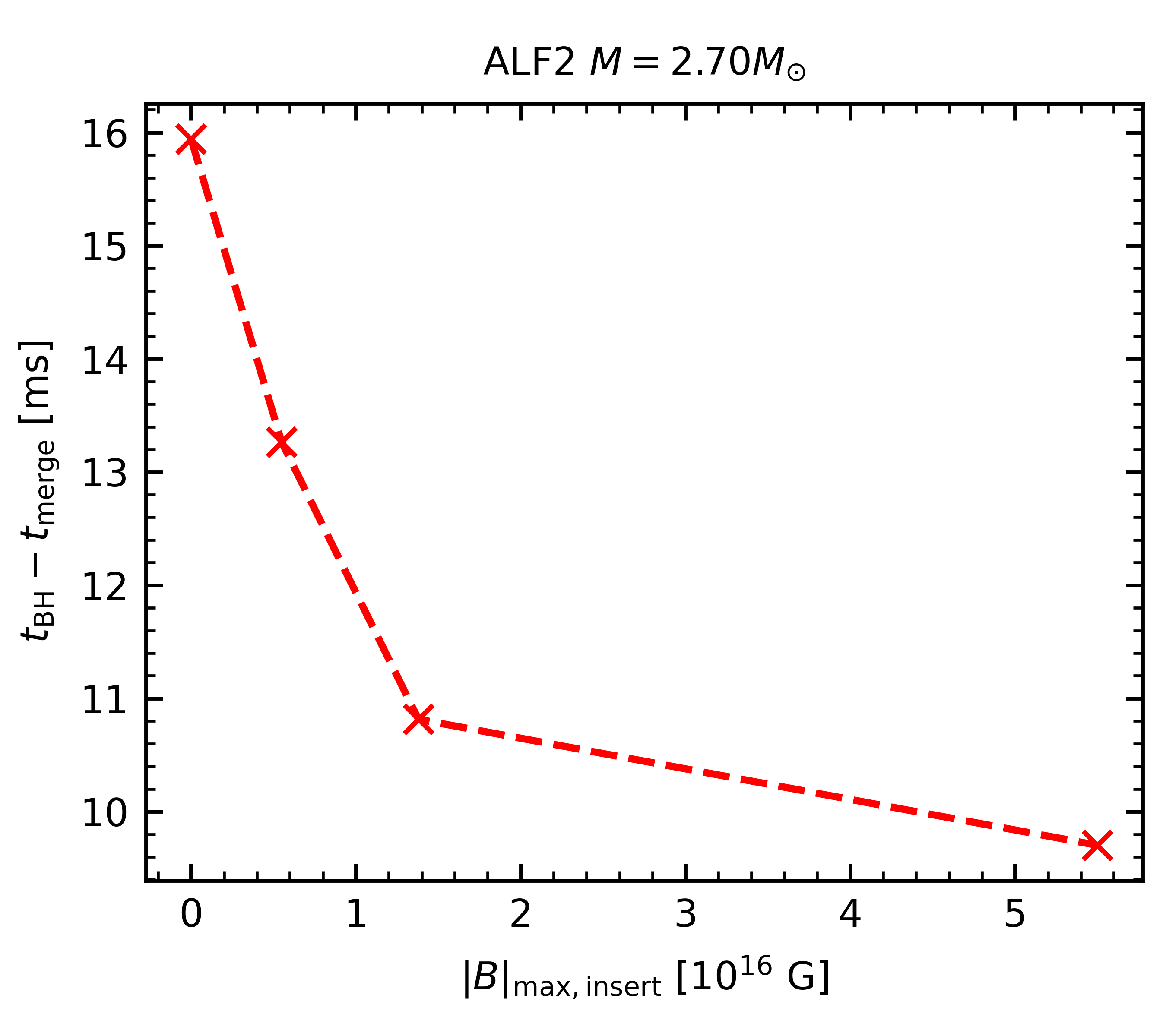}
    \caption{Lifetime of the NS remnant given by $t_{\textup{BH}} - t_{\textup{merge}}$ vs. magnetic field, where $t_{\textup{BH}}$ is the formation time of the BH, for the first four ALF2 $M=2.70M_{\odot}$ cases.}
    \label{fig:t_BH_vs_B}
\end{figure}

The clearest trend observable in the time domain signals is in the ALF2 $M=2.70M_{\odot}$ cases for $\vert B \vert_{\textup{max},\textup{insert}} = 0$ to $5.5\times10^{16}G$, where the lifetime of the HMNS remnant, given by $t_{\textup{BH}} - t_{\textup{merge}}$ where $t_{\textup{BH}}$ denotes the time of BH formation, decreases for an increasing magnetic field strength, also shown in Fig.~\ref{fig:t_BH_vs_B}. This can be attributed to the redistribution of angular momentum in the remnant due to magnetic braking and the effective viscosity due to magnetic turbulance \cite{Sun:2018gcl,Ruiz:2021qmm} (see \ref{subsec:M_processes}), which drives the star towards uniform rotation and causes the total angular momentum in the core to decrease. As a result the core of the star contracts and ultimately collapses to a BH once the central density crosses the turning point threshold. The ALF2 $M=2.70M_{\odot}$ case with the largest magnetic field is also very quickly driven towards uniform rotation, however in this case sufficient mass is ejected that the mass of the remnant falls below the supramassive limit, allowing it to persist as a uniformly rotating supramassive star (a more detailed discussion of this case is given in section \ref{sec:dyn}). One can also see that cases with the strongest magnetic fields (fifth row) show a faster decay in GW amplitude compared to the cases with smaller magnetic fields due to the stronger damping due to magnetic braking and magnetic effective turbulent viscosity (see Section \ref{sec:visc}). 

\begin{figure*}
    \centering
    \includegraphics[width=\linewidth]{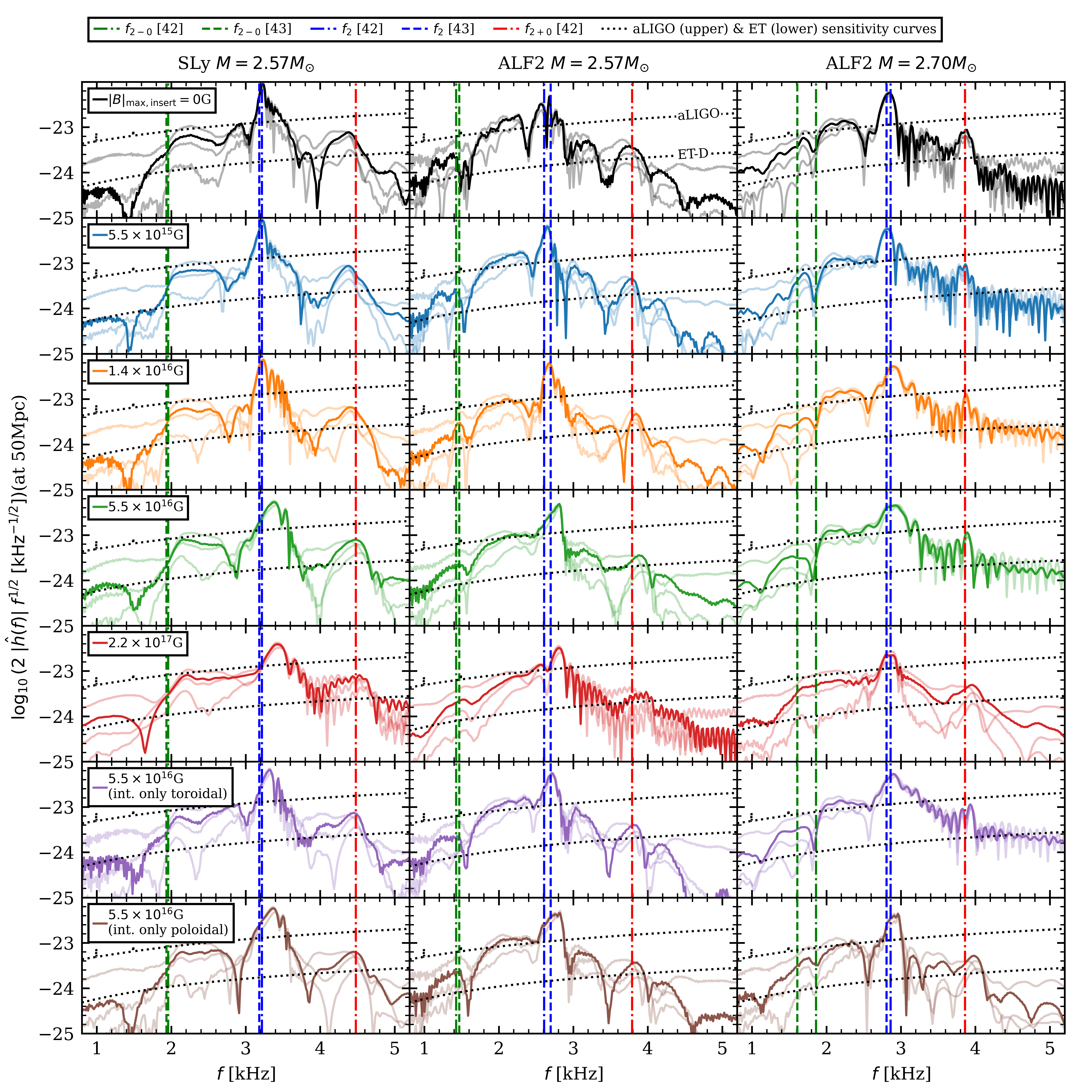}
    \caption{Amplitude spectral density (ASD) of the postmerger ($t > t_{\textup{merge}}$) GW signal for an optimally oriented merger at a distance of 50 Mpc. To illustrate how the power spectrum curve varies with the window function used with the FFT, we show multiple curves corresponding to $\delta = 0.01, 0.05, 0.1, 0.25$, where $\delta$ is the parameter of the Tukey window function, with the $\delta = 0.05$ curve shown in bold as a fiducial choice. For comparison the characteristic frequencies of the ASD peaks obtained by past authors for simulations without magnetic fields are shown as vertical lines. Dashed vertical lines show $f_{2-0}$ and $f_2$ given by the empirical relations in \cite{Vretinaris:2019spn}, Eqns. (6) and (4). The dot-dashed vertical lines show the $f_{2\pm 0}$ and $f_2$ frequencies taken from \cite{Rezzolla:2016nxn}, with $f_2,f_0$, interpolated from \cite{Rezzolla:2016nxn} TABLE II, and $f_{2\pm 0}$ estimated as $f_{2\pm 0} \approx f_2 \pm f_0$. Dotted curves show the sensitivity curves for aLIGO \cite{aLIGO:2020wna} (upper curve) and the Einstein Telescope in the D topology \cite{Hild:2010id} (lower curve).}
    \label{fig:GW_fft_combined}
\end{figure*}
\begin{figure*}
    \centering
    \includegraphics[width=\linewidth]{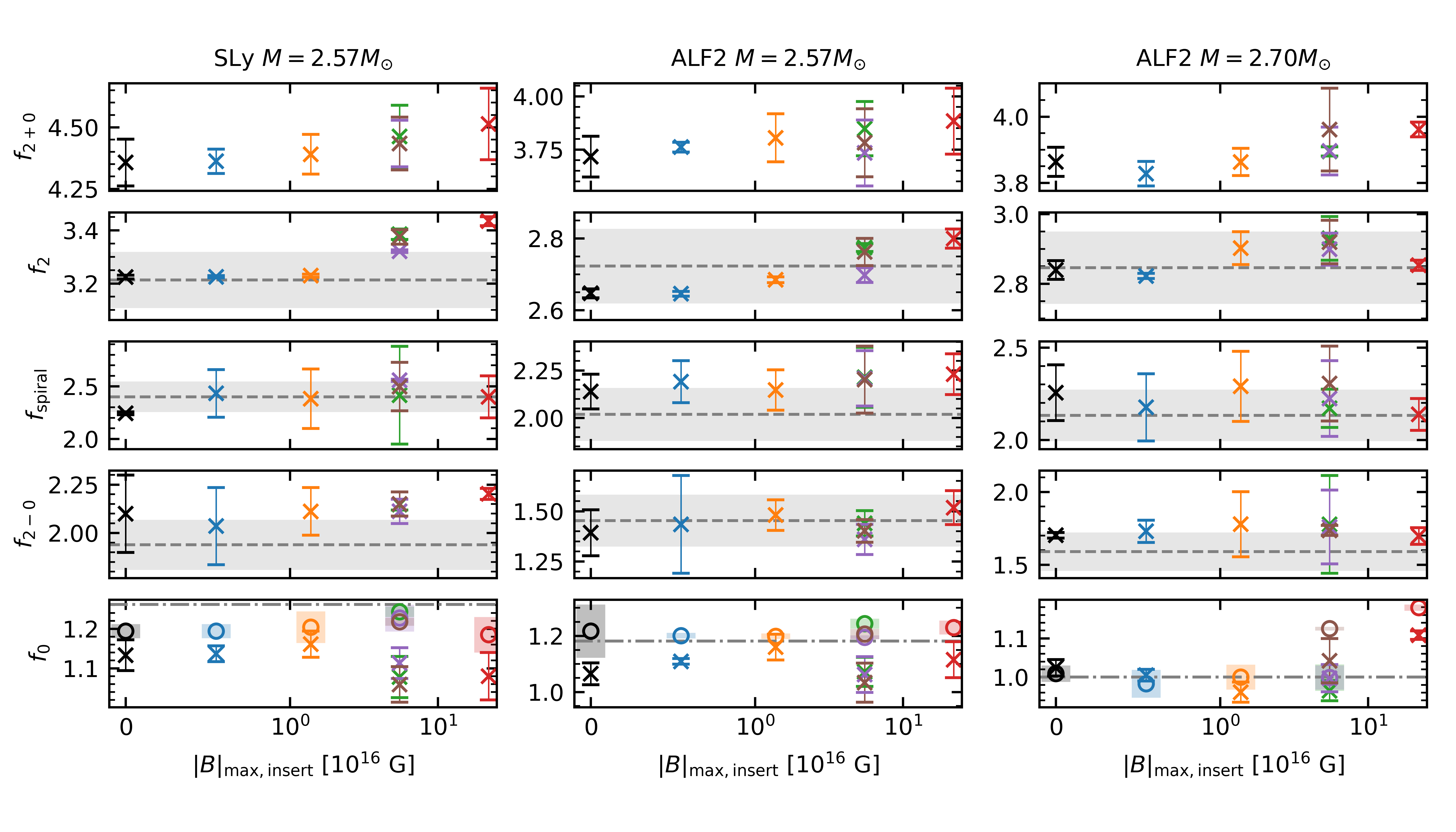}
    \caption{Characteristic frequencies in kHz measured from the power spectrum of the postmerger ($t > t_{\textup{merge}}$) signal. Values for $f_{2-0},f_{\textup{spiral}},f_2,f_{2+0}$ were obtained by fitting Gaussians to the peaks shown in Fig. \ref{fig:GW_fft_combined}. To give an estimate of the error due to choices of window function this procedure was repeated for the power spectra obtained with Tukey parameter $\delta = 0.01, 0.05, 0.1, 0.25$. The colors used for each case are the same in those as Figs. \ref{fig:GW_combined} and \ref{fig:GW_fft_combined}. In the second, third and fourth rows we show the prediction from the empirical quasi-universal relations obtained by Vretinaris et al. \cite{Vretinaris:2019spn} as a grey dashed line, with the $2\sigma$ error shown as a grey box. The bottom shows the frequency of the fundamental quasi-radial mode, $f_0$, obtained from $f_{2+0}-f_2$ (crosses and error bars) and from the power spectrum of the minimum lapse, $\alpha_{\textup{min}}(t)$ (circles and faded boxes). Finally, we also include $f_0$ values interpolated from the data for hydro-only simulations given in \cite{Rezzolla:2016nxn} Table II, shown as a horizontal dot-dashed line.}
    \label{fig:GW_f_vs_B}
\end{figure*}

\subsubsection{Frequency domain}

A GW detector will observe a signal of the form $h(t) = F_{\cross} h_{\cross} + F_+ h_+$ where $F_{\cross},F_+$ are functions of the orientations of the detector and source. The signal to noise ratio (SNR) for a wave extracted with an optimum Wiener filter is then given by 
\begin{equation}
    \textup{SNR} = \langle h \vert h \rangle^{1/2},
\end{equation}
for inner product
\begin{equation}
    \langle h_1 \vert h_2 \rangle \equiv 4\;\textup{Re}\left\{\int^{\infty}_0 \frac{\tilde{h}_1^*(f)\tilde{h}_2(f)}{S_h(f)}\dd f\right\},
\end{equation}
where $S_h(f)$ is the noise power spectral density of the detector \cite{Moore:2014lga,Rezzolla:2016nxn} and a ``$\;\tilde{\;\;}$\;" denotes the one-sided Fourier transform
\begin{equation}
    \Tilde{h}(f) \equiv \int^{\infty}_0 h(t) e^{-i2\pi f t} \dd t.
\end{equation}
The maximum SNR across all orientations is then approximately given by \cite{Rezzolla:2016nxn}
\begin{equation}
    \left[\int^{\infty}_0 \dd (\ln f) \frac{\textup{ASD}^2}{S_h(f)}\right]^{1/2},
\end{equation}
where the amplitude spectral density (ASD) is defined as 

\begin{align}
    \textup{ASD} \equiv& \;2 f^{1/2} \vert \hat{h}(f) \vert, \\
    \vert\hat{h}(f)\vert \equiv& \;\sqrt{\frac{\vert \Tilde{h}_+(f)\vert^2 + \vert \Tilde{h}_{\times}(f)\vert^2}{2}}.
\end{align}
We compute the ASD for the postmerger signal for $t > t_{\textup{merge}}$ using a Tukey window function $W(t;\delta)$ \cite{Tukey:1967} for a event at $50$Mpc distance, with the results shown in Fig. \ref{fig:GW_fft_combined}, along with the noise curves $\sqrt{S_h(f)}$ for the design sensitivity of Advanced LIGO \cite{aLIGO:2020wna} and the third generation Einstein Telescope detector (D topology) \cite{Hild:2010id}. A window function is necessary to isolate the postmerger part of the signal and reduce spectral leakage, however the choice of window can alter the precise shape of the power spectrum. To get a measure of the variation we used four different values for the Tukey window function parameter $\delta = 0.01, 0.05, 0.1, 0.25$, with the ASD for a fiducial choice of $\delta = 0.05$ shown via a bold curve and the others shown as faded curves. 

The strongest peak, corresponding to the $f_2$ fundamental quadrupolar mode, can be identified in all cases between $2.5$ and $3.5$kHz. In most cases one can also identify the $f_{2-0}, f_{2+0}$ side peaks, with the $f_{2+0}$ peak being more prominent. The $f_{\textup{spiral}}$ peak is more difficult to discern; however, in most of the spectra one can roughly identify it with the broad humped peak between $f_{2-0}$ and $f_2$ which overlaps with the $f_{2-0}$ peak. For comparison, we also plot as vertical lines the values for the characteristic frequencies obtained by previous authors for these EOSs and binary masses without magnetic fields. We see that the peaks for our hydro only simulations (top row in Fig.~\ref{fig:GW_fft_combined}) broadly line up with these previous values. However, as the magnetic field increases we see that some of characteristic frequencies deviate from their hydro-only values, with the $f_2$ peaks shifting to the right (most noticable for the SLy $M=2.57 M_{\odot}$ cases). 

To quantify this shift we measure the $f_{2-0},f_{\textup{spiral}},f_2$ and $f_{2+0}$ frequencies by fitting Gaussians to the corresponding ASD peaks shown in Fig. \ref{fig:GW_fft_combined}, with the results shown in the first four rows of Fig. \ref{fig:GW_f_vs_B} The error bars show the $2\sigma$ frequency spread across the four different window function parameters, and the colors correspond to those used in Figs. \ref{fig:GW_combined} and \ref{fig:GW_fft_combined} (so the cases started with interior-only toroidal and poloidal fields are shown in purple and brown respectively). We also show the values from the quasi-universal relations derived from hydro-only simulations of Vretinaris,
Stergioulas \& Bauswein (2020) \cite{Vretinaris:2019spn} as dashed horizontal lines with the associated $2\sigma$ uncertainties shown as a grey bar. The frequency of the quasi-radial mode, $f_0$, can be inferred either from $f_{2+0} - f_2$ (shown with crosses with capped error bars in Fig. \ref{fig:GW_f_vs_B} bottom row) or from the oscillations of the minimum of the lapse function $\alpha_{\textup{min}}(t)$ (shown as circles with thick solid error bars). For comparison, we also show the value from $f_0$ interpolated from the data given in Table II of \cite{Rezzolla:2016nxn} for hydro-only simulations. The numerical values for the measured characteristic frequencies are given in App. \ref{app:I0} Table \ref{tab:GW_freqs}. The results can be summarized as follows:

\begin{enumerate}

\item The $f_2$ frequency increases with magnetic field strength, with the exception of the ALF2 $M=2.70M_{\odot}$, $\vert B \vert_{\textup{max},\textup{insert}} = 2.2\times 10^{17}G$ case, with a maximum shift of $\sim +200$Hz. 

\item The $f_{2+0}$ frequency also increases with magnetic field strength, which we can attribute primarily to the increase in the $f_2$ mode. The frequency difference $f_{2+0} - f_2$ remains approximately consistent with the $f_0$ value inferred from oscillations in $\alpha_{\textup{min}}$ (some deviation is expected as the mode combination is only quasi-linear \cite{Vretinaris:2019spn}), with some deviations from the hydro-only value for large magnetic fields. In particular, the ALF2 $M=2.70M_{\odot}$ case with the strongest magnetic field shows an increase in $f_0$ of $>100$Hz compared to the hydro-only case. 

\item The $f_2$ values for the $|B|_{\textup{max}} = 5.5\times 10^{16}$G pulsar-like and interior-only poloidal cases agree within the 2$\sigma$ uncertainties, while the $f_2$ values for the interior-only toroidal cases are $\sim 30-60$Hz lower for the $M=2.57 M_{\odot}$ cases. This is within the $2\sigma$ uncertainty for the ALF2, $M=2.70M_{\odot}$ cases and $\sim 1\sigma-2\sigma$ outside it for $M=2.57M_{\odot}$ cases.

\item There is less evidence for a consistent trend with increasing magnetic field for $f_{2-0}$ and $f_{\textup{spiral}}$, although this is partly due to the increased uncertainties due to the difficulty in accurately identifying these overlapping peaks. 

\end{enumerate}

The shift in the $f_2$ peak with magnetic field is the most important result, as the $f_2$ dominant quadrupolar mode has the largest amplitude, and is therefore easiest for both current and future detectors to resolve (see Fig. \ref{fig:GW_combined}). A shift in $f_2$ of $200$Hz is larger than the measurement uncertainties of $\delta f_2 \lesssim 50$Hz achievable with Advanced LIGO or third generation detectors \cite{Clark:2015zxa,Yang:2017xlf}, and therefore in principle detectable.

\subsubsection{Mismatch}

Instead of just examining the frequency of a single oscillation mode, we can also quantify the overall difference between the GW waveforms of the hydro-only and magnetized cases by computing the match function defined as
\begin{equation}
    \mathcal{M}_{\textup{GW}} = \max_{(\phi_c,t_c)}\frac{\langle h_1 \vert h_2(\phi_c,t_c)\rangle}{\sqrt{\langle h_1 \vert h_1 \rangle \langle h_2 \vert h_2 \rangle}}, \label{eq:match}
\end{equation}
for two waveforms $h_1,h_2$, where in \eqref{eq:match} the expression is maximized over all possible phase shifts $\phi_c$ and time shifts $t_c$ between $h_1$ and $h_2$. The match gives the fraction of SNR recovered when searching for signal $h_1$ with $h_2$ as the filter rather than $h_1$ \cite{Harry:2018hke}, so the decrease in SNR squared is 
\begin{equation}
    \delta (\textup{SNR}^2) = \textup{SNR}^2(1 - \mathcal{M}^2_{\textup{GW}}).
\end{equation}
A rule-of-thumb used in \cite{Harry:2018hke,Ruiz:2021qmm} is that two waveforms can be distinguished only if $\delta (\textup{SNR}^2) > 1$. The GW amplitude decreases with distance from the source $r$ as $h \propto 1/r$. Let $\overline{\textup{SNR}} = \langle rh \vert rh \rangle^{1/2}$, then the criterion to distinguish is 
\begin{equation}
    \delta (\textup{SNR}^2) = \frac{1}{r^2}\overline{\textup{SNR}}^2(1 - \mathcal{M}^2_{\textup{GW}}) > 1,
\end{equation}
which we can rearrange as 
\begin{equation}
    r < r_{\textup{max}} = \overline{\textup{SNR}}\sqrt{1 - \mathcal{M}^2_{\textup{GW}}}, \label{eq:r_max}
\end{equation}
giving a maximum source distance for distinguishing the waveforms. Using \eqref{eq:r_max} we compute the maximum source distance at which one could distinguish the waveforms for the magnetized cases from the corresponding pure-hydro waveform with the same EOS and mass. We assume an optimum orientation of source and detector such that $h(t) \approx h^{22}_+(t)$ and extend the waveforms from our NR simulations to the early inspiral / low frequency regime by appending a\texttt{IMRPhenomD\_NRTidalv2} model \cite{Khan:2015jqa,Dietrich:2019kaq} fitted to the inspiral part of the numerical waveform. 
%\texttt{IMRPhenomD} does not include NS finite size effects, but as the early inspiral portion of the waveforms prior to magnetic field insertion is, by construction, identical between the magnetized and hydro-only cases this should not make a substantial difference to our results. 

The closest NSNS signal observed to date was GW170817 with a luminosity distance of $40^{+8}_{-14}$Mpc \cite{LIGOScientific:2017vwq}. Using Advanced LIGO we find $r_{\textup{max}} \sim 7-10$ Mpc, while for the Einstein Telescope (D topology) $r_{\textup{max}}$ is $\sim 60 - 80$ Mpc.  Our results therefore confirm the findings in \cite{Ruiz:2021qmm} that current ground-based GW observatories will be unlikely to detect the postmerger imprint of a magnetic field on the waveform, but more sensitive third generation detectors may be able to distinguish magnetic field effects at GW170817 distances. It should be noted that this analysis ignores potential degeneracies between the magnetic field and changes in all other parameters of the binary. A full analysis would require marginalization over all other parameters, including the unknown EOS, in order to compare the magnetized waveform with its closest possible hydro-only equivalent, however we leave this analysis for a future work. 

\subsection{Dynamics}
\label{sec:dyn}

\begin{figure*}
\begin{tabular}{ccc}
  \includegraphics[width=0.33\textwidth]{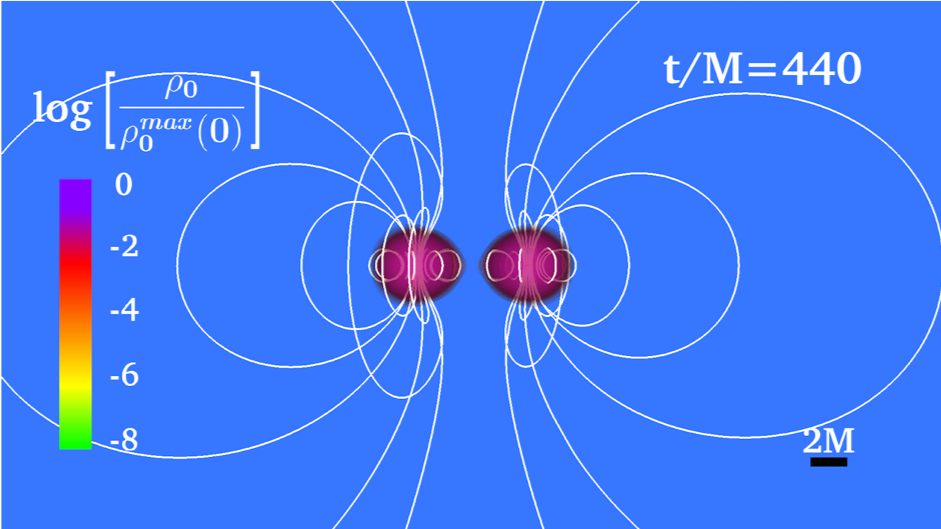} & \includegraphics[width=0.33\textwidth]{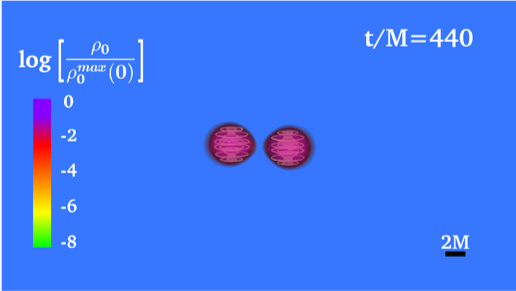} & \includegraphics[width=0.33\textwidth]{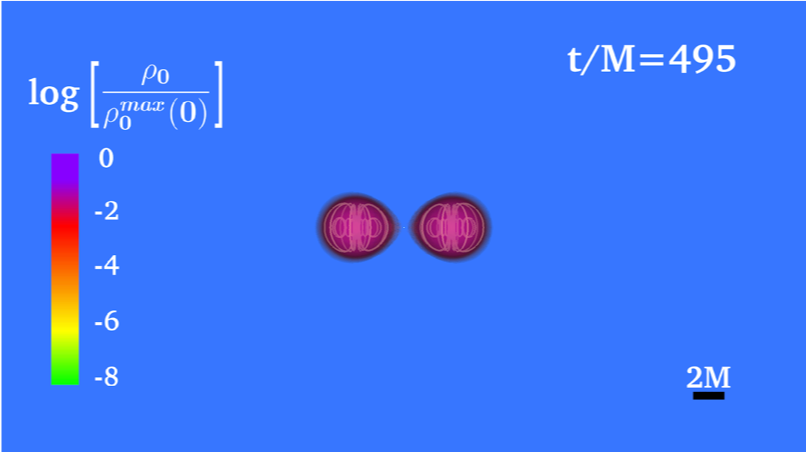}   \\
 \includegraphics[width=0.33\textwidth]{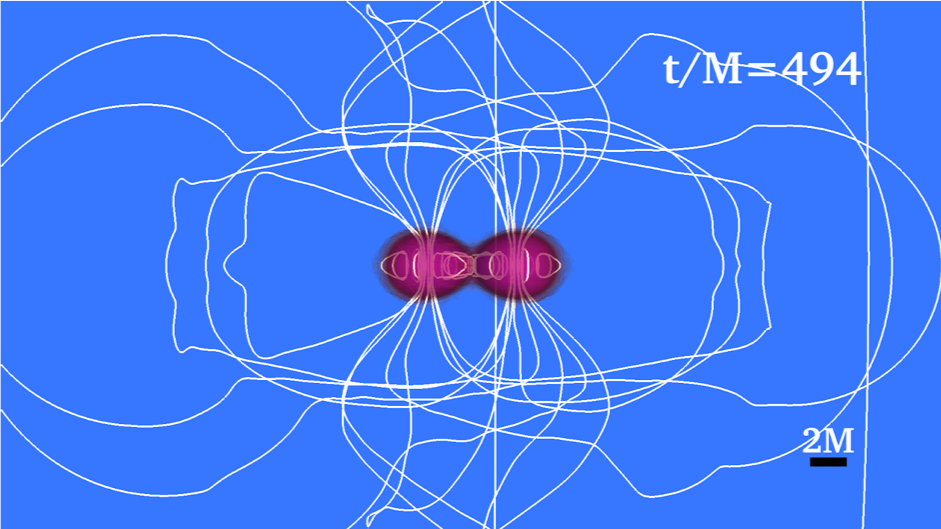} & \includegraphics[width=0.33\textwidth]{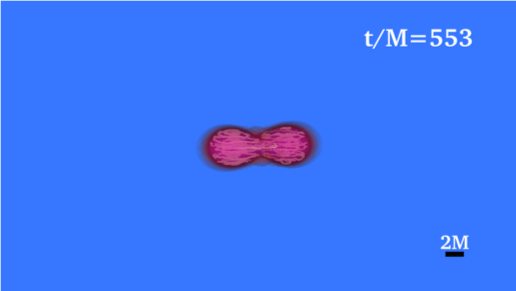} & \includegraphics[width=0.33\textwidth]{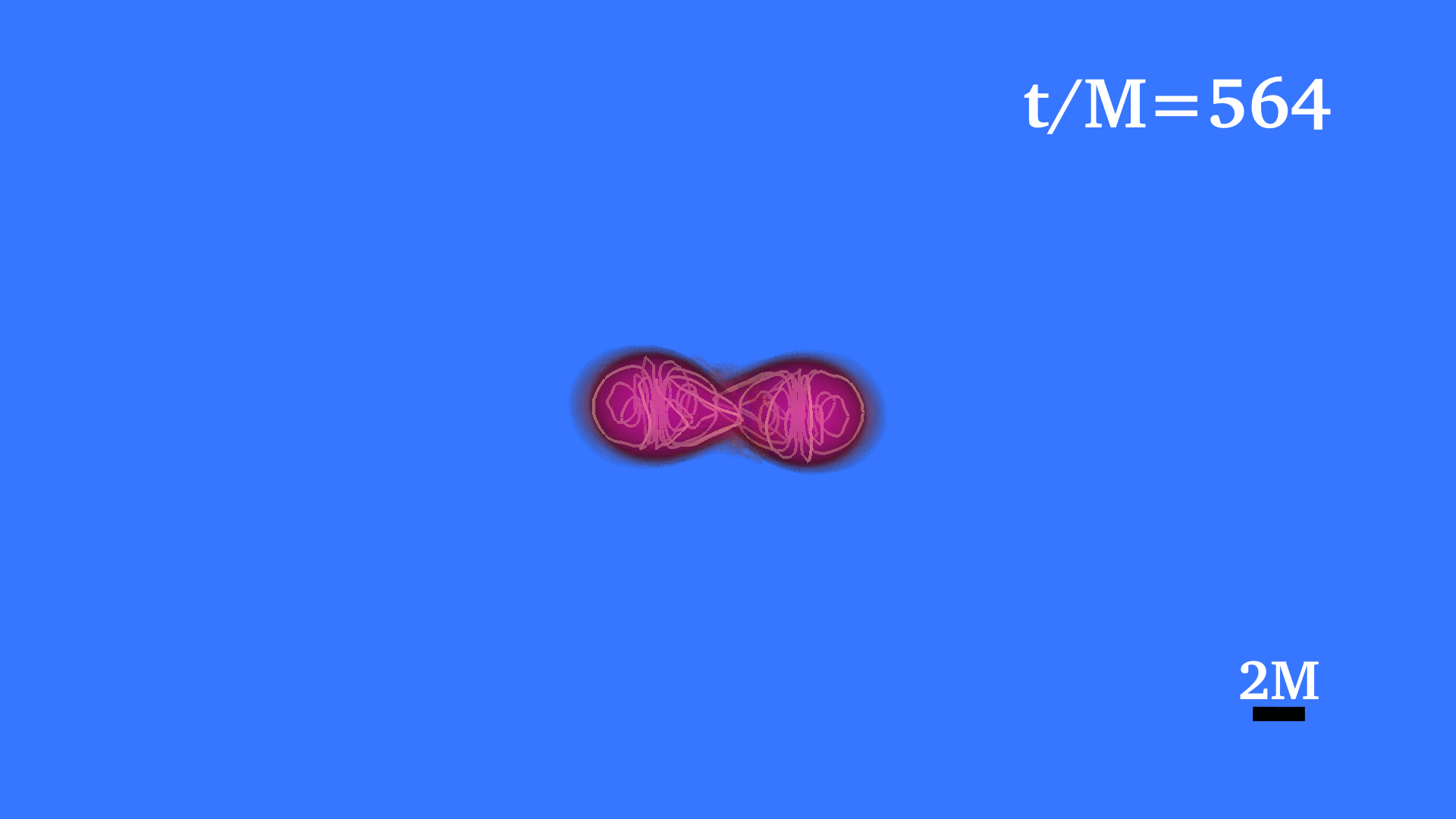}   \\
\includegraphics[width=0.33\textwidth]{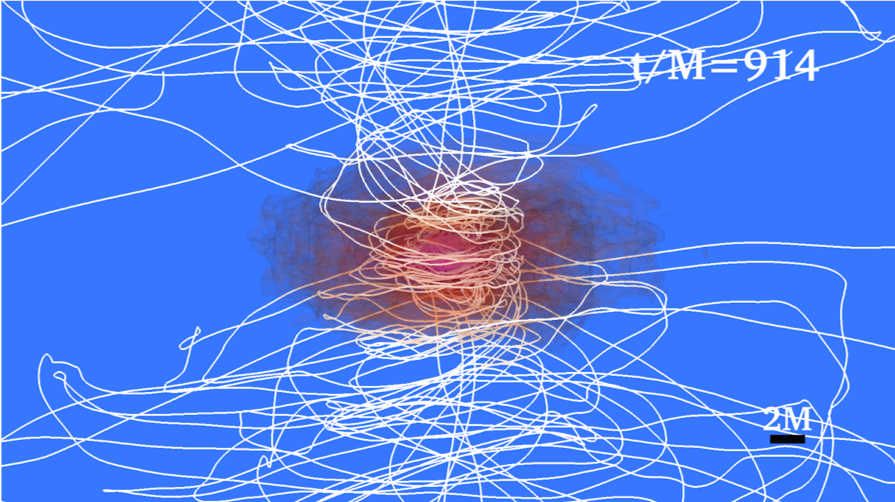} & \includegraphics[width=0.33\textwidth]{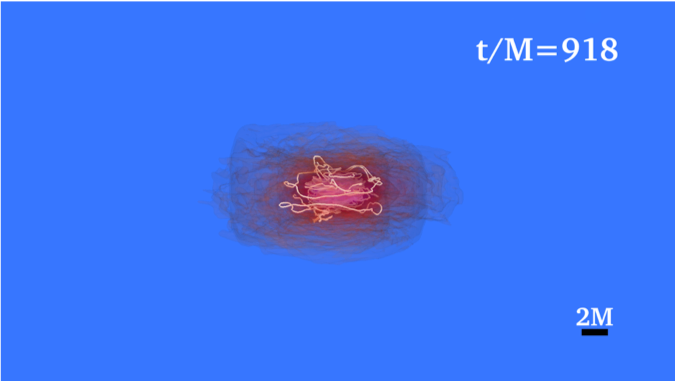} & \includegraphics[width=0.33\textwidth]{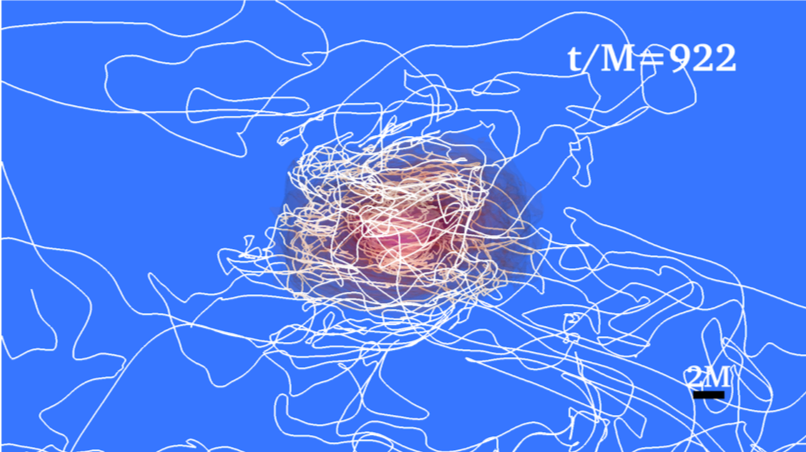}   \\
\includegraphics[width=0.33\textwidth]{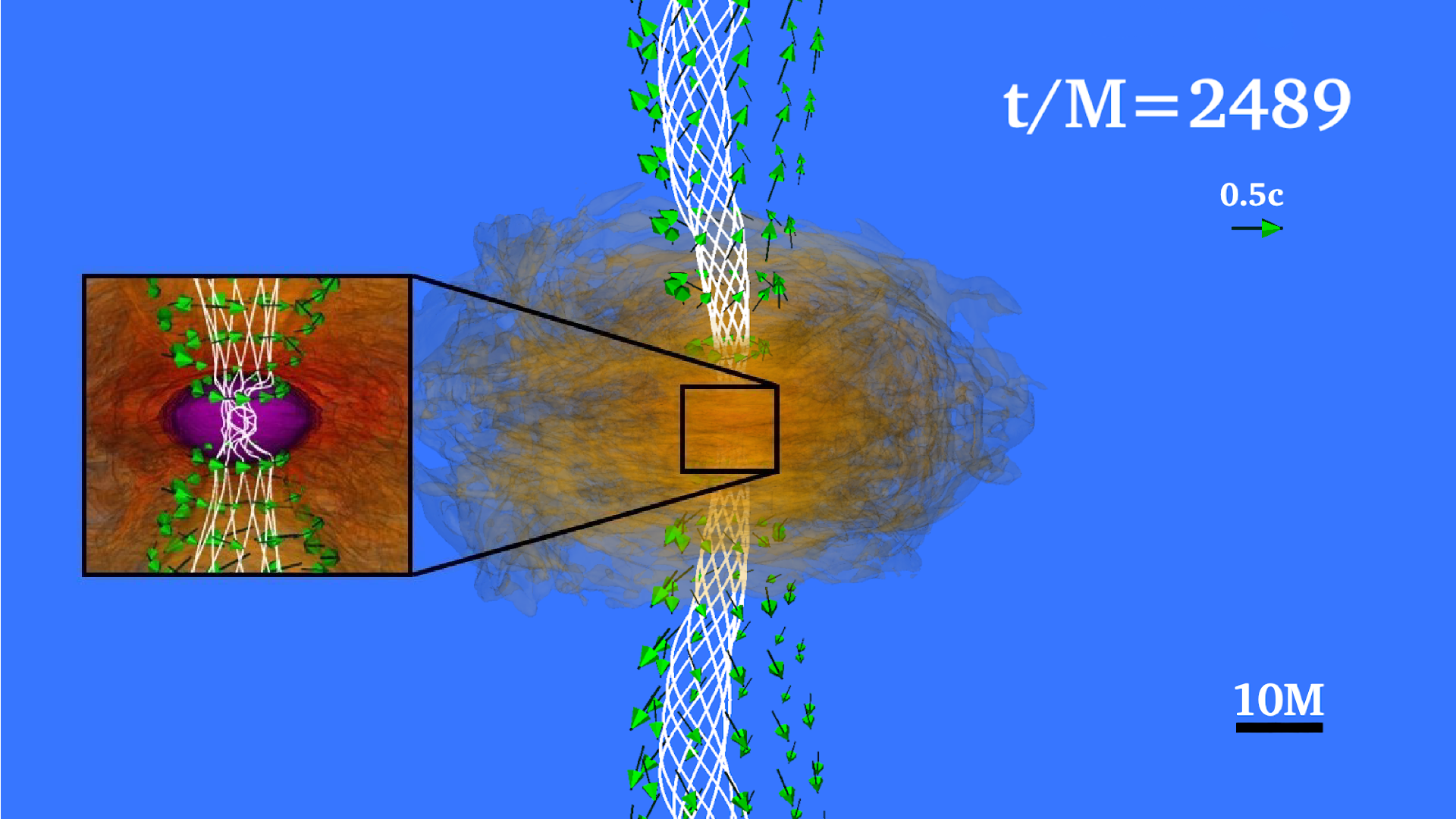} & \includegraphics[width=0.33\textwidth]{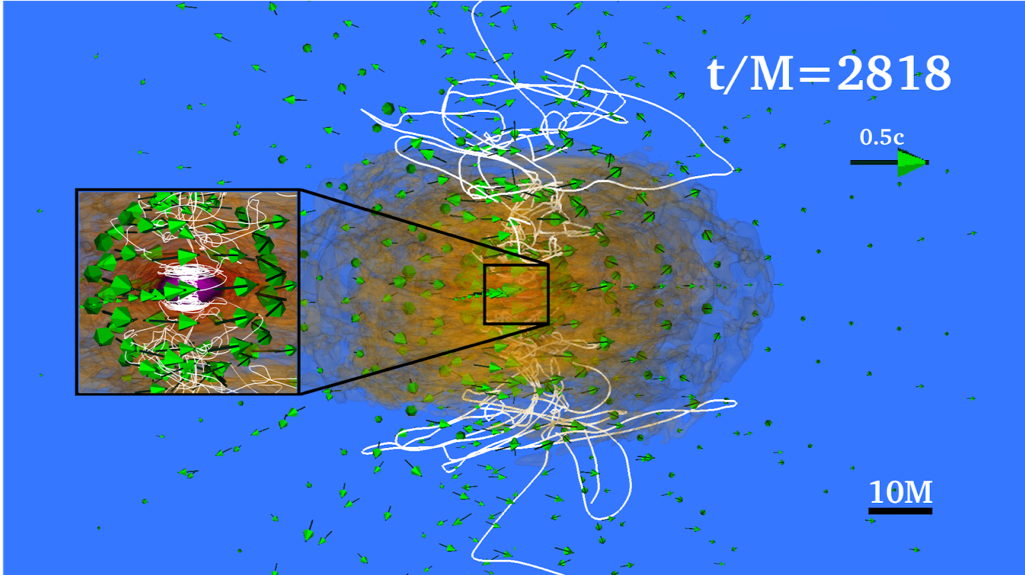} & \includegraphics[width=0.33\textwidth]{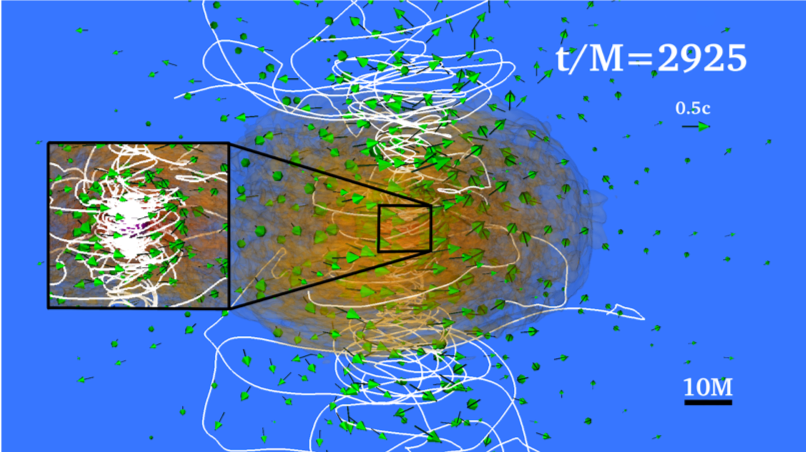}   
\end{tabular}
\caption{3D snapshots of the evolution of the rest-mass density for the SLy $M=2.57$ cases with $\vert B \vert_{\textup{max},\textup{insert}}=5.5\times 10^{16}$G (see Sec.~\ref{subsec:Initialdata}) at four instances in time. The case with the initially pulsar-like dipolar magnetic field is shown on the left, the interior-only toroidal case is shown in the middle, and the interior-only poloidal case on the right. The images on the first three rows show the baryonic rest-mass density (for $\rho_0/\rho^{\textup{max}}_0(t=0)\geq 10^{-2}$) and the magnetic field lines at magnetic field insertion, when the stars make contact, and shortly after the merger, respectively. The bottom row of images show the torus of matter that forms around the central SMNS remnant at $\sim 25$ms, near the end of the simulation. The yellow torus shows the region with $\rho_0/\rho^{\textup{max}}_0(t=0)\gtrsim 10^{-6}$, the white lines show the magnetic field lines from the remnant poles and the green arrows show the fluid velocity. In the case with an initial pulsar-like topology the field lines have become tightly wound in a helical funnel, with collimated outflow, while for the cases with no initial exterior fields the field lines are still disorganised and there is little evidence of collimated outflow. We also show an insert with the front-half cut away on the meridional $xz$ plane, showing the remnant itself in purple and the higher density parts of the torus in red-orange. Here $M =2.57 M_{\odot}$ and $M_{\odot} = 4.9\times 10^{-3}\textup{ms} = 1.4$km.
}
\label{fig:3D_plots}
\end{figure*}

The general evolution of the NSNS binaries matches the description from our previous works (see e.g. \cite{Bamber:2024kfb,Ruiz:2021qmm,Ruiz:2016rai}), as depicted for the SLy $M=2.57M_{\odot}$, $\vert B \vert_{\textup{max},t=0} = 5.5\times 10^{16}$ cases shown in Fig. \ref{fig:3D_plots}. The stars start in quasi-circular orbits, and inspiral due to the loss of angular momentum and enegy via gravitational radiation. At $\sim 1$ms before they merge (corresponding to $t/M \sim 440$ for the SLy $M=2.57M_{\odot}$ cases) the magnetic fields are added. The stars then plunge and make contact, forming a shear surface between them. This surface is broken up into vortices due to the effect of the KHI, and the two NS cores then merge and bounce until they coallesce into a rotating ellipsoidal remnant. The outer layers gain angular momentum due to torques from the rotation of the double core structure, orbital angular momentum convection and magnetic viscosity, for magnetized cases, producing an outflow of ejecta. Finally, the ellipsoidal remnant either collapses to a BH, or gradually settles down to a axisymmetric star, surrounded by a lower density torus of bound material (visible as an orange cloud in the bottom row of Fig. \ref{fig:3D_plots}). Beyond this generic picture we see differences due to both the initial strength of the magnetic field and its topology, as we discuss in the follow sections.  

\begin{figure*}
    \centering
    \includegraphics[width=\linewidth]{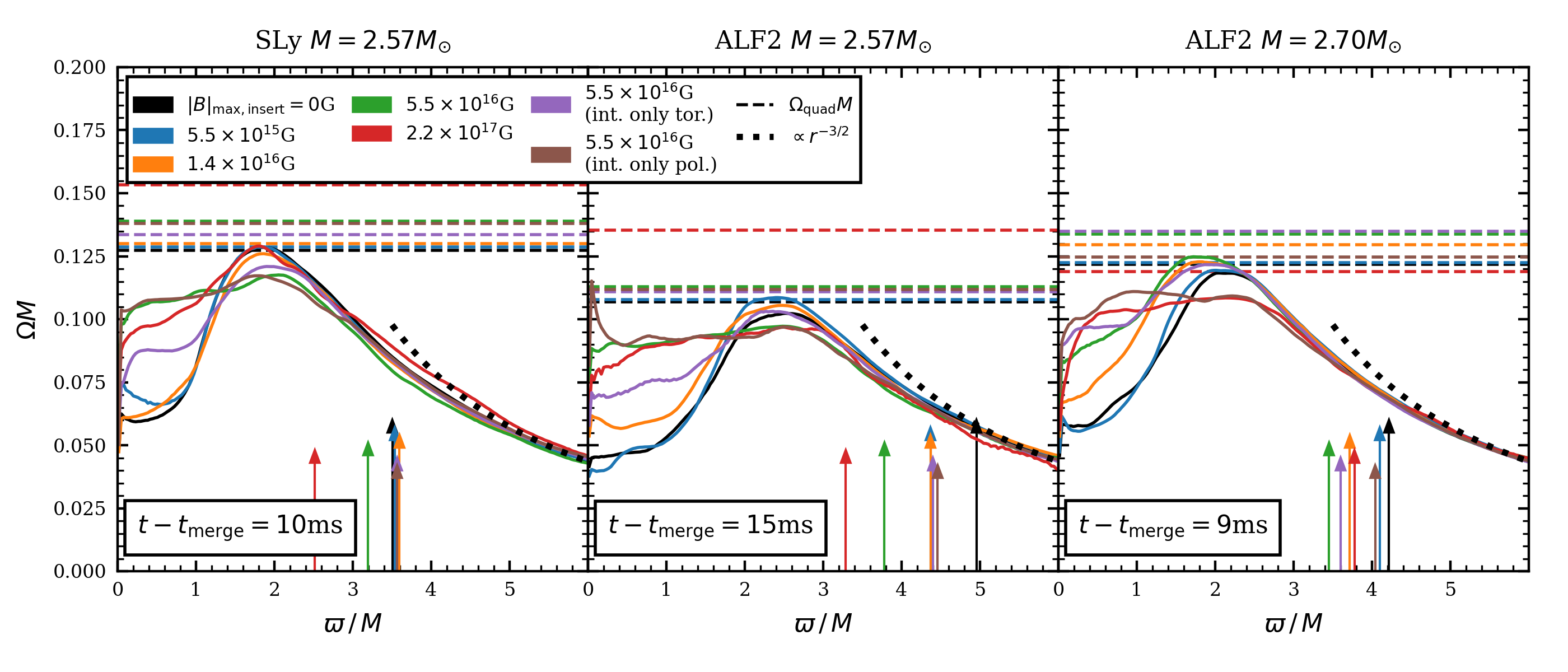}
    \caption{Average azimuthal angular velocity profiles for the different cases in the equatorial plane. The times shown are the latest times before collapse to a BH. We also plot with dashed horizontal lines the angular velocity of the quadrupole from the instantaneous $f_2$ frequency $\Omega_{\textup{quad}} := \Omega_2/2 = \pi f_{2} \approx \pi f_{\textup{GW}}$. The vertical arrows denote the radius at which $\rho_0/\rho_0^{\rm max} = 0.01$, which indicates the size of the remnant.}
    \label{fig:Ang_vel}
\end{figure*}

\subsubsection{Rotation profiles}

Fig.~\ref{fig:Ang_vel} shows the averaged rotation profile $\Omega(r,t)$, \cite{Hanauske:2016gia,Kastaun:2014fna} (see Eq. (2) in \cite{Ruiz:2019ezy}) for the different cases. For each EOS and binary mass combination we plot the profiles at the latest time before any of the cases have started to collapse to BHs, when they are still quasi-stationary HMNSs or SMNSs. The vertical arrows show the radius at which the average density is $1\%$ of the maximum density in order to give an idea of the stellar size. We see that the hydro-only cases (the black curves in Fig. \ref{fig:Ang_vel}) maintain a diferentially rotating profile consistent with those described in \cite{Kastaun:2014fna,Hanauske:2016gia,Uryu:2017obi,Ruiz:2019ezy,Cassing:2024dxp}, with a slower rotating core, an increase in $\Omega$ up to a maximum $\Omega=\Omega_{\textup{max}}$, followed by a Keplerian $\Omega \propto r^{-3/2}$ decay for $r \gtrsim 3M \sim 10$km. However, as discussed in Section \ref{subsec:M_processes}, for the cases with a magnetic field the magnetic winding and magnetic effective turbulent viscosity acts to damp the differential rotation, driving the remnant stars to uniform rotation, as reported originally in e.g. \cite{Shapiro:2000zh,Duez:2006qe} and many subsequent works. We can see that this effect is stronger for stronger initial magnetic field strengths, as one would expect, although the SLy $\vert B \vert_{\textup{max},\textup{insert}}=2.2\times 10^{17}$G case retained a sloped rotation profile, increasing almost linearly with radius up to $r \sim 2M$ as seen in Fig. \ref{fig:Ang_vel}, as it collapses to a BH (see below). We also see that the cases with an initial interior-only toroidal field remain closer to the hydro-only rotation profile than the same-strength interior-only poloidal and pulsar like cases, which is likely because the magnetic winding is driven by the radial $B^{\varpi}$ component of the magnetic field. For the cases that reach near-uniform rotation before collapse to a BH or the end of the simulation angular velocity of the fluid corresponds to rotation period of $\sim 0.75$ms. 

The instantaneous angular velocity of the $m=2$ quadrupole moment, $\Omega_{\textup{quad}} = \pi f_2$, is shown with a horizontal dashed line. We note that while the hydro-only cases roughly obey the $\Omega_{\max} = \Omega_{\textup{quad}}$ correspondance reported in \cite{Kastaun:2014fna,Hanauske:2016gia}, after $\sim 8-10$ms postmerger this no longer holds for the cases with strong magnetic fields, as the quadrupole and ellipsoid shape rotates faster than the fluid itself. We also see that the magnetised remnants are more compact than for the hydro-only cases, despite the fact the angular velocity of the core is larger. This is likely because the transport and loss of angular momentum from the remnant interior is also much larger with stronger magnetic fields (as shown in section \ref{sec:remnant_fate}).

\subsubsection{Fate of the remnants}
\label{sec:remnant_fate}

\begin{figure}
    \centering
    \includegraphics[width=\linewidth]{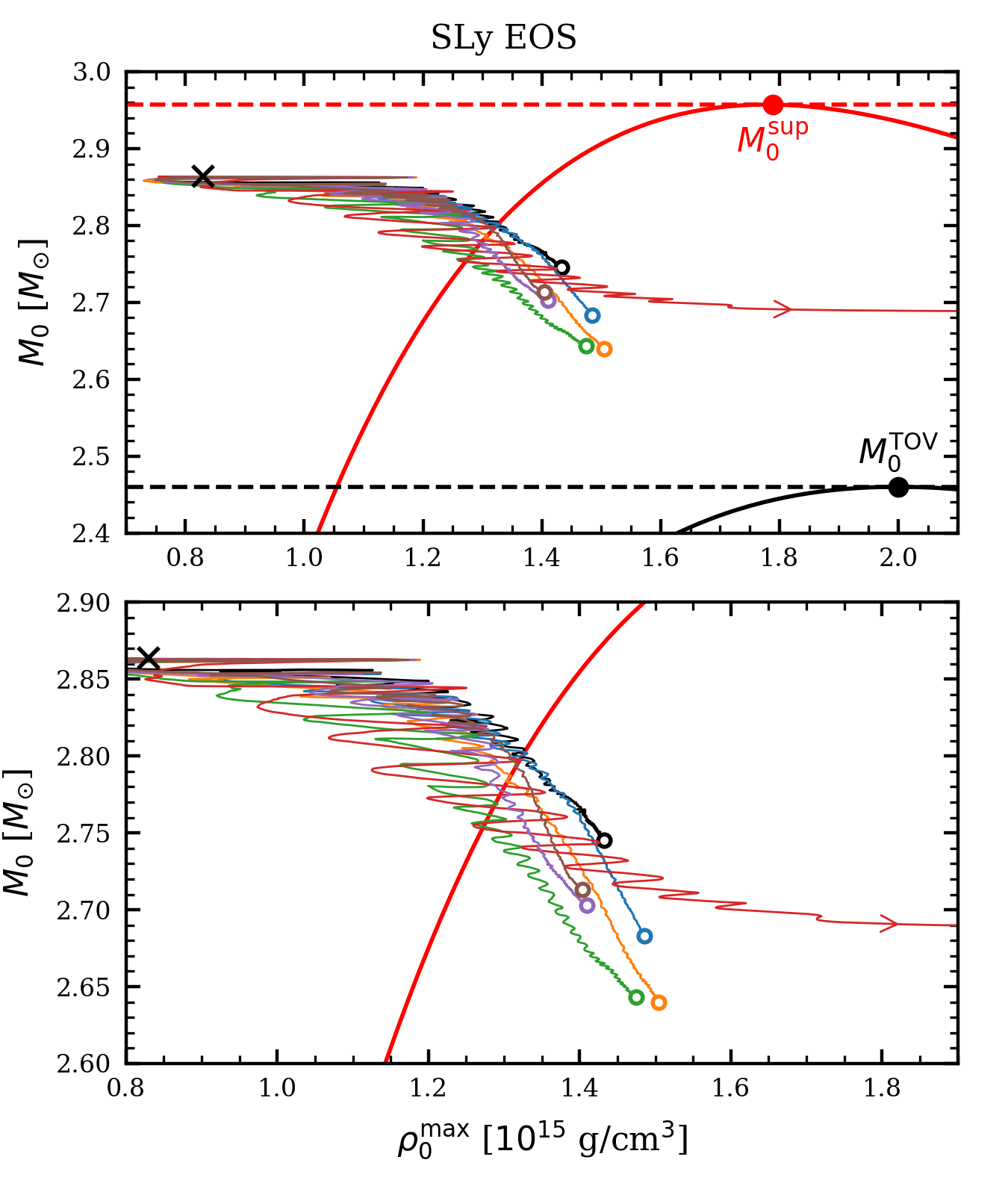} \\
    \caption{Trajectories of the rest mass $M_0$ of the NSs/NS merger remnant vs. the maximum rest-mass density $\rho^{\textup{max}}_0$ for the SLy cases. The colors for the different cases correspond to those used in Figs. \ref{fig:GW_combined} and \ref{fig:GW_fft_combined}, and the hollow circles show the end point of the simulations for the cases which do not collapse to BHs. We also plot the rest mass vs. central density curves for cold, spherical stars (thick black curve) with the maximum spherical mass, $M^{\textup{TOV}}_0$, shown as a horizontal black dashed line and black filled circle. The thick red curve shows the rest mass vs. central rest-mass density curve for cold, uniformly rotating stars at the Keplerian limit, with the supramassive limit $M^{\textup{sup}}_0$ shown with a red dashed line. The black crosses denote the initial $M_0$ and $\rho^{\textup{max}}_0$ at the start of the simulations. The bottom panel shows a closer view of the main region of interest.}
    \label{fig:M0_vs_rho_c_SLy}
\end{figure}

\begin{figure}
    \centering
    \includegraphics[width=\linewidth]{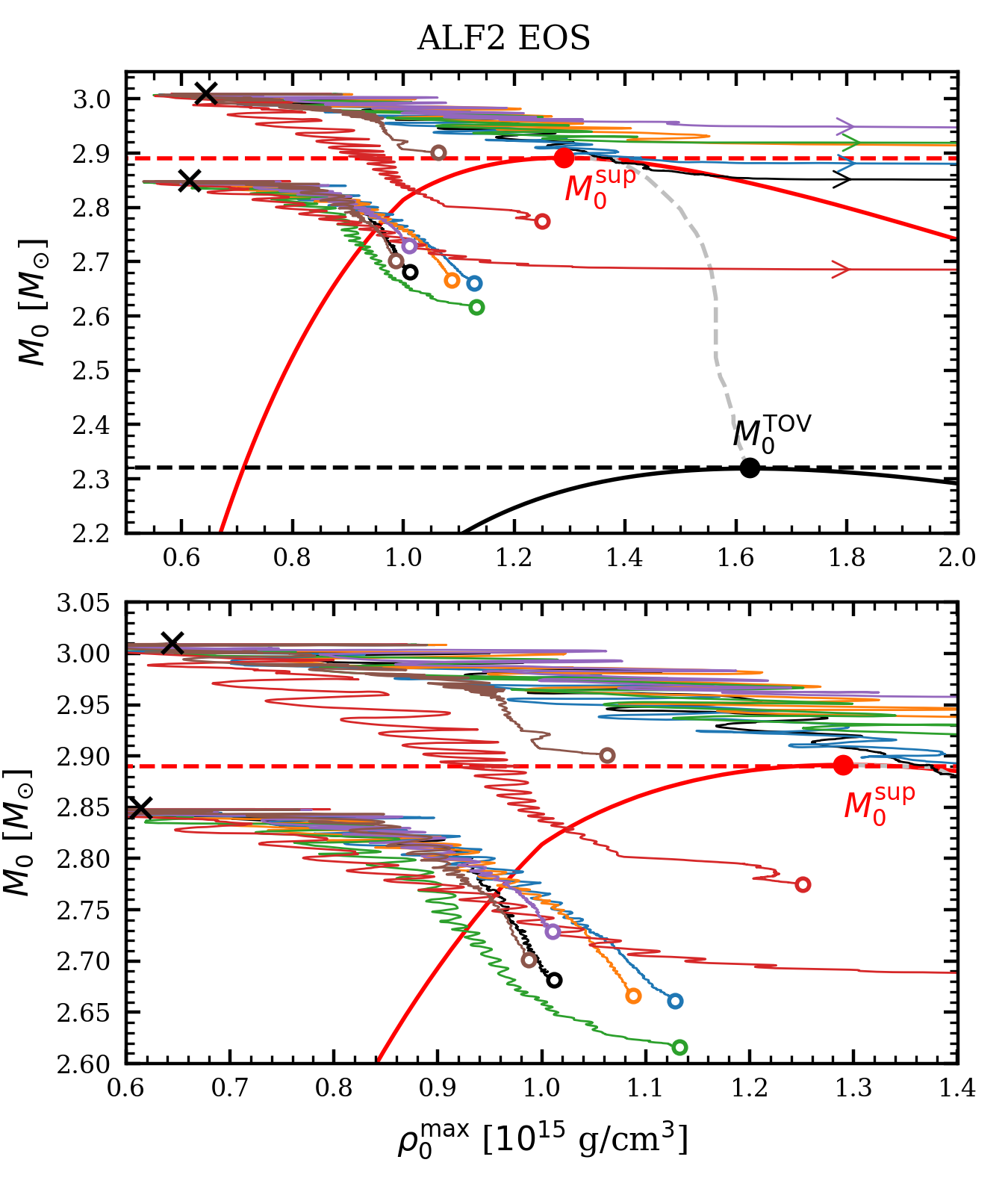} \\
    \caption{Same as Fig. \ref{fig:M0_vs_rho_c_SLy} but for the ALF2 cases. Here the locus of turning points for sequences of uniformly rotating stars is shown with a grey dashed line, and on each panel the lines for the cases with $M=2.70M_{\odot}$ are located above the lines for the $M=2.57M_{\odot}$ cases}
    \label{fig:M0_vs_rho_c_ALF2}
\end{figure}

\begin{figure}
    \centering
    \includegraphics[width=\linewidth]{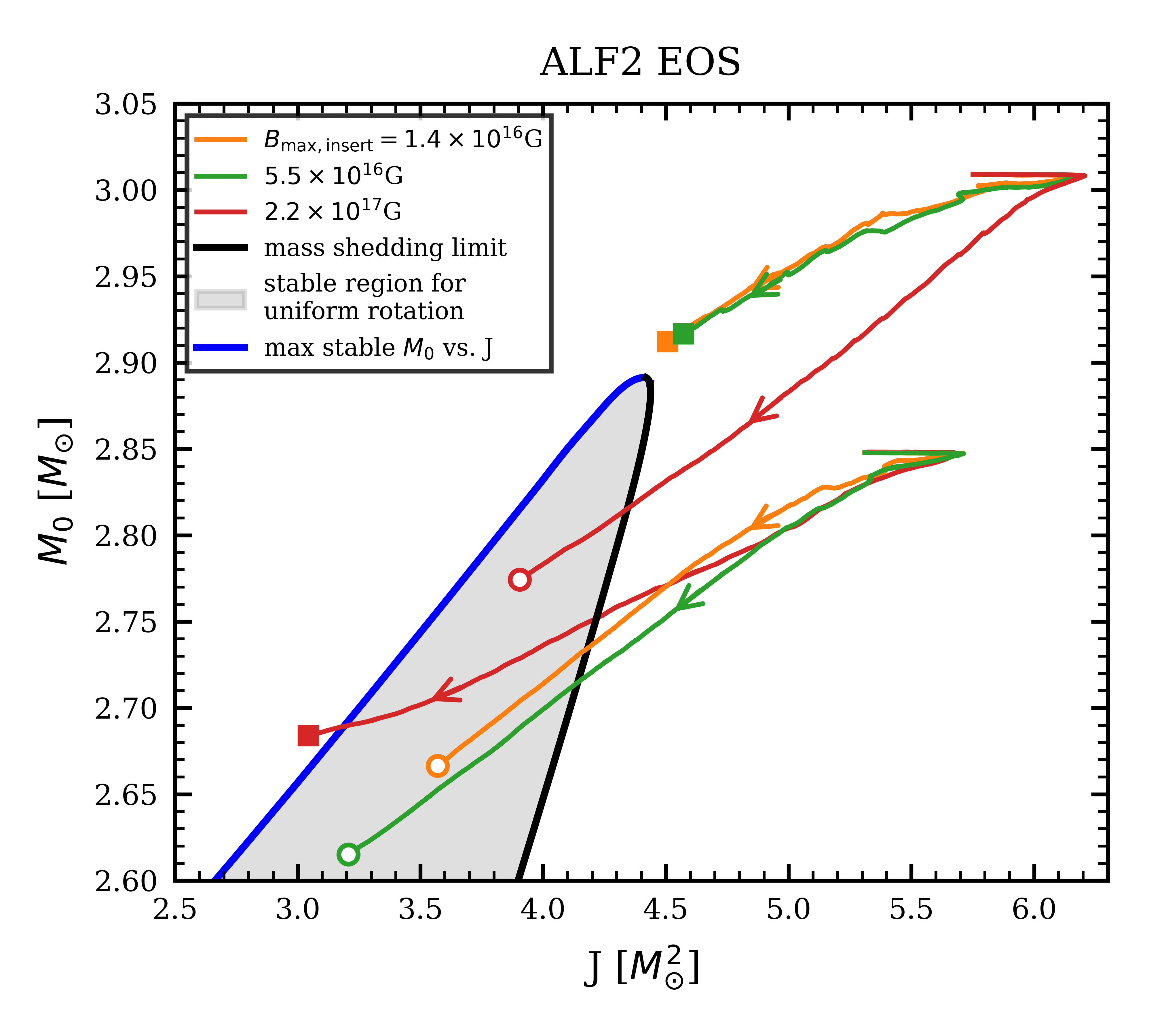}
    \caption{Trajectories of rest mass $M_0$ and angular momentum $J$ of the merger remnants for cases with the ALF2 EOS and pulsar-like initial magnetic fields, with $\vert B \vert_{\textup{max},\textup{insert}} =  1.4\times 10^{16}\textup{G},5.5\times 10^{16}$G and $2.2\times 10^{17}$G. Collapse to a BH is indicated with a solid square, while stable NS remnants at the end point of the simulation are shown with hollow circles.}
    \label{fig:M0_vs_J}
\end{figure}

The trajectories of the rest mass of the stars/remnant (defined in eq. \eqref{eq:M0_def}) vs. the maximum rest-mass density are shown in Figs. \ref{fig:M0_vs_rho_c_SLy} and \ref{fig:M0_vs_rho_c_ALF2} (the color coding for different magnetic fields is the same as in Fig. \ref{fig:GW_combined}), along with the equilibrium curves for cold spherical stars (thick black line), uniformly rotating stars at the mass-shedding (Keplerian) limit (thick red line) and the critical masses $M^{\textup{TOV}}_0$ and $M^{\textup{sup}}_0$. Remnants above the red dashed line are hypermassive, while those below the red dashed line but above the black dashed line are supramassive. 

For a uniformly rotating star to be stable it must be between the thick black and red curves, with a density to the left of the turning point for its angular momentum sequence (the locus of turning points for the ALF2 EOS is shown with a grey dashed line). This allows us to explain the fate of many of the remnants: for the cases with $M=2.57$ and $\vert B \vert_{\textup{max},\textup{insert}}\leq 5.5\times 10^{16}$ the final mass and maximum density is firmly within the stable region for uniformly rotating stars, ensuring the remnants do not collapse even as the magnetic field drives them toward uniform rotation. Conversely, most of the cases with masses that remain above the supramassive limit cannot survive as stable uniformly rotating stars, so collapse to BHs (indicated by the maximum density going beyond the right hand side of the plots) after the central density crosses the turning point threshold. The exception is the ALF2 $M=2.70$ case with initial interior-only poloidal magnetic fields (the upper brown line in Fig. \ref{fig:M0_vs_rho_c_ALF2}), which remains as a NS remnant by the end of the simulation at $25$ms postmerger, despite having a rest mass slightly above $M^{\textup{sup}}_0$. This may be because the thermal energy generated by the MRI induced turbulence is substantial, with $P_{\textup{th}}/P \sim 2-3\%$ in the core by the end of the simulation, which could increase the effective $M^{\textup{sup}}_0$ just enough for the remnant to remain stable temporarily against collapse. The ALF2 $M=2.70M_{\odot}$, $\vert B \vert_{\textup{max},\textup{insert}} = 0$ and $\vert B \vert_{\textup{max},\textup{insert}} = 5.5\times 10^{15}$G cases (the top thin black and blue lines in Fig. \ref{fig:M0_vs_rho_c_ALF2}) collapse to BHs despite having final rest masses below $M^{\textup{sup}}_0$. However, one can see that by the time the rest mass drops below $M^{\textup{sup}}_0$ the maximum density is too large for the star to survive in the stable region. 

The most interesting cases are those with the very largest magnetic field strength of $\vert B \vert_{\textup{max},\textup{insert}}=2.2\times 10^{17}$G (the thin red lines). One can see that, contrary to expectations, the higher mass ALF2 $M=2.70$ case ends up as a stable, long lived SMNS while the lower mass $M=2.57$ cases collapse to BHs. To explain this we need to examine the evolution of the angular momentum. In Fig.~\ref{fig:M0_vs_J} we show the trajectories of rest mass vs. angular momentum (defined in Eq.~\eqref{eq:J_def}) for the cases with the ALF2 EOS and the three largest initial pulsar-like magnetic fields. We also show the mass-shedding limit with a black line, and with a dark blue line the maximum rest mass for a stable uniformly rotating star with that angular momentum. The region between these curves, shown in grey, is the location of stable, cold, uniformly rotating stars. Collapse to a BH is shown with a solid square, while the end points of NS remnants that do not collapse are shown with hollow circles. 

From Figs.~\ref{fig:M0_vs_J} and \ref{fig:M0_vs_rho_c_ALF2} we see that the remnant in the ALF2 $M=2.57$, $\vert B \vert_{\textup{max},\textup{insert}}=2.2\times 10^{17}$G case starts with a rest mass below the supramassive limit and larger angular momentum than any stable equilibrium configuration. After merger it begins to lose both mass and angular momentum. The angular momentum is lost both via a redistribution to the outer disk due to magnetic effective turbulent viscosity, the loss of dynamical ejecta, and the emission of EM and GW radiation (see Fig. \ref{fig:MJ_conservation}), until it crosses the blue curve and ends up in the unstable region. By this point it has near-uniform rotation (with an almost flat rotation profile within $\sim 2.4M$, as shown in the right-hand plot of Fig. \ref{fig:Ang_vel}). As it is no longer supported by differential rotation, it undergoes catastrophic collapse to a BH. Conversely, the cases with smaller magnetic fields, and the $M=2.70$ cases with $\vert B \vert_{\textup{max},\textup{insert}}=2.2\times 10^{17}$G, maintain sufficient angular momentum to remain in the grey stable region, and consequently persist as SMNSs to the end of the simulation. 

We note that the idea that efficient loss of angular momentum could cause SMNS merger remnants to collapse on ms timescales was proposed in Ma et al. (2018) \cite{Ma:2017yva}, but excluded for hydro-only numerical relativity simulations by Radice et al. (2018) \cite{Radice:2018xqa}. Our results show that a NSNS binary with a baryonic mass well below the supramassive limit can in fact produce a remnant that collapses to a BH on ms timescales if the remnant has a very strong magnetic field. However, such magnetic fields are likely to be unrealistic for real astrophysical NSNS mergers, and would likely produce a EM counterpart with a luminosity much larger than that observed for the sGRB associated with GW170817 (see the discussion in section \ref{sec:EM_jets}). Therefore, unless an analogous source of efficient angular momentum loss can be established to occur in nature, the conclusion of \cite{Ruiz:2017due} that a hypermassive remnant is required for collapse on short timescales, remains valid. 

\subsubsection{Viscosity}
\label{sec:visc}

In the $\alpha$-disk model of a steady-state thin accretion disk the shear viscosity is given by 
\begin{equation}
    \nu = \alpha_{\nu} c^2_s \Omega^{-1}, 
\end{equation}
where $c_s$ is the sound velocity and $\alpha_{\nu}$ a dimensionless constant (see e.g. \cite{Shakura:1973boa,Shapiro:1983,Shibata:2017xht}). Radice (2017) \cite{Radice:2017zta} conducted binary NSNS merger simulations where turbulent viscosity was modelled using a general relativistic large-eddy formulation. In that work the addition of viscosity produced three competing effects: increased rotation of the central core, as the remnant is driven towards uniform rotation, causing it to expand; loss of angular momentum from the outer envelope, causing the remnant to contract; and thermal heating in the core from viscous dissipation, causing it to expand. As discussed above, our results show similar features due to the magnetic effective turbulent viscosity, except that the loss of angular momentum largely dominates over the other two effects.  This may be because angular momentum loss is driven not just by the turbulent viscosity by also by the magnetic braking, which induces angular momentum transport without associated thermal heating.

\begin{figure}
    \centering
    \includegraphics[width=\linewidth]{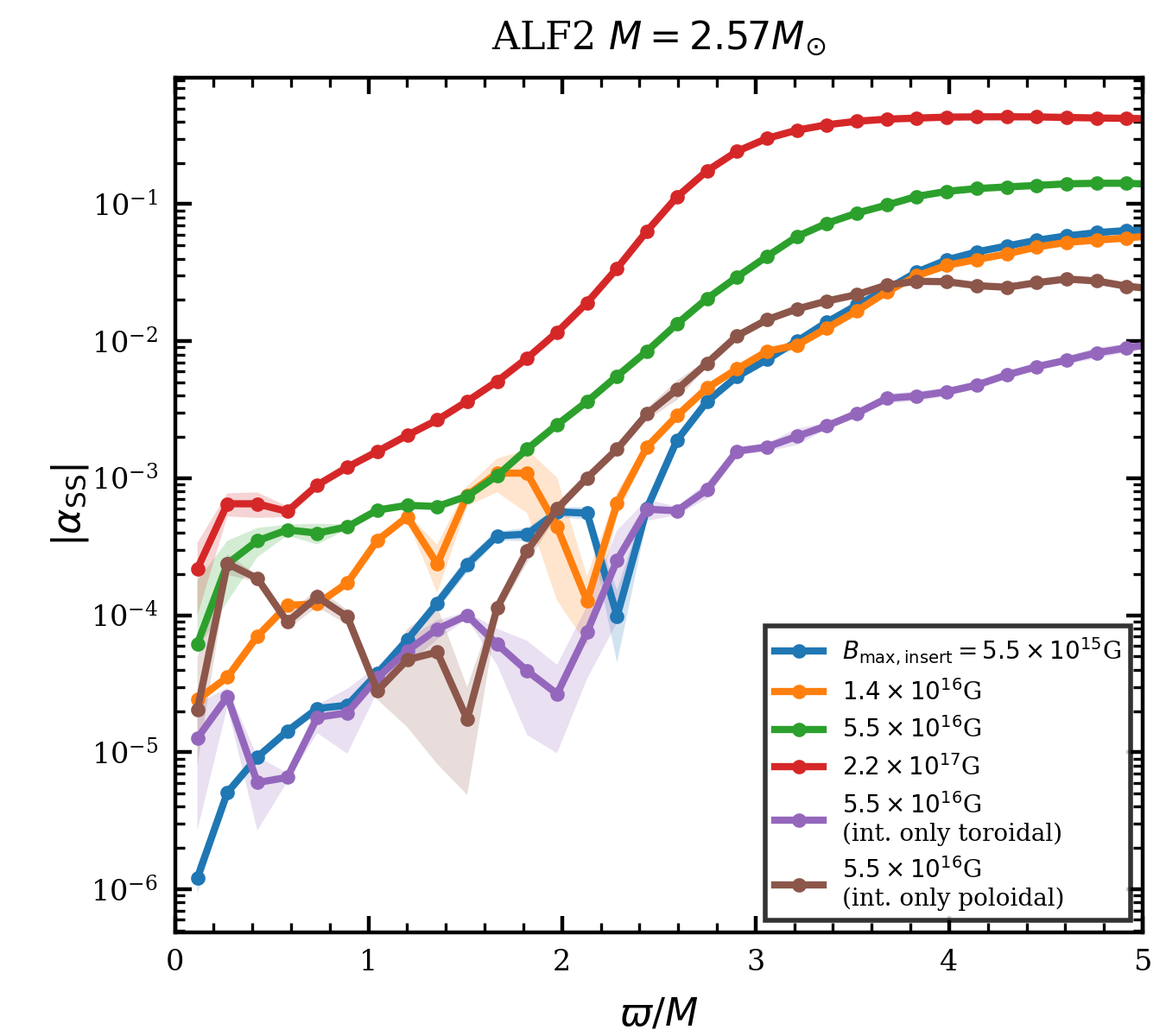}
    \caption{Shakura–Sunyaev viscosity parameter $\alpha_{\textup{SS}}$, computed using an circular average at each cylindical radius $\varpi$, and a time average over $\Delta t \sim 0.2$ms, at $\sim 14$ms postmerger. The shaded region shows the approximate variation around the average.}
    \label{fig:alpha_SS}
\end{figure}

Stresses due to the magnetic field effective viscosity can be approximated by the Shakura–Sunyaev parameter $\alpha_{\nu} = \alpha_{\textup{SS}}$ which is shown for the ALF2, $M=2.57$ cases as a function of cylindrical radius, averaged over azimuthal angle and $\vert z \vert <30M_{\odot}\sim 10 M$, in Fig. \ref{fig:alpha_SS} at $\sim 14$ms postmerger when all of the cases form a quasi-equilibrium NS remnant. We see that the viscosity increases for stronger magnetic fields, and spans a large range from $<10^{-3}$ at the center of the remnant to $\sim 10^{-2}$ to $0.4$ in the outer, low-density torus. We also see that $\alpha_{\textup{SS}}$ is smaller for the cases with initial interior only toroidal and poloidal magnetic fields, which we can attribute to reduced growth in the magnetic field for these cases compared to the pulsar-like ones with smaller $\vert B \vert_{\textup{max},\textup{insert}}$ (see the discussion of the evolution of the magnetic field in Section \ref{sec:mag_ev}). Mass-weighted spatial average values of $\alpha_{\textup{SS}}$ for all cases with magnetic fields are given in Table \ref{tab:alpha_SS_data}. Note that $\alpha_{\rm SS}$ is sensitive to numerical resolution, as low-resolution simulations tend to underestimate turbulence and angular momentum transport, leading to a lower $\alpha_{\rm SS}$, while higher resolutions better capture MRI and magnetic field dynamics, resulting in more accurate values with typical values ranging  between $0.01$ and $0.1$. The cases which collapse to BHs as SMNS remnants, with $M=2.57M_{\odot}$ and $\vert B \vert_{\textup{max},\textup{insert}} = 2.2\times 10^{17}$G, have $\langle \alpha_{\textup{SS}}\rangle \sim 0.06-0.07$, larger than the $\sim 0.01$ values considered in e.g. \cite{Shibata:2017xht,Shibata:2017jyf} but well within the range of values observed in numerical simulations \cite{Shi:2015mvh}, although the large $\sim 0.4$ value of $\alpha_{\rm SS}$ observed in the outer torus remain unrealistically large. 

\begin{table}
\begin{ruledtabular}
     \begin{tabular}{l|ccc}
       Magnetic topology  & SLy & ALF2 & ALF2 \\
      \& $\vert B \vert_{\textup{max},\textup{insert}}$ & $M=2.57M_{\odot}$ & $M=2.57M_{\odot}$ & $M=2.70M_{\odot}$ \\
      \hline
      PU $5.5\times 10^{15}$ & 0.007 & 0.010 & 0.002 \Tstrut \\
PU $1.4\times 10^{16}$ & 0.011 & 0.010 & 0.004 \\
PU $5.5\times 10^{16}$ & 0.017 & 0.020 & 0.013 \\
PU $2.2\times 10^{17}$ & 0.066 & 0.058 & 0.023 \\
IT $\,\;5.5\times 10^{16}$ & 0.004 & 0.001 & 0.003 \\
IP $\,\;5.5\times 10^{16}$ & 0.006 & 0.007 & 0.008
     \end{tabular}
\end{ruledtabular}
\caption{Mass-weighted average Shakura–Sunyaev viscosity parameter $\langle \alpha_{\textup{SS}} \rangle$ for all cases with magnetic fields at times $10$, $14$ and $9$ms postmerger for the SLy $M=2.57M_{\odot}$, ALF2 $M=2.57M_{\odot}$ and ALF2 $M=2.70M_{\odot}$ cases respectively. The magnetic field initial topology is denoted ``PU" for pulsar-like, ``IT" for interior-only toroidal and ``IP" for interior-only poloidal with the initial maximum field, as measured by normal observers in G.}
\label{tab:alpha_SS_data}
\end{table}

\subsubsection{Evolution of the magnetic field}
\label{sec:mag_ev}

\begin{figure}
    \centering
    \begin{tabular}{cc}\includegraphics{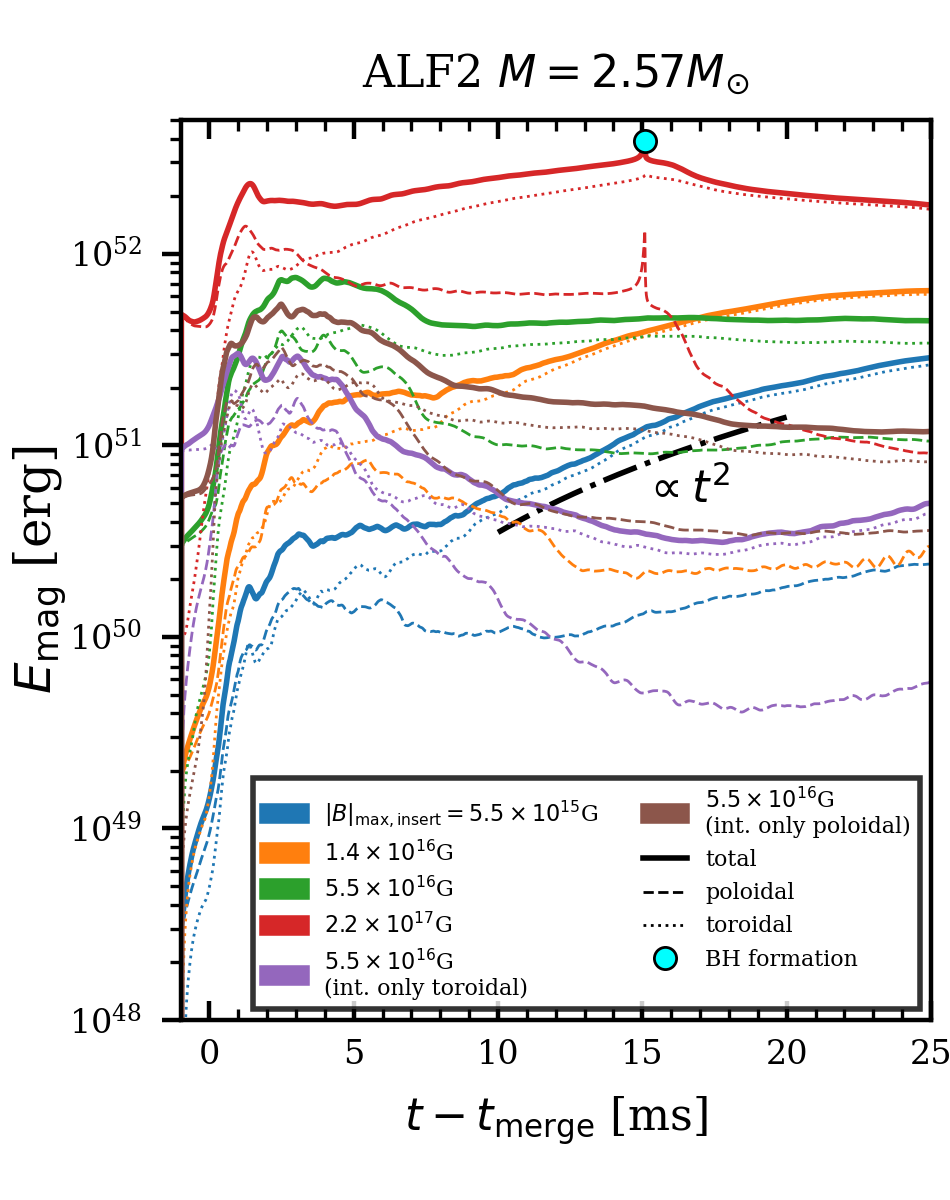} & \includegraphics{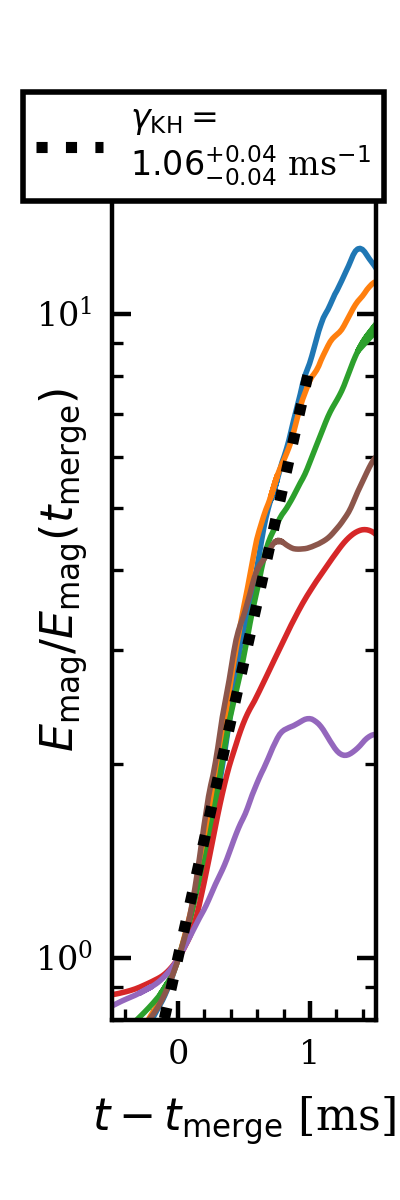} 
    \end{tabular}
    \caption{Left: total electromagnetic energy $E_{\rm mag}$ the ALF2 $M=2.57M_{\odot}$ cases along with its toroidal and poloidal components vs. time. For the case that collapses to a BH we exclude the energy within the apparent horizon and denote BH formation with a cyan circle. Growth proportional to $(t - t_\textup{merge})^2$, as expected for magnetic winding, is shown with a black dot-dashed line. Right: exponential growth of $E_{\rm mag}$ in the first ms after merger. The black dotted line shows a $E_{\textup{mag}}(t) = E_{\textup{mag}}(t_{\textup{merge}})\exp(2\gamma_{\textup{KH}}(t-t_{\textup{merge}}))$ trendline, fitted to the data.}
    \label{fig:EM_energy_ALF2_2.57}
\end{figure}

The total electromagnetic energy, along with its toroidal and poloidal components, for the ALF2 $M=2.57M_{\odot}$ cases (displayed as a representative set) is shown in Fig. \ref{fig:EM_energy_ALF2_2.57}. Although the initial maximum magnetic field strength is the same for all the $\vert B \vert_{\textup{max},\textup{insert}} = 5.5\times 10^{16}$ cases, the different magnetic field topology start with different total energies, with the largest being the interior-only toroidal case. As in our previous works \cite{Bamber:2024kfb} and those of other authors \cite{Kiuchi:2023obe} we see rapid exponential growth in the magnetic field energy in the first $\sim 1$ms after merger due to the KHI instability, shown in detail in the right-hand plot. For all the cases with initial poloidal-only fields we see the same growth rate in the linear regime of $\gamma_{\textup{KH}}\sim 1\;\textup{ms}^{-1}$, independent of magnetic field strength and in close agreement with the results of \cite{Bamber:2024kfb}. This is to be expected as the growth rate due to the KHI is dependent on the resolution (see section \ref{subsec:M_processes}) and we use the same resolution for all these cases and those in \cite{Bamber:2024kfb}. The case with initially toroidal interior-only magnetic fields shows a markedly slower growth rate. This may be because for a toroidal topology the strongest magnetic field is not located at the core of the star, where the KHI-induced turbulence is most active; also, magnetic winding does not operate for a toroidal field. From $\sim 1-3$ms the magnetic field energy continues to grow, likely due to a combination of KHI, magnetic winding and MRI-driven turbulence (and possibly an $\alpha\Omega$ dynamo as suggested in \cite{Kiuchi:2023obe}), until the magnetic field reaches saturation values or the MRI is terminated in the main body of the remnant as the rotation profile evolves such that $\partial_{\varpi}\Omega \geq 0$ (see Fig. \ref{fig:Ang_vel}). The outer region with a Keplerian rotation profile still satisfies $\partial_{\varpi}\Omega < 0$; however the weaker poloidal magnetic field and lower spatial resolution in these regions may mean we are not resolving the MRI in this regime. The decline in electromagnetic energy for the $\vert B \vert_{\textup{max},\textup{insert}} = 5.5\times 10^{16}$ cases after $\sim 3$ms postmerger was observed in \cite{Bamber:2024kfb,Palenzuela:2021gdo,Aguilera-Miret:2023qih} and may be due to magnetic field reconnection, as suggested by \cite{Hayashi:2022cdq}. The pulsar-like cases with $\vert B \vert_{\textup{max},\textup{insert}} = 5.5\times 10^{15}$G and $1.4\times 10^{16}$G show clear evidence of growth due to magnetic winding after $\sim 7$ms postmerger, with the toroidal component of the electromagnetic energy growing dramatically as $\propto t^2$ \cite{Kiuchi:2023obe} as the toroidal field grows as $B^T \propto t$ (Eq. \eqref{eq:BT_winding}) \cite{Shibata:2006hr}. The absence of similar behavior for the $\vert B \vert_{\textup{max},\textup{insert}} = 5.5\times 10^{16}$ cases may be because these have less differential rotation in the main part of the remnant at $t \gtrsim 7$ms postmerger, as shown in the middle panel of Fig. \ref{fig:Ang_vel}. The case with the strongest magnetic field does show evidence of winding despite reaching a near-uniform rotation profile within $r/M \lesssim 3$; however we have verified that this growth, unlike the cases with smaller magnetic fields, is entirely driven by the outer differentially rotating disk, as the electromagnetic energy contained within the region with $\rho_0/\rho^{\textup{max}}_{0,t=0} > 10^{-3}$ declines from $\sim 2$ms postmerger up till the collapse to a BH at $\sim 15$ms postmerger. 

\subsubsection{Effect of magnetic field topology}

In Fig. \ref{fig:3D_plots} we can see differences due to the different initial topologies. By the end of the simulation at $\sim 25$ms postmerger (bottom row) the case with initial pulsar-like magnetic fields has formed tightly wound helical magnetic field structures extending from the poles of the remnant, with mildly relativistic outflow along a magnetically-dominated evacuated funnel. By contrast, for the cases with initially interior-only magnetic fields the exterior magnetic field remains disorganized and weak by the end of the simulation, without the tightly collimated mildly relativistic outflow. The magnetic fields in the case with an initial toroidal interior-only topology also remain confined within the star for a longer time compared to the poloidal interior-only case. From our past studies \cite{Ruiz:2021qmm} and the works of other authors \cite{Kiuchi:2023obe,Ciolfi:2020hgg} it is likely that the cases with initially interior-only magnetic fields will eventually also produce large-scale helical magnetic fields and magnetically driven outflow, as the magnetic fields are amplified by magnetic winding and dynamo effects. However, these results show that the initial magnetic topology, and the presence of premerger external fields, has an important impact on the timescales of the formation of incipient jets, which will be discussed further in section \ref{sec:EM_jets}. 

\subsection{Incipient jets and luminosity}
\label{sec:EM_jets}

\subsubsection{Incipient jets}

 \begin{figure*}
    \centering
     \includegraphics[width=0.8\linewidth]{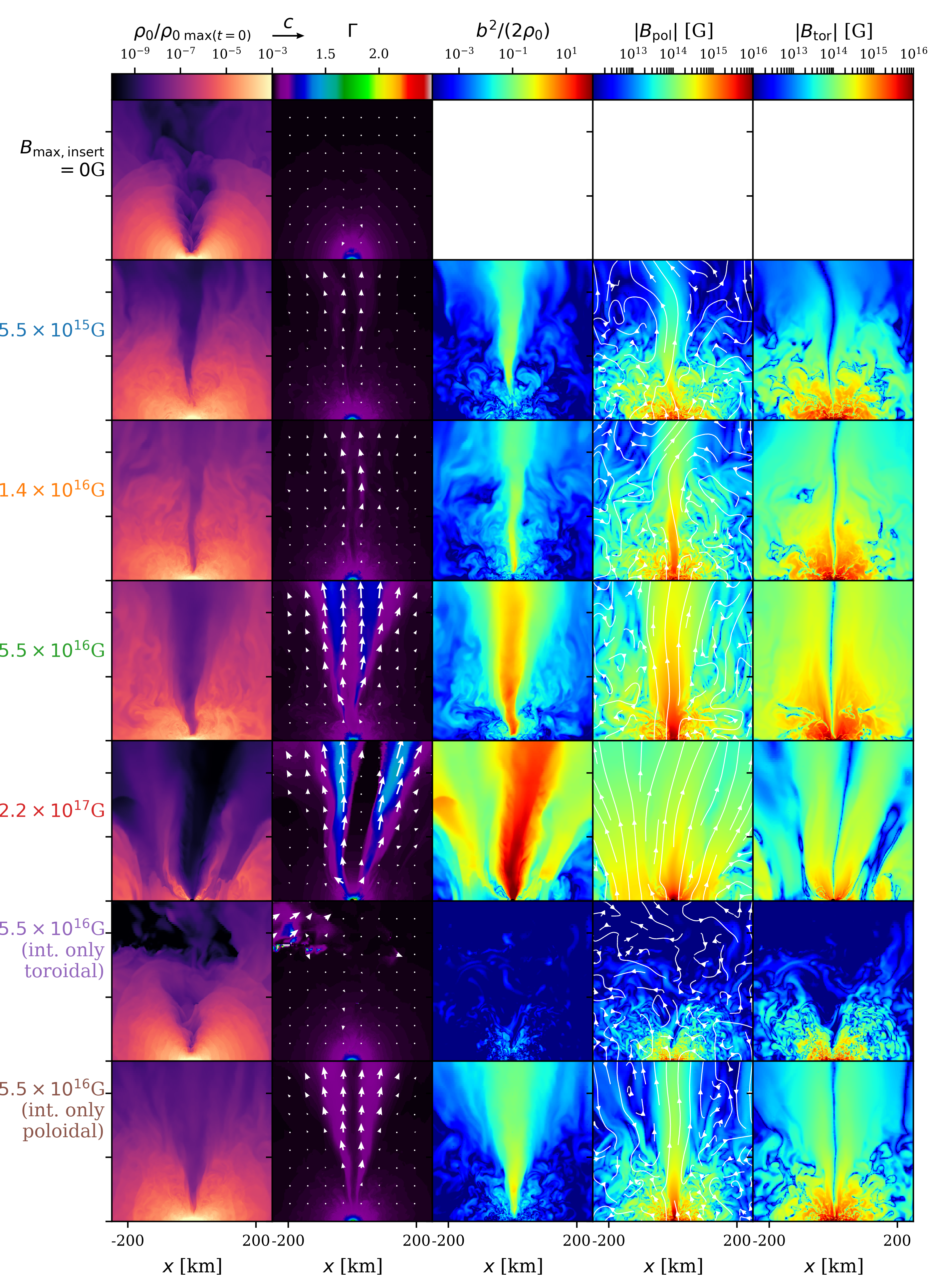}
     \caption{2D plots in the meridional $xz$ plane at $\sim 25$ms postmerger for the simulations with the ALF2 EOS and $M=2.57 M_{\odot}$. We show the rest-mass density $\rho_0$ on a log scale relative to the initial maximum rest-mass density (first column), the Lorentz factor (second column), $b^2/(2\rho_0)$ which approximately corresponds to the magnetization $\sigma$ (third column), and the strength of the poloidal and toroidal magnetic field (fourth and fifth columns), with the magnetic field lines shown in the poloidal field plots. The same scale is used for the $x$ and $z$ axes. The arrows in the Lorentz factor plot (second column) indicate the flow velocities, and the arrow labelled ``c'' at the top left of the second column indicates the magnitude of the speed of light compared to the white velocity arrows. The plots of $b^2/(2\rho_0)$, $B_{\textup{pol}}$ and $B_{\textup{tor}}$ for the hydro-only $\vert B \vert_{\textup{max},\textup{insert}} = 0$ case are left blank.}
     \label{fig:2D_plots}
\end{figure*}

 \begin{figure*}
    \centering
     \includegraphics[width=0.8\linewidth]{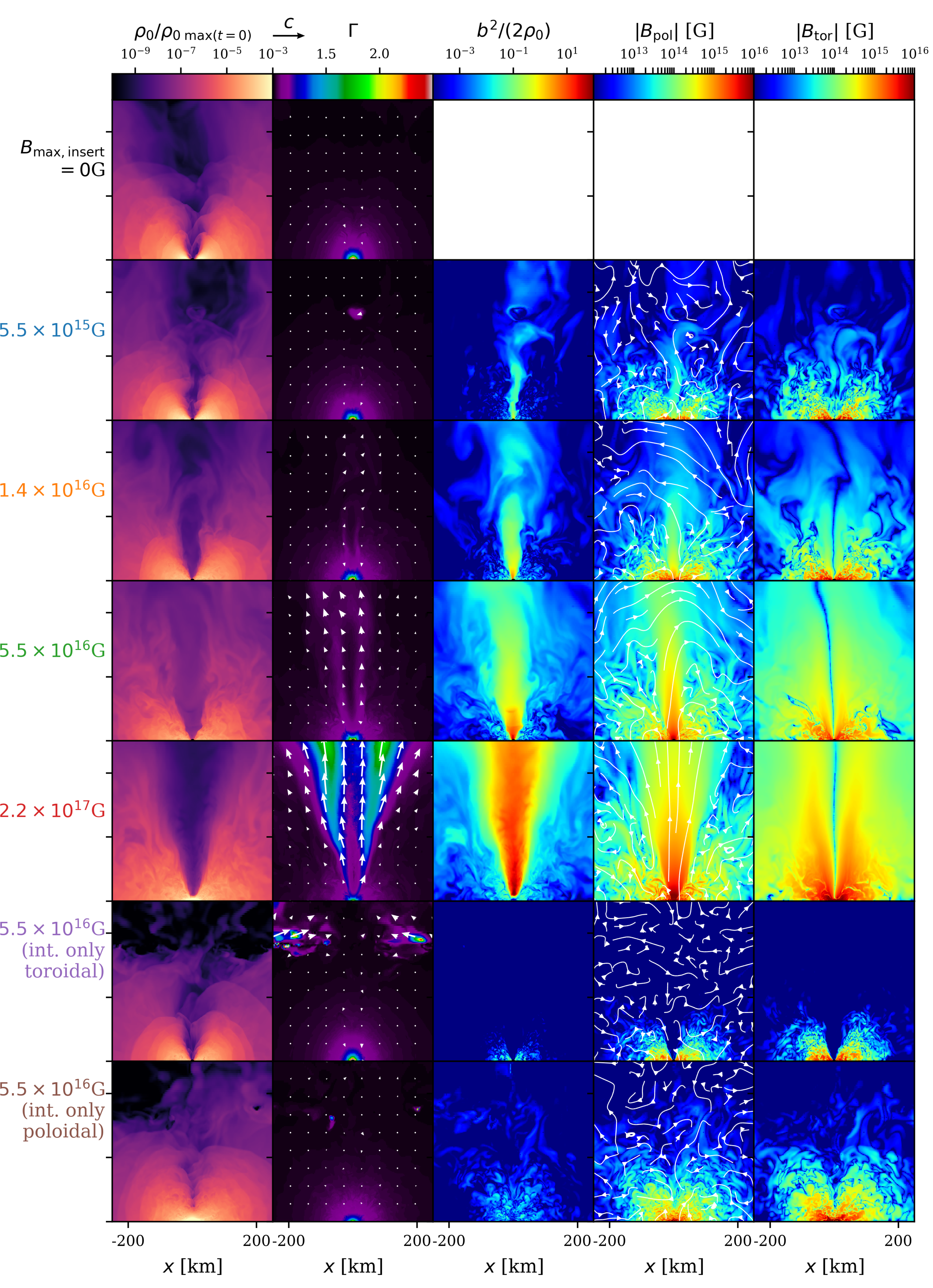}
     \caption{2D plots in the meridional plane at $\sim 25$ms postmerger for the simulations with the ALF2 EOS and $M=2.70 M_{\odot}$. Labelling is the same as in Fig. \ref{fig:2D_plots}.}
     \label{fig:2D_plots_2}
\end{figure*}

Along with the GW signal, another key observable related to the magnetic field is the (potential) EM counterpart associated with the NSNS merger. The only event with a confirmed EM counterpart observed to date is GW170817~\cite{LIGOScientific:2017vwq}, where short $\gamma$-ray burst (sGRB) GRB 170817A was detected $1.734 \pm 0.054$s after GW170817~\cite{LIGOScientific:2017ync,Connaughton:2017,Goldstein:2017mmi,Savchenko:2017ffs}. This was followed by observations across the electromagnetic spectrum (see e.g.~\cite{Coulter:2017wya,DAvanzo:2018zyz,Resmi:2018wuc} including UV/optical/infrared emission consistent with a kilonova or macronova \cite{Cowperthwaite:2017dyu,Smartt:2017fuw,Kasliwal:2017ngb}: a mostly-isotropic thermal transient powered by the radioactive decay in the neutron-rich non-relativistic ejecta \cite{Metzger:2019zeh}, with an ejecta mass of $0.04 \pm 0.01 M_{\odot}$ ($\sim 1.4\%$ of the total binary mass) with velocities of $\sim 0.1c$ to $\sim 0.3c$. The $\gamma-$ray emission has been attributed to a narrow collimated relativistic jet (where by ``jet" we mean a collimated outflow of EM fields and plasma along the rotational axis of the source \cite{DeYoung:1991}) with the $\gamma-$rays produced by either internal \cite{Rees:1994nw,Piran:2004ba} or external \cite{Katz:1993fn,Veres:2018trt} shocks. While the observed isotropic equivalent $\gamma$-ray luminosity, $L_{\gamma,\textup{iso}} \sim 10^{47}$erg s$^{-1}$, was unusually low for a sGRB this has been attributed to its jet being viewed off-axis at $\sim 20-30^{\circ}$ from the line of sight (see e.g. \cite{Lazzati:2017zsj,Ghirlanda:2018uyx,Troja:2020pzf,Fraija:2017aev,Mooley:2018qfh}). Its on-axis emission likely had an isotropic-equivalent $\gamma-$ray luminosity of $\geq 10^{51}$ erg s$^{-1}$ powered by a kinetic energy outflow of $\gtrsim 10^{53}$ erg s$^{-1}$, consistent with the typical values of observed sGRBs of $\sim 10^{49}-10^{54}$ erg s$^{-1}$ \cite{Li:2016pes,Beniamini:2020adb,Shapiro:2017cny} corresponding to a real $\gamma$-ray luminosity of $10^{47}-10^{52}$ erg s$^{-1}$ \cite{Zhu:2023nkx}. 

In order to produce this emission the jet must be able to accelerate the outflow of plasma up to asymptotic Lorentz factors of at least $\Gamma \gtrsim 20$ \cite{Ghirlanda:2017opl} with typical values being $\Gamma \sim O(10^2)$ \cite{Granot:2005ye,Zou:2010,Paczynski:1986px,Ghirlanda:2017opl}. There is still debate about the nature of the central engine powering the jets, as discussed in \cite{Bamber:2024kfb}. There is strong evidence that accretion onto a spinning BH can power an ultrarelativistic jet through the Blandford-Znajek (BZ) mechanism \cite{Blandford:1977ds,Kyutoku:2017voj} and efficiently accelerate the outflow out to large distances and up to ultrarelativistic speeds ($\Gamma_{\infty} \gg 100$ where $\Gamma_{\infty}$ is the asymptotic Lorentz factor) \cite{Paschalidis:2014qra,McKinney:2005zx,Gottlieb:2022tkb,Gottlieb:2022sis,Gottlieb:2023est}. The highly magnetized accretion disk that forms around the BH after the collapse of a HMNS creates the optimal conditions for a BZ-driven jet \cite{Ruiz:2016rai,Usov:1992zd}. However, it is less clear whether a magnetar central engine (i.e. a highly magnetized long-lived SMNS remnant immersed in a cloud of tidal debris) can also produce jets consistent with sGRBs \cite{Usov:1992zd,Duncan:1992hi,Zhang:2000wx,Metzger:2007cd,Metzger:2011,Bucciantini:2011kx,Metzger:2013cha,Ruiz:2017due,Ciolfi:2020hgg}. Plateaus in the soft X-ray emission of some sGRBs have been cited as evidence for continuous energy injection on longer timescales than BH accretion, and therefore as evidence for the spin-down radiation from a NS engine \cite{Ciolfi:2018tal,Rowlinson:2013ue,Zhang:2000wx,Jordana-Mitjans:2022gxy}. However, no such plateau was observed in the afterglow of the sGRB GW170817A and alternative explanations have been proposed for these X-ray features observed in other sGRBs (e.g. \cite{Strang:2019piq,Oganesyan:2020,Dereli-Begue:2022clf,Ciolfi:2014yla}) which are compatible with a BH engine. The additional energy injection from a magnetar remnant increases the energy of the quasi-isotropic ejecta and the associated kilonova, producing stronger radio emission at late-times \cite{Metzger:2013cha}. The non-detection of such late-time radio emission has been used to rule out a magnetar remnant central engine for some sGRBs \cite{Metzger:2013cha,Horesh:2016dah,Fong:2016orv}. Conversely, other authors have argued that, given the uncertainties in the physical parameters, radio observations of sGRB afterglows remain broadly compatible with magnetar remnants \cite{Liu:2019fgl}. The largest problem with the magnetar central engine model is greater baryon pollution, i.e.~the higher baryon density in the polar regions, compared to BH systems, which may limit the asymptotic Lorentz factor to $O(10)$ or lower \cite{Sarin:2020gxb}, and GRMHD simulations have produced mixed results (see e.g. \cite{Ciolfi:2020hgg,Mosta:2020hlh,Kiuchi:2023obe,Aguilera-Miret:2024cor} and \cite{Bamber:2024kfb} for a more detailed discussion). 

\begin{figure}
    \centering
    \includegraphics[width=\linewidth]{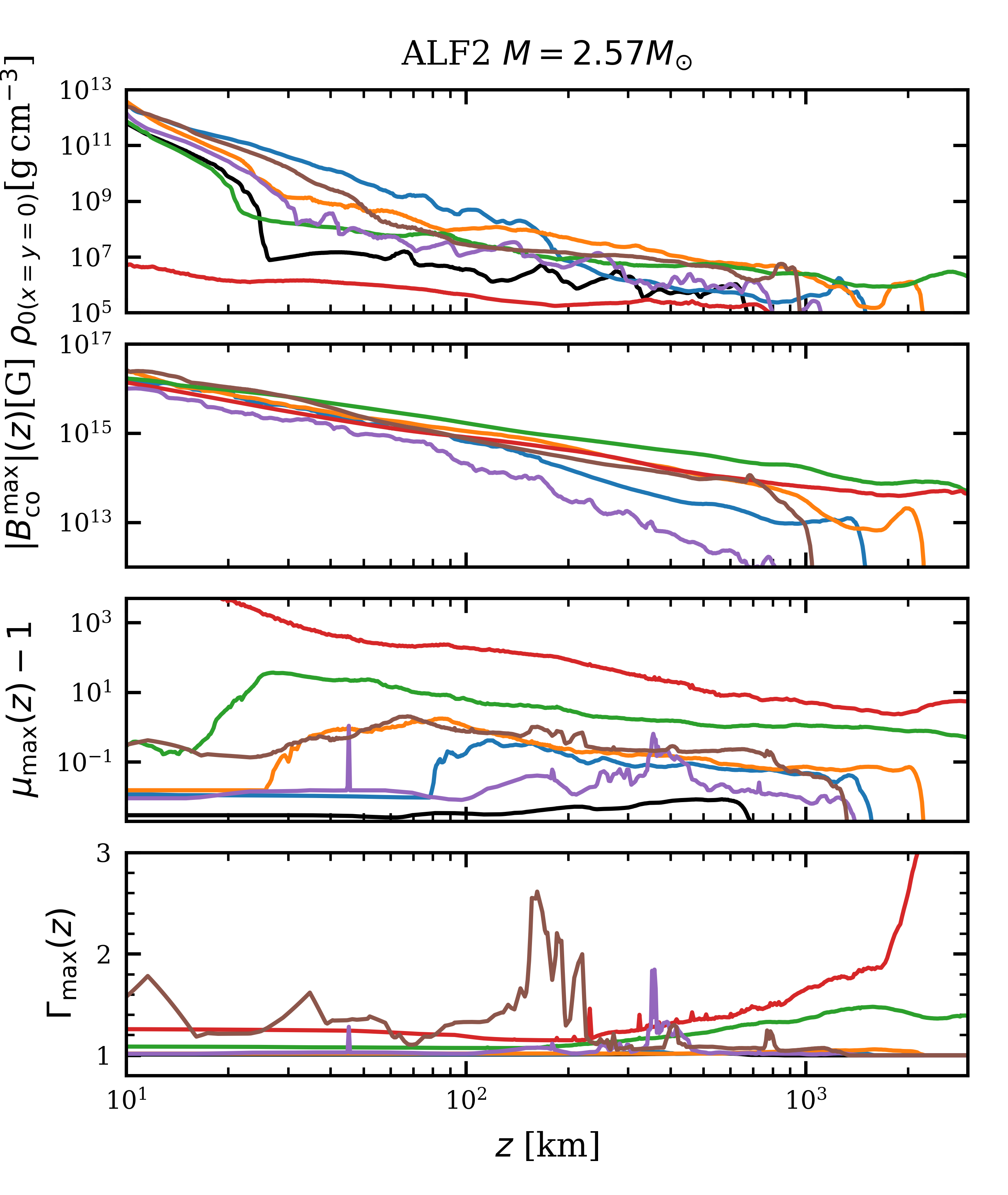}
    \caption{Change in outflow properties vs. height for the ALF2 $M=2.57 M_{\odot}$ cases. We show the rest-mass density along the $z$ axis (top), maximum comoving magnetic field $\vert B_{\textup{co}}\vert = \sqrt{4\pi b^2}$ at each height (second from top), maximum energy (minus rest-mass energy) to mass flux ratio ($\mu-1$) at each height (second from bottom), and maximum Lorentz factor at each height (bottom). The values are averaged over the last $2$ms of each simulation up to $25$ms postmerger. The color coding for each initial magnetic field is the same as all the other figures (e.g. Fig. \ref{fig:2D_plots}).}
    \label{fig:ALF2_2.57_vs_z}
\end{figure}

The set of simulations presented in this work are likely too limited in time and spatial range to fully determine the nature of any potential jet and sGRB emission from these systems (see the discussion in \cite{Bamber:2024kfb} and section \ref{sec:jet_theory}). However, we can still examine the outflow at the end of the simulations at $\sim 25$ms postmerger to assess whether they have the features of \textit{incipient} jets, which we define \cite{Paschalidis:2014qra,Ruiz:2021qmm,Bamber:2024kfb} as a tightly collimated, mildly relativistic outflow which is driven by a tightly wound, helical, force-free magnetic field, with the ratio of magnetic energy to rest-mass density in the comoving frame $b^2/(2\rho_0) \gg 1$. In Figs. \ref{fig:2D_plots}-\ref{fig:2D_plots_2} we show plots in the meridonal $xz$ plane of the rest-mass density, Lorentz factor, $b^2/(2\rho_0)$ and poloidal and toroidal components of the magnetic field for the ALF2 $M=2.57 M_{\odot}$ cases (the SLy $M=2.57 M_{\odot}$ cases give similar results) and the  ALF2 $M=2.70 M_{\odot}$ cases, respectively. For the ALF2 $M=2.57 M_{\odot}$ cases we also show how key outflow properties vary with height along the $z$ axis in Fig. \ref{fig:ALF2_2.57_vs_z}. A summary of the physical properties of all the NSNS merger remnants and their final fate is shown in Table \ref{tab:Results_data}.  

We can see from the first four rows of Figs. \ref{fig:2D_plots} and \ref{fig:2D_plots_2} that, as we might expect, increasing the initial strength of the magnetic field increases the value of the force-free parameter $b^2/(2\rho_0)$ in the central funnel along the polar axis at 25ms postmerger, from a maximum of $\sim 0.1$ for the $\vert B \vert_{\textup{max},\textup{insert}}=5.5\times 10^{15}$G case to values of well over 100 for the $\vert B \vert_{\textup{max},\textup{insert}}=2.2\times 10^{17}$G case, with values  $b^2/(2\rho_0)>10$ out to $>500$km from the central engine. In Fig. \ref{fig:2D_plots} one can see that the poloidal magnetic field for the $\vert B \vert_{\textup{max},\textup{insert}}=2.2\times 10^{17}$G case (row five), with a BH central engine, is actually weaker at 25ms postmerger than for the pulsar-like $\vert B \vert_{\textup{max},\textup{insert}}=5.5\times 10^{16}$G case (row four), which has a NS remnant. However, the matter density along the funnel is also much smaller with less baryon pollution, as you can see from the plots in the first column and in the top plot of Fig. \ref{fig:ALF2_2.57_vs_z}, which accounts for the larger $b^2/(2\rho_0)>10$. This pattern of systems with NS central engines having stronger magnetic fields but more baryon pollution compared to systems with BHs was also observed in \cite{Bamber:2024kfb}. The larger magnetic acceleration per unit mass also produces faster outflow, with the maximum Lorentz factor in the simulation region at 25ms postmerger, $\Gamma_{\textup{max}}$, reaching only slightly more than 1 for most of the low $\vert B \vert_{\textup{max},\textup{insert}}$ cases and more than 5 for the ALF2 $M=2.57M_{\odot}$, $\vert B \vert_{\textup{max},\textup{insert}}=2.2\times 10^{17}$G case (see Table~\ref{tab:Results_data} and Fig.~\ref{fig:ALF2_2.57_vs_z}). Let us now look at the cases with initial magnetic fields confined to the stellar interior. Both the $M=2.57$ and $M=2.70$ cases with initial toroidal interior-only fields (row six Figs. \ref{fig:2D_plots} and \ref{fig:2D_plots_2}) show no sign of producing jetlike features at 25ms postmerger, with the magnetic field largely confined to the dense matter torus, rather than the partially evacuated polar funnel. The ALF2 $M=2.57$ poloidal interior-only case (bottom row Fig. \ref{fig:2D_plots}) does show the formation of a magnetically dominated funnel and a large-scale exterior poloidal field from the poles, however the corresponding ALF2 $M=2.70M_{\odot}$ case does not. It is possible that the large amount of baryonic matter in the $M=2.70$ case makes it more difficult for the magnetic field to break out into the exterior regions. As in \cite{Bamber:2024kfb} we see in Fig. \ref{fig:ALF2_2.57_vs_z} (third row) that $\mu_{\textup{max}}$ not perfectly conserved along the outflow axis, which is likely either an indication that we have not reached ideal steady-state outflow, or an indication of mixing between the outflow in the funnel and the surrounding torus.

\begin{figure}
    \centering
    \includegraphics[width=\linewidth]{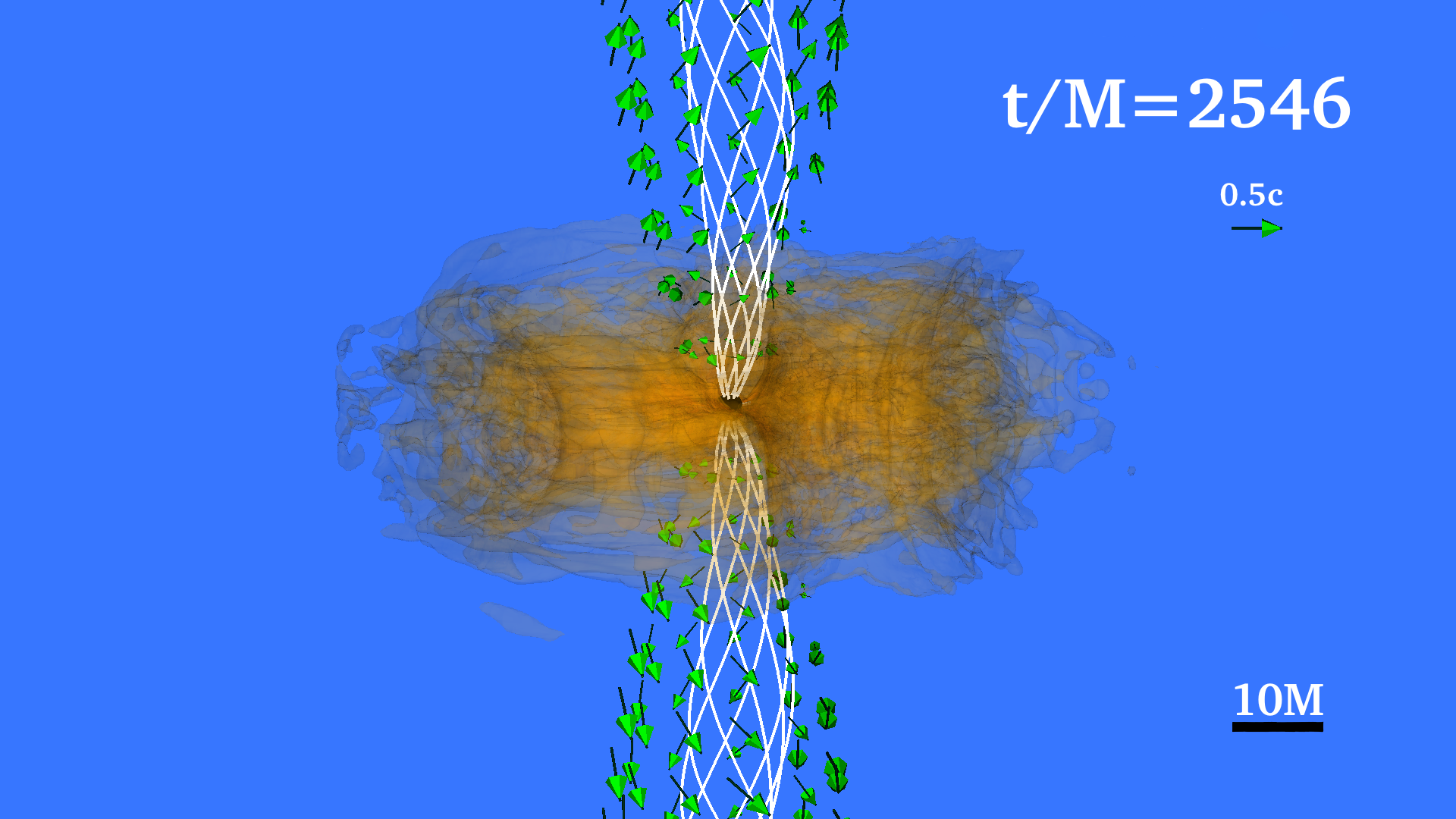}
    \caption{Final state of the ALF2 $M=2.57M_{\odot}$, $\vert B \vert_{\textup{max},\textup{insert}}=2.2\times 10^{17}$G case. As in Fig. \ref{fig:3D_plots} the orange disk shows the region with $\rho_0/\rho^{\textup{max}}_0(t=0)\gtrsim 10^{-6}$, the white lines show the magnetic field lines from the remnant poles and the green arrows show the fluid velocity. The black hole is shown in black.}
    \label{fig:ALF2_221017}
\end{figure}

The cases with $M=2.57M_{\odot}$ and $\vert B \vert_{\textup{max},\textup{insert}}=2.2\times 10^{17}$G are of particular interest as they collapse to BHs as supramassive remnants, and therefore could potentially mimic the electromagnetic emission of a short-lived HMNS remnant, as discussed in section \ref{sec:dyn}. As they show high $b^2/(2\rho_0)$, moderately large $\Gamma_{\textup{max}}$, collimated outflow and tightly wound magnetic helical field lines (shown in a 3D image of the final state of the ALF2 case in Fig. \ref{fig:ALF2_221017}) we can say these cases produce incipient jets. However, we caution that such high magnetic fields are likely unrealistic. Based on our previous works (e.g. \cite{Bamber:2024kfb,Ruiz:2021qmm,Ruiz:2016rai}) it is likely that the cases with $\vert B \vert_{\textup{max},\textup{insert}}=5.5\times 10^{16}$G would also continue to collimate and produce incipient jets, or at least jetlike structures, if allowed to progress further. It is possible the cases with smaller initial magnetic fields will eventually do the same (as in e.g. \cite{Kiuchi:2023obe}) once they develop sufficient magnetic pressure to overcome the ram pressure of the baryonic debris, but longer evolutions are required to verify this hypothesis. 

 \begin{table*}
  \begin{adjustbox}{width =\linewidth, center}
 %\begin{ruledtabular}
 \begin{tabular}{lll|cccccccccccccccc}
 \toprule \Tstrut
 EOS & M & $\vert B \vert_{\textup{max},\textup{insert}}$ & $t_{\textup{sim}}$ & $M_0^{\textup{rem}}$ & $J_{\textup{rem}}$ & $\Delta t_{\textup{BH}}$ & $M_{\textup{BH}}$ & $\chi_{\textup{BH}}$ & $M_{\textup{disk}}$ & $M_{\textup{esc}}$ & $E_{\textup{GW}}/M$ & $ J_{\textup{GW}}/J$ & $L_{\textup{EM}}$ & $L_{\textup{fluid}}$ & $\Gamma_{\textup{max}}$ & $L_{\textup{knova}}$ & $v_{\mathrm{eje}}$ & Fate. \\
 \hline
SLy & 2.57 &   0 & 40.3 & 2.74 & 3.68 &    &    &    & 0.22 & 0.0009 & 2.8\% & 30.2\% & $0$ & $10^{50.9}$ & 1.03 & $10^{40.0}$ & 0.06 & SMNS\Tstrut \\
SLy & 2.57 & \textcolor{C0}{PU $5.5\times 10^{15}$} & 34.2 & 2.68 & 3.14 &    &    &    & 0.27 & 0.0014 & 2.8\% & 30.0\% & $10^{49.8}$ & $10^{51.3}$ & 1.04 & $10^{40.3}$ & 0.11 & SMNS \\
SLy & 2.57 & \textcolor{C1}{PU $1.4\times 10^{16}$} & 34.0 & 2.64 & 2.84 &    &    &    & 0.30 & 0.0059 & 2.7\% & 29.1\% & $10^{50.6}$ & $10^{51.6}$ & 1.09 & $10^{40.6}$ & 0.17 & SMNS \\
SLy & 2.57 & \textcolor{C2}{PU $5.5\times 10^{16}$} & 31.6 & 2.64 & 2.93 &    &    &    & 0.23 & 0.0263 & 2.5\% & 27.0\% & $10^{52.3}$ & $10^{53.0}$ & 1.42 & $10^{41.1}$ & 0.37 & SMNS$^*$ \\
SLy & 2.57 & \textcolor{C3}{PU $2.2\times 10^{17}$} & 36.2 & 2.69 & 2.80 & 11.7 & 2.33 & 0.51 & 0.09 & 0.0444 & 1.9\% & 22.5\% & $10^{53.8}$ & $10^{54.0}$ & 5.02 & $10^{41.2}$ & 0.54 & BH$^{\dagger}$ \\
SLy & 2.57 & \textcolor{C4}{IT $5.5\times 10^{16}$} & 36.2 & 2.70 & 3.45 &    &    &    & 0.25 & 0.0019 & 2.6\% & 28.1\% & $10^{48.5}$ & $10^{51.3}$ & 1.12 & $10^{40.3}$ & 0.10 & SMNS \\
SLy & 2.57 & \textcolor{C5}{IP $5.5\times 10^{16}$} & 37.9 & 2.71 & 3.46 &    &    &    & 0.24 & 0.0011 & 2.7\% & 28.6\% & $10^{47.2}$ & $10^{51.1}$ & 2.60 & $10^{40.1}$ & 0.07 & SMNS \\
\hline
ALF2 & 2.57 &   0 & 37.3 & 2.68 & 3.96 &    &    &    & 0.27 & 0.0004 & 1.3\% & 20.0\% & $0$ & $10^{50.5}$ & 1.02 & $10^{40.0}$ & 0.08 & SMNS\Tstrut \\
ALF2 & 2.57 & \textcolor{C0}{PU $5.5\times 10^{15}$} & 34.9 & 2.66 & 3.45 &    &    &    & 0.28 & 0.0010 & 1.7\% & 24.0\% & $10^{49.6}$ & $10^{51.0}$ & 1.03 & $10^{40.2}$ & 0.11 & SMNS \\
ALF2 & 2.57 & \textcolor{C1}{PU $1.4\times 10^{16}$} & 36.3 & 2.67 & 3.57 &    &    &    & 0.27 & 0.0027 & 1.6\% & 22.8\% & $10^{50.5}$ & $10^{51.4}$ & 1.06 & $10^{40.5}$ & \red{0.16} & SMNS \\
ALF2 & 2.57 & \textcolor{C2}{PU $5.5\times 10^{16}$} & 34.6 & 2.62 & 3.21 &    &    &    & 0.26 & 0.0192 & 1.4\% & 20.5\% & $10^{52.3}$ & $10^{52.9}$ & 1.48 & $10^{41.0}$ & \red{0.41} & SMNS$^*$ \\
ALF2 & 2.57 & \textcolor{C3}{PU $2.2\times 10^{17}$} & 40.4 & 2.68 & 3.05 & 15.1 & 2.34 & 0.54 & 0.09 & 0.0437 & 1.1\% & 17.2\% & $10^{53.9}$ & $10^{53.9}$ & 5.35 & $10^{41.3}$ & \red{0.57} & BH$^{\dagger}$ \\
ALF2 & 2.57 & \textcolor{C4}{IT $5.5\times 10^{16}$} & 35.1 & 2.73 & 4.09 &    &    &    & 0.21 & 0.0007 & 1.7\% & 23.4\% & $10^{47.4}$ & $10^{50.9}$ & 1.58 & $10^{40.1}$ & \red{0.08} & SMNS \\
ALF2 & 2.57 & \textcolor{C5}{IP $5.5\times 10^{16}$} & 35.9 & 2.70 & 3.87 &    &    &    & 0.23 & 0.0007 & 1.6\% & 22.8\% & $10^{49.4}$ & $10^{50.7}$ & 2.94 & $10^{40.2}$ & \red{0.11} & SMNS$^*$ \\
\hline
ALF2 & 2.70 &   0 & 27.8 & 2.85 & 4.70 & 15.9 & 2.45 & 0.69 & 0.26 & 0.0021 & 1.8\% & 23.2\% & $0$ & $10^{50.5}$ & 1.02 & $10^{40.3}$ & \red{0.10} & BH\Tstrut \\
ALF2 & 2.70 & \textcolor{C0}{PU $5.5\times 10^{15}$} & 33.0 & 2.88 & 4.39 & 13.3 & 2.52 & 0.73 & 0.19 & 0.0007 & 2.0\% & 24.4\% & $10^{48.4}$ & $10^{51.1}$ & 1.02 & $10^{40.1}$ & \red{0.08} & BH \\
ALF2 & 2.70 & \textcolor{C1}{PU $1.4\times 10^{16}$} & 33.3 & 2.91 & 4.51 & 10.8 & 2.56 & 0.72 & 0.13 & 0.0013 & 2.3\% & 26.5\% & $10^{49.4}$ & $10^{51.0}$ & 1.02 & $10^{40.2}$ & \red{0.10} & BH \\
ALF2 & 2.70 & \textcolor{C2}{PU $5.5\times 10^{16}$} & 32.8 & 2.92 & 4.57 & 9.7 & 2.57 & 0.73 & 0.11 & 0.0032 & 2.2\% & 26.0\% & $10^{50.6}$ & $10^{51.6}$ & 1.78 & $10^{40.5}$ & \red{0.17} & BH$^{\dagger}$ \\
ALF2 & 2.70 & \textcolor{C3}{PU $2.2\times 10^{17}$} & 33.9 & 2.77 & 3.90 &    &    &    & 0.24 & 0.0250 & 0.9\% & 15.3\% & $10^{52.7}$ & $10^{53.2}$ & 2.10 & $10^{41.0}$ & \red{0.34} & SMNS$^*$ \\
ALF2 & 2.70 & \textcolor{C4}{IT $5.5\times 10^{16}$} & 32.8 & 2.95 & 4.70 & 10.8 & 2.59 & 0.75 & 0.10 & 0.0008 & 2.4\% & 27.8\% & $10^{44.9}$ & $10^{51.7}$ & 3.21 & $10^{40.1}$ & \red{0.09} & BH \\
ALF2 & 2.70 & \textcolor{C5}{IP $5.5\times 10^{16}$} & 36.8 & 2.90 & 4.86 &    &    &    & 0.19 & 0.0004 & 1.3\% & 18.8\% & $10^{47.0}$ & $10^{50.5}$ & 1.58 & $10^{39.9}$ & \red{0.06} & HMNS \\
\botrule
 \end{tabular}
% \end{ruledtabular}
\end{adjustbox}
 \caption{Summary of the key outcomes for our NSNS merger simulations. The ADM mass $M$ is given in solar masses, and the magnetic field initial topology is denoted ``PU" for pulsar-like, ``IT" for interior-only toroidal and ``IP" for interior-only poloidal with the initial maximum field, as measured by normal observers, given in G. The total simulation time is $t_{\textup{sim}}$ [ms], and $\Delta t_{\textup{BH}} = t_{\textup{BH}} - t_{\textup{merge}}$ [ms], the time between peak GW amplitude and BH formation for the cases which form a BH. $M_0^{\textup{rem}}$ $[M_{\odot}]$ and $J_{\textup{rem}}$ $[M^2_{\odot}]$ are the rest mass and angular momentum of the NS remnants defined as in Eqns. \eqref{eq:M0_def}\eqref{eq:J_def}, evaluated either just before collapse to a BH or at $t_{\textup{sim}}$ if the NS remnant does not collapse. $M_{\textup{BH}}$ and $\chi_{\textup{BH}}$ are the mass and spin of the black holes at $t_{\textup{sim}}$, calculated using the isolated horizon formalism~\cite{Dreyer:2002mx}. The rest mass in the accretion disk is denoted $M_{\textup{disk}}$, also evaluated at $t_{\textup{sim}}$, while $M_{\textup{esc}}$ denotes the rest mass which escapes as ejecta, calculated via Eq. \eqref{eq:esc_mass}. The fractions of the total energy $M$ and angular momentum $J$ carried off by GWs are $E_{\textup{GW}}/M$ and $ J_{\textup{GW}}/J$ respectively. $L_{\textup{EM}}$ [erg/s] and $L_{\textup{fluid}}$ [erg/s] denote the Poynting and fluid luminosity, respectively, averaged over the last $\sim$5ms of the simulation and $\Gamma_{\textup{max}}$ denotes the maximum Lorentz factor observed for the fluid within the simulation box at $t_{\textup{sim}}$. $L_{\textup{knova}}$[erg/s] denotes the estimated peak EM luminosity in $\textup{erg}\;\textup{s}^{-1}$ of the potential kilonova arising from the sub-relativistic ejecta,  calculated from the ejecta mass and the ejecta velocity $v_{\textup{eje}}$ averaged over $500$km around the peak of the outflow. The final column shows the fate of the merger. A $\dagger$ symbol denotes cases which show evidence of an incipient jet by the end of the simulation, powered by the BZ mechanism, which are likely precursors of sGRBs. An asterisk denotes cases which show incipient jets with SMNS central engines by the end of the simulation, for which the connection to sGRBs is more uncertain (see \cite{Bamber:2024kfb}).}
 \label{tab:Results_data}
 \end{table*}

\subsubsection{Luminosity}

\begin{figure}
    \centering
    \includegraphics[width=\linewidth]{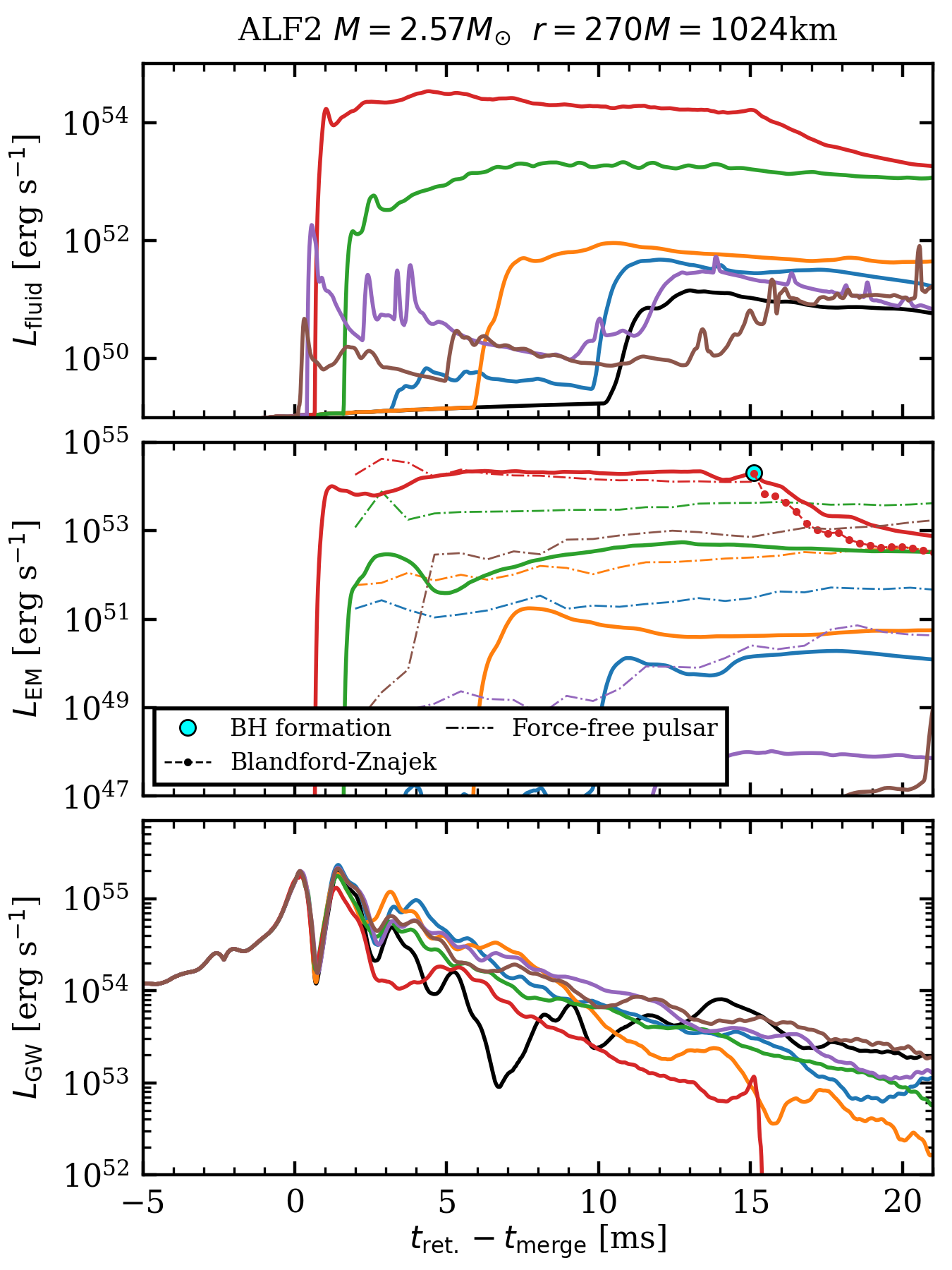}
    \caption{Luminosity via fluid outflow, EM radiation and  GWs, vs. retarded time since $t_{\textup{merge}}$ (defined as the time of peak GW emission) extracted over a sphere at $r_{\textup{ext}} = 270M = 1024\textup{km}$ for the ALF2 $M=2.57M_{\odot}$ cases. The predicted EM luminosity for the BZ mechanism (equation \ref{eq:BZ_lum}) is shown with large connected dots, while the predicted EM luminosity for a force-free, rigidly rotating spherical pulsar (Eq. \ref{eq:pulsar_lum}) is shown with thin dot-dashed lines. The color coding for the different magnetic fields is the same as in the other figures (e.g. Fig. \ref{fig:2D_plots})}
    \label{fig:combined_luminosity}
\end{figure}

The expected luminosity for a jet powered by the steady-state BZ mechanism can be estimated as \cite{Thorne:1986} 
\begin{equation}
\begin{split}
    L_{\textup{BZ}} \sim& \frac{1}{128} \chi^2 r^2_H B^2_p \\
                           \sim& 10^{52} \left(\frac{\chi}{0.64}\right)^2\left(\frac{M_{\textup{BH}}}{2.5M_{\odot}}\right)^2\left(\frac{B_{p}}{10^{16}\textup{G}}\right)^2\textup{erg}\,\textup{s}^{-1},
    \label{eq:BZ_lum}
\end{split}
\end{equation}
where $B_{p}$ is the poloidal magnetic field measured at the BH pole, $\chi$ the dimensionless spin of the BH and $r_H$ the equatorial BH radius. For a rigidly rotating oblate pulsar with a force-free magnetosphere in flat space time with angular velocity $\Omega$ and magnetic dipole moment $\mu_B$ Ruiz et al. (2015) \cite{Ruiz:2014zta}, consistent with the results of \cite{Spitkovsky:2006np}, found the luminosity can be given by
\begin{equation}
\begin{split}
    L_{\textup{pulsar}} \approx& 1.02 \mu^2_B \Omega^4\\
                                      \sim& 10^{51} \left(\frac{B_{p}}{10^{16}\textup{G}}\right)^2 \left(\frac{R}{10\textup{km}}\right)^6 \left(\frac{P}{\textup{ms}}\right)^{-4} \textup{erg}\,\textup{s}^{-1}, \label{eq:pulsar_lum}
\end{split}
 \end{equation}
where $R$ is the stellar radius and $P$ the rotation period. In Fig. \ref{fig:combined_luminosity} we show the evolution of the fluid luminosity (the outflow of fluid kinetic and internal energy), the EM luminosity, and the GW luminosity, for the set of ALF2 $M=2.57M_{\odot}$ simulations, evaluated at a distance of $260M = 1024\textup{km}$. In the middle plot we also show the predicted EM luminosity from \eqref{eq:BZ_lum} for cases which have collapsed to a BH and \eqref{eq:pulsar_lum} when there is a rotating NS remnant. For a magnetic dipole in spherical coordinates $\{r,\theta,\phi\}$, and weak gravity, the orthonormal poloidal magnetic field is $B_r = 2\mu_B\cos(\theta)/r^3, B_{\theta} = \mu_B\sin(\theta)/r^3$ so $\mu_B = r^3\sqrt{(B_r/2)^2 + B^2_{\theta}}$. The postmerger magnetic field of our NS remnants is not a perfect dipole, so we calculate an approximate effective dipole moment in the outflow region with the largest $b^2/(2\rho_0)$ (and thererfore closest to force-free) as 
 \begin{equation}
 	\mu_{B} \approx \frac{1}{V} \int_{V} r^3\sqrt{(B_r/2)^2 + B^2_{\theta}} \dd^3 x,
 \end{equation}
where volume $V$ is a cylinder of radius $R_*$ around the $z$ axis and above the NS pole, at a height $2R_* < z < 4R_*$ where $R_*=6M_{\odot}$ is the approximate radius of the stars. We estimate $\Omega$ as a mass-weighted average over $r \leq R_*$ in the $xy$ plane. 

From Fig.~\ref{fig:combined_luminosity} one can see that pulsar formula \eqref{eq:pulsar_lum}, shown with a thin dot-dashed line, approximately matches the observed precollapse EM luminosity for the case with the largest magnetic field of $\vert B \vert_{\textup{max},\textup{insert}}=2.2\times 10^{17}$G (red lines), but significantly overestimates the luminosity for the cases with smaller magnetic fields. This can be explained both by the fact that a significant fraction of the EM emission has been converted to kinetic energy of the fluid by the time it reaches the extraction radius, as discussed in section~\ref{sec:jet_theory}, and the fact that the NS remnants are not in a force-free environment but are surrounded by significant baryon pollution, as shown in Fig.~\ref{fig:2D_plots}, with more baryon pollution for cases with smaller initial magnetic fields. The case with the largest magnetic field is closest to force-free and the case with the initial interior-only toroidal magnetic field (brown lines) is furthest from being force-free, so we see~Eq.~\eqref{eq:pulsar_lum} provides the best prediction for the former case and the worst (by several orders of magnitude) for the latter. For the $\vert B \vert_{\textup{max},\textup{insert}}=2.2\times 10^{17}$G case that forms a BH the BZ formula matches the observed EM luminosity on BH formation (shown with a cyan circle) but subsequently somewhat underpredicts the observed emission, although it tracks the decrease in luminosity over time. This is may be because at 25ms postmerger we have not yet reached steady-state, and the environment is not highly force-free, even for this case with the strongest magnetic field.
We can see in the top panel that the fluid luminosity shows many of the same trends as the EM luminosity, with smaller magnetic fields associated with smaller and more delayed outflows. 

The electromagnetic luminosity $L_{\textup{EM}}$ can be related to the isotropic-equivalent $\gamma-$ray luminosity for a jet observed head-on as \cite{Bamber:2024kfb}
\begin{equation}
    L_{\gamma,\textup{iso}} = \frac{1}{C_{\textup{col}}}\eta_{\textup{EM}}L_{\textup{EM}},
\end{equation}
where $C_{\textup{col}}$ is a factor that accounts for $\gamma$-ray collimation \cite{Khan:2018ejm}, and $\eta_{\textup{EM}}$ corresponds to the efficiency of converting the outgoing Poynting flux in the simulation region to $\gamma-$ray photons in the emission zone. For perfect collimation, where all the emission is contained within a homogenous jet of half-opening angle $\theta$, we have $C_{\textup{col}} = 1 - \cos \theta \sim 10^{-2}-10^{-1}$ for $\theta \sim 10^{\circ}-30^{\circ}$ \cite{Escorial:2022nvp}. In our simulations the half-opening angle for the EM Ponyting flux is $\sim 10^{\circ}$ at $r=1024$km \cite{Bamber:2024kfb}.

The prompt $\gamma-$ray emission mechanism is still an open question \cite{Parsotan:2022uur} making the efficiencies highly uncertain \cite{Beniamini:2016hzc}, although $\sim 10^{-1}$ has been taken as a fiducial value~\cite{Khan:2018ejm} for the EM Poynting luminosity for outflow from a black hole central engine. For $\eta_{\textup{EM}} \sim 10^{-1}$ the cases with $M=2.57M_{\odot}$, $\vert B \vert_{\textup{max},\textup{insert}}=2.2\times 10^{17}$G which collapse to BHs as SMNS remnants have $L_{\textup{EM}} \sim 10^{54} \textup{erg}\;\textup{s}^{-1}$ giving $L_{\gamma,\textup{iso}} \sim 10^{54}-10^{55}\textup{erg}\;\textup{s}^{-1}$, which is above the range of isotropic equivalent luminosities for observed sGRBs \cite{Li:2016pes,Beniamini:2020adb,Shapiro:2017cny}. However, given the uncertainties it is possible the efficiency is lower, $\eta_{\textup{EM}} \lesssim 10^{-2}$, in which case even the cases with the strongest magnetic fields could potentially be compatible with the upper limit from observations.  

The dynamical ejected mass varies from $< 10^{-3}M_{\odot}$ to $\sim 0.04 M_{\odot}$, as shown in Table \ref{tab:Results_data}, with velocities $v_{\textup{eje}}$ from $\sim 0.1c$ to $\sim 0.5c$, with more mass ejected at higher velocity at 25ms postmerger for cases with stronger magnetic fields. This ejecta would likely produce a kilonova transient \cite{Metzger:2019zeh,Radice:2018pdn}. Using the analytic model derived in \cite{Perego:2021dpw} and discussed in further detail in \cite{Ruiz:2021qmm}, Sec. III C (see Eq. (8)-(10) in \cite{Ruiz:2021qmm}) we estimate the peak luminosity $L_{\textup{knova}}$, the peak time of the kilonova emission $\tau_{\textup{peak}}$, and the effective black-body temperature $T_{\textup{peak}}$. We find the estimated luminosities ($L_{\textup{knova}} \sim 10^{40}\textup{erg}\;\textup{s}^{-1}$ for the pure-hydro cases, up to $L_{\textup{knova}} \sim 10^{41}\textup{erg}\;\textup{s}^{-1}$ for cases with strong magnetic fields) are consistent with those measured for kilonovae in general \cite{Ascenzi:2018mbh,Rastinejad:2021nev} and the magnetized cases are consistent with the $L_{\textup{knova}}\sim 10^{41}$erg s$^{-1}$ reported for kilonova AT2017gfo associated with GRB 170817A \cite{Villar:2017wcc}. The peak times of $\tau_{\textup{peak}} \sim 2 - 7$ days and ejecta masses are also consistent with observations. The ejecta velocity for the cases with the strongest magnetic fields is larger than the typical $\sim 0.1c - 0.3c$ usually quoted (e.g. \cite{Metzger:2019zeh}), however it has been suggested that the fastest component of AT2017gfo reached $0.4-0.45c$ \cite{Sneppen:2024gnt}. The estimated effective temperatures of  $T_{\textup{peak}} \sim 10^{3.0 - 3.5}$K correspond to wavelengths of $\lambda_{\textup{peak}} \sim 800-2000$nm, with the hottest, bluest emission produced by the cases with the largest magnetic fields. We note, however, that the overprediction of electromagnetic luminosity, and the difficulty in generating such a strong magnetic field from realistic pre-merger neutron star fields, means that the cases with the strongest initial magnetic fields ($\vert B \vert_{\textup{max},\textup{insert}}=2.2\times 10^{17}$G) are unlikely to be realistic models for the GW1701817 or similar events, even if the predicted kilonova luminosity or other observables are compatible with observations.

\begin{figure}
    \centering
    \includegraphics[width=\linewidth]{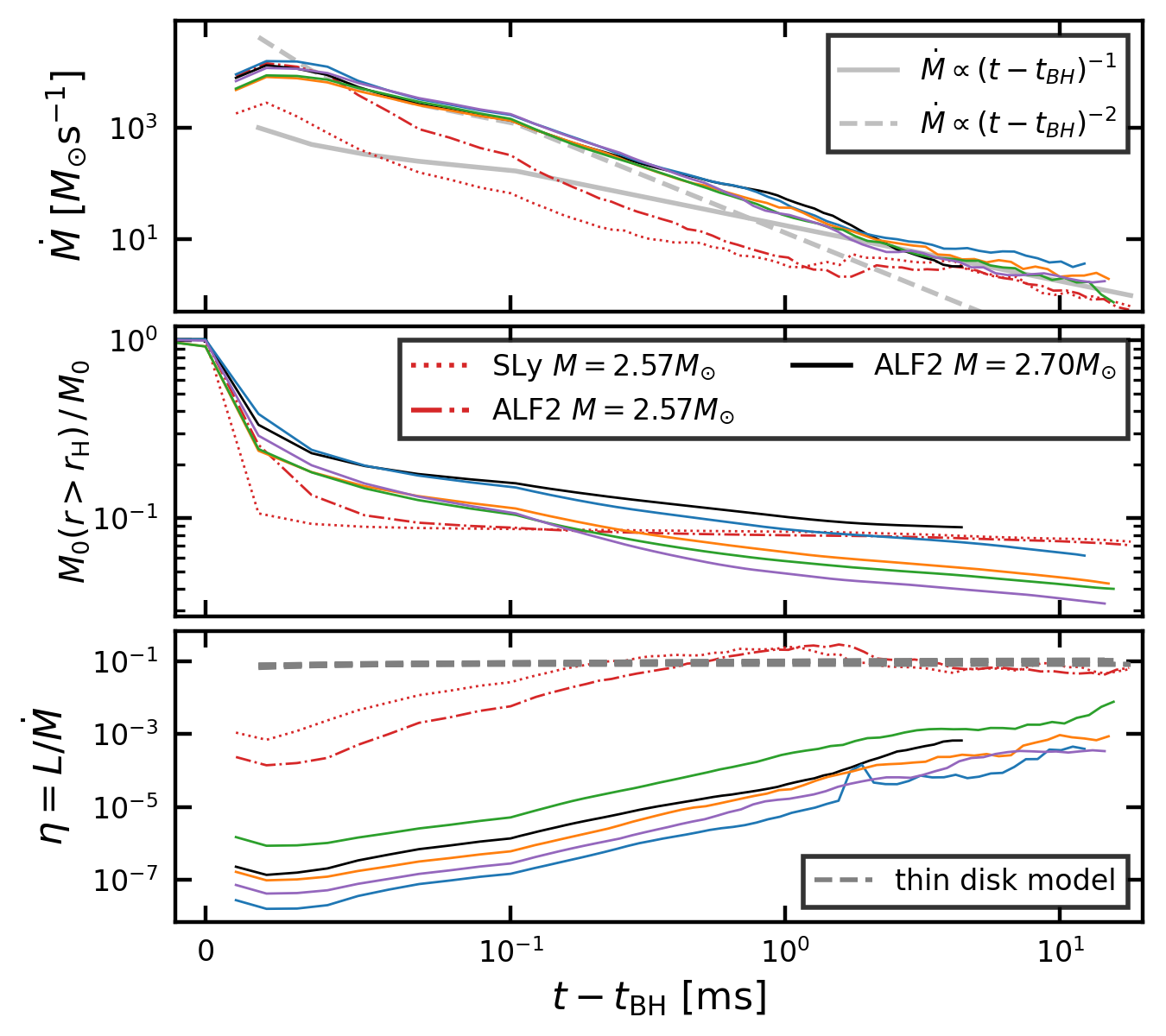}
    \caption{The accretion rates onto the BH (top), fraction of the total rest mass outside the apparent horizon (middle), and the accretion efficiency (bottom) for the cases which form a BH before the end of the simulation. The SLy $M=2.57M_{\odot}$ case is shown with a dotted line, the ALF2 $M=2.57M_{\odot}$ case with a dot-dashed line and the ALF2 $M=2.70M_{\odot}$ with solid lines, with the same color coding for different magnetic fields as in all the other figures. In the top plot we also show $\dot{M} \propto (t-t_\textup{BH})^{-2}$ and $\dot{M} \propto (t-t_\textup{BH})^{-1}$ trendlines. In the bottom panel the accretion efficiency for an ideal Novikov-Thorne thin disk \cite{Shapiro:2004ud} is shown with a black dashed line.}
    \label{fig:BH_accretion}
\end{figure}

The accretion rate $\dot{M}$, rest mass outside the apparent horizon and accretion efficiency $\eta = (L_{\textup{fluid}} + L_{\textup{EM}})/\dot{M}$ are shown in Fig. \ref{fig:BH_accretion}. We see the accretion rate is fairly consistent accross all the ALF2, $M=2.70 M_{\odot}$ cases which have the same ADM mass and EOS. At early times, $t - t_{\textup{BH}} \lesssim 0.5$ms, the accretion rate decays more quickly, closer to $\dot{M} \propto (t-t_{\textup{BH}})^{-2}$ the expected scaling for highly magnetized disks \cite{Gottlieb:2023est}, while for $t - t_{\textup{BH}} \gtrsim 2$ms it transitions to a scaling closer to $\dot{M} \propto (t-t_{\textup{BH}})^{-1}$, the scaling for hydrodynamically-dominated accretion driven by shocks and spirals in the disk \cite{Gottlieb:2023est}. The ALF2, $M=2.57M_{\odot}$, $\vert B \vert_{\textup{max},\textup{insert}}=2.2\times 10^{17}$G case (red dot-dashed line) shows a noticeably steeper decline in the accretion rate and maintains the $\propto (t-t_{\textup{BH}})^{-2}$ scaling for longer, as we might expect for a more highly magnetized accretion disk. Interestingly by the end of the simulation both cases with $M=2.57M_{\odot}$, $\vert B \vert_{\textup{max},\textup{insert}}=2.2\times 10^{17}$G (red lines) have approximately the same accretion rates and mass outside the AH. These are also the only two cases which reach the $\eta_{\textup{EM}} \sim 0.1$  efficiency expected for a Novikov-Thorne thin disk \cite{Shapiro:2004ud}.

\section{Conclusions}
\label{sec:Conclusions}

It has been suggested that the kHz GW signal from NSNS mergers holds great potential for informing our knowledge of the nuclear EOS (see e.g. \cite{Kruger:2021zta}), and as a result a large amount of work has been put into predicting this signal, and deriving quasi-universal relations between its characteristic frequencies and fundamental NS properties \cite{Shibata:2005ss,Shibata:2006nm,Stergioulas:2011gd,Takami:2014tva,Bauswein:2014qla,Rezzolla:2016nxn,Vretinaris:2019spn}. A common assumption of these works is that the effect of the magnetic field on the postmerger signal and remnant evolution can be neglected, allowing the use of simpler hydrodynamics-only simulations \cite{Shibata:2017xht}. However, the magnetic field is significantly amplified by several mechanisms during the postmerger regime, as shown by high resolution GRMHD simulations \cite{Kiuchi:2014hja}, causing the NS remnant and surrounding material to become highly magnetized. Our group has conducted self-consistent GRMHD simulations of both magnetized black hole-neutron star mergers \cite{Paschalidis:2014qra,Ruiz:2018wah} and NSNS mergers \cite{Ruiz:2016rai,Ruiz:2017inq,Ruiz:2019ezy,Ruiz:2020via,Ruiz:2021qmm,Bamber:2024kfb} finding indications that the presence of a magnetic field could leave a measurable imprint in the postmerger GW signal \cite{Ruiz:2021qmm}, and that the initial magnetic field topology has an important impact on the formation of incipient relativistic jets and the lifetime of NS remnants undergoing delayed collapse to a BH.  

In this work we sought to characterize this imprint by conducting GRMHD simulations of NSNS mergers with a range of different seed magnetic fields. We used two different equations of state SLy and ALF2, both broadly compatible with current observational constraints, and two different binary ADM masses: $M=2.57M_{\odot}$ which produces long-lived SMNS remnants with no magnetic field and $M=2.70M_{\odot}$ which produces short-lived HMNS remnants that collapse to BHs on ms timescales with no magnetic field. For the seed magnetic fields we used three different initial magnetic topologies: pulsar-like dipolar poloidal fields extending from the interior to the exterior, poloidal-only fields confined to the stellar interior, and toroidal-only fields also confined to the star. As pure poloidal and toroidal fields are thought to be unstable on Alfv\'{e}n timescales we insert the magnetic field at 1ms before merger to ensure the initial field remains close to its inserted topology in the inspiral regime and therefore distinguish their influence. We evolve the interior-only poloidal and interior-only toroidal topologies with our canonical magnetic field strength of $\vert B \vert_{\textup{max},\textup{insert}}=5.5\times 10^{16}$G, the same as that used in \cite{Bamber:2024kfb}. Along with hydrodynamics-only simulations with no magnetic field we use four different magnetic field strengths for the pulsar-like topology, from an initial maximum magnetic field measured by normal observers of $\vert B \vert_{\textup{max},\textup{insert}}=5.5\times 10^{15}$G (10 times smaller than that used in \cite{Bamber:2024kfb}) to $\vert B \vert_{\textup{max},\textup{insert}}=2.2\times 10^{17}$G (4 times larger than that used in \cite{Bamber:2024kfb}). 

We find the effective magnetic viscosity and magnetic braking causes angular momentum transport which both drives the NS remnants towards uniform rotation and causes them to lose mass and angular momentum, in addition to the mass and angular momentum loss via EM and GW radiation. For most of the cases the loss of angular momentum causes the remnant star to contract and the angular velocity of the quadrupole density oscillation to increase (rotating faster than the NS fluid itself), thereby causing the $f_2$ frequency of the dominant postmerger GW mode to shift upwards over time. Furthermore, for the ALF2 $M=2.70M_{\odot}$ case with the strongest magnetic field the remnant loses sufficient mass that it persists as a long-lived SMNS star, despite the loss of angular momentum. With the exception of this case, increasing the initial magnetic field strength also causes a decrease in the lifetime of the $M=2.70M_{\odot}$ hypermassive remnants.

The overall $f_2$ frequencies extracted from the whole postmerger signal show an increase of up to $\sim 200$Hz compared to the hydro-only simulation, with similar shifts for the simulations with initial pulsar-like and interior-poloidal topologies of the same $\vert B \vert_{\textup{max},\textup{insert}}$, although the cases with initial interior-only toroidal topologies show smaller frequency shifts. The quasi-radial oscillation frequency $f_0$ and the subdominant $f_{\textup{spiral}}$ frequency associated with the rotation of the outer spiral bulges show less evidence of a consistent trend with increasing magnetic field, and, as the $f_2$ mode has the largest amplitude, it is the most important frequency for the purposes of EOS constraints. 
Advanced LIGO and third generation detectors can resolve the characteristic postmerger frequencies to within $\sim O(100)$ Hz or less
\cite{Clark:2015zxa,Yang:2017xlf}, therefore the imprint of the magnetic field is in principle detectable. In practice though, a measurement of the postmerger GW spectrum may not be enough, and a more complete observational campaign would be needed for concrete conclusions \cite{Tsokaros:2024wgb}. 

We also find that very strong magnetic fields (with $\vert B \vert_{\textup{max},\textup{insert}}\sim 2.2\times 10^{17}$G, the largest seed magnetic field we explored) can cause SMNS remnants with masses well below the supramassive limit $M_{\textup{sup}}$ to lose so much angular momentum that they collapse to BHs, on the short ms timescales usually associated with HMNS remnants. Many of the constraints on the NS maximum mass, and therefore EOS, from GW170817 were derived previously with the assumption that the merger had to have formed a HMNS remnant, or at least a SMNS remnant with a mass very close to $M_{\textup{sup}}$, to produce a jet powered by a BH + accretion disk central engine (e.g. \cite{Ruiz:2017due,Rezzolla:2017aly}) and therefore that the supramassive limit had to be less than the total mass of the GW170817 binary. Here we present for the first time numerical GRMHD simulations demonstrating a SMNS merger remnant collapsing to a BH on timescales of $\sim 10$ms, and demonstrate that these cases go on to produce magnetically dominated incipient jets, comparable to those produced after the delayed collapse of HMNS remnants in \cite{Ruiz:2016rai,Ruiz:2021qmm,Bamber:2024kfb}. 

However, it should be noted that the associated magnetic fields are likely unrealistically large, and produced EM luminosities of $L_{\textup{EM}} \sim 10^{54}\textup{erg}\;\textup{s}^{-1}$, corresponding to isotropic equivalent $\gamma-$ray luminosity of $L_{\gamma,\textup{iso}} \sim 10^{54}-10^{55}\textup{erg}\;\textup{s}^{-1}$, above the range observed in typical sGRBs \cite{Li:2016pes,Beniamini:2020adb,Shapiro:2017cny}. Therefore it is highly unlikely that our $\vert B \vert_{\textup{max},\textup{insert}}\sim 2.2\times 10^{17}$G simulations can be a direct model for GW170817, or similar events, so the results of \cite{Ruiz:2017due,Rezzolla:2017aly,Nathanail:2021tay} and similar works are likely still valid. 

%Nonetheless, the values for the mass-averaged magnetic $\alpha-$viscosity parameter for these cases of $\alpha_{\textup{SS}} \sim 0.06-0.07$, while large, lie within the range observed in simulations of MHD turbulence \cite{Shi:2015mvh}.

%, so if any equivalent source of high viscosity and efficient angular momentum loss exists in nature it is possible that the assumptions about GW170817 and associated constraints will need to be revised. 

Our current simulations do not include neutrino radiation transport, although we have included neutrino emission in a previous work \cite{Sun:2022vri}. Since neutrino emission is relatively inefficient at removing angular momentum \cite{Sun:2022vri,Baumgarte:1998sn} we expect our main results will still hold even when neutrinos are present. Further studies are needed with neutrino radiation and full tabulated finite-temperature EOSs to provide more definitive answers. While we see similar results in this work to those of our previous studies which used a lower resolution \cite{Ruiz:2021qmm} (see Appendix \ref{sec:conv}), some resolution dependence is unavoidable due to the difficulty in achieving asymptotic convergence for the magnetic instabilities and numerical viscosity. Therefore simulations at even higher resolutions, beyond those achievable with our computational resources for this work, will also be beneficial.

Movies and additional 3D visualizations highlighting
our simulations can be found at \cite{website}.

\section*{Acknowledgements}
We thank members of our Illinois Relativity Undergraduate Research Team (Seyed Dastgheib, Jonah  Doppelt, Yuheng Guo, Yinuan Liang, Rohan Narasimhan, Cody Olson and Eric Yu) for assistance with some of the 3D visualizations. This work was supported in part by National Science Foundation (NSF) Grants PHY-2308242, OAC-2310548 and PHY-2006066 to the University of Illinois at Urbana-Champaign. 
M.R. acknowledges support by the Generalitat Valenciana Grant CIDEGENT/2021/046 and by the Spanish Agencia Estatal de Investigaci\'on (Grant PID2021-125485NB-C21). 
A.T. acknowledges support from the National Center for Supercomputing Applications (NCSA) at the University of Illinois at Urbana-Champaign through the NCSA Fellows program. 
This work used Stampede2 at TACC and Anvil at Purdue University through allocation MCA99S008,
from the Advanced Cyberinfrastructure Coordination Ecosystem: Services \& Support (ACCESS) program, which is supported by National Science Foundation grants \#2138259, \#2138286, \#2138307, \#2137603, and \#2138296.
This research also used Frontera at TACC through allocation AST20025. Frontera is made possible by NSF award OAC-1818253. 

The authors thankfully acknowledge the computer resources at MareNostrum and the technical support provided by the Barcelona Supercomputing Center (AECT-2023-1-0006).

\appendix

\section{Convergence test}
\label{sec:conv}

\begin{figure}
    \centering
    \includegraphics[width=\linewidth]{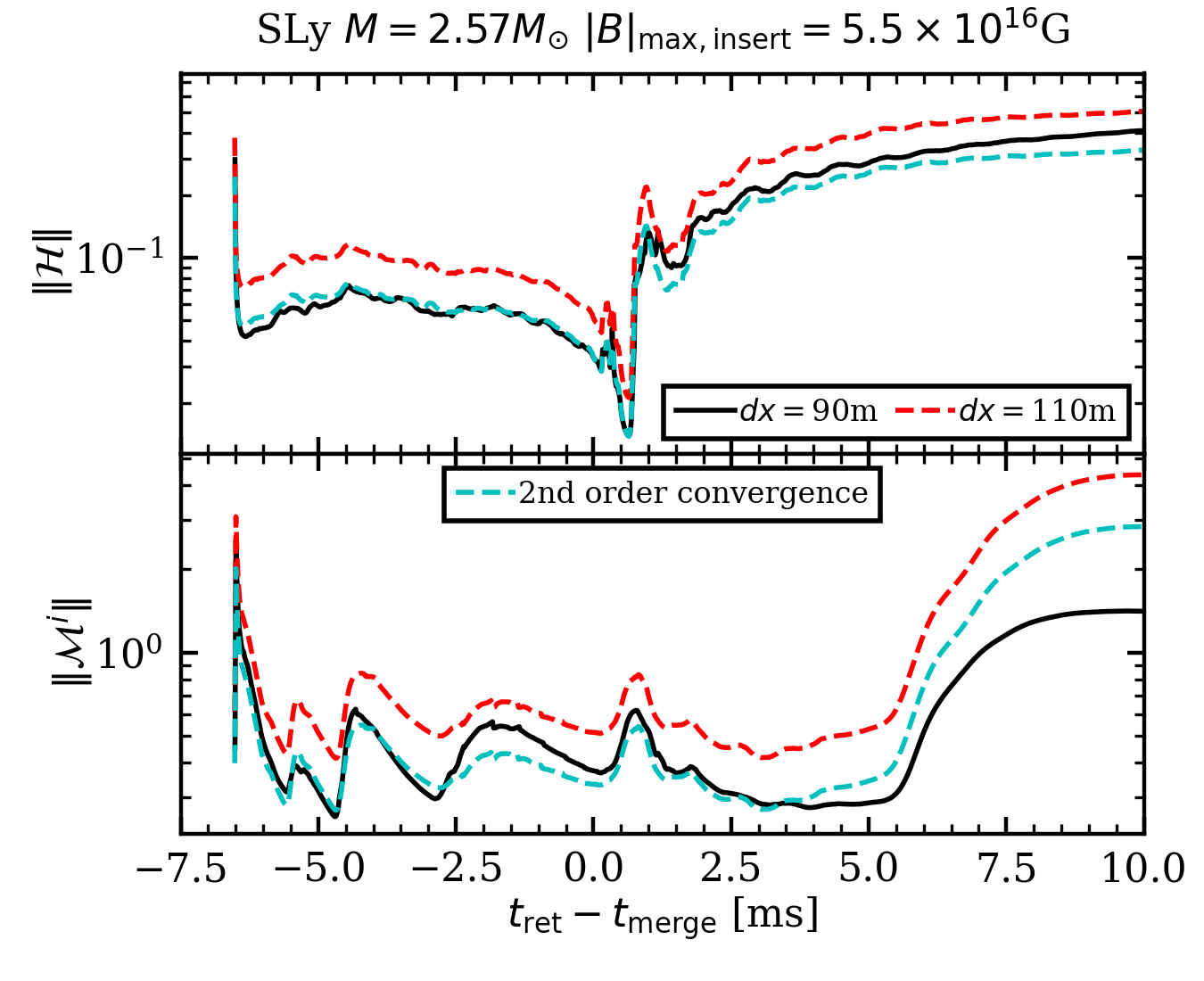}
    \caption{Simple convergence test for the SLy $M=2.57M_{\odot}$, $\vert B \vert_{\textup{max},\textup{insert}}=5.5\times 10^{16}$G case with a pulsar-like initial magnetic field.}
    \label{fig:convergence_test}
\end{figure}

In Fig. \ref{fig:convergence_test} we show a simple convergence test for our code for one of the cases considered in this work: the SLy $M=2.57M_{\odot}$, $\vert B \vert_{\textup{max},\textup{insert}}=5.5\times 10^{16}$G case with an initial pulsar-like magnetic field topology. We plot normalized Hamiltonian (top) and Momentum (bottom) constraints (see e.g. \cite{Etienne:2007jg}) for simulations of resolution $\Delta x_{\textup{min}} = 90$m (black), as used in this work, and $\Delta x_{\textup{min}} = 110$m (red), as used in \cite{Ruiz:2021qmm}, along with the expected constraint value for the high resolution simulation if we had perfect second order convergence (cyan), given by $\Vert \mathcal{H} \Vert^{\textup{ideal}}_{\textup{90m}} = (90/110)^2 \Vert \mathcal{H} \Vert_{\textup{110m}}$. We see that the true high resolution result achieves second order convergence (as in e.g. \cite{Etienne:2015cea} and as expected for HLL fluxes with PPM+ reconstruction \cite{Duez:2005sf}) up to about 5ms postmerger, where it transitions to first order convergence in the Hamiltonian constraint, which is a result of how the simulation code handles shocks during the merger. 

One can also compare with the results of our previous work \cite{Ruiz:2021qmm}, which used a resolution of $\dd x_{\textup{finest}}=110$m. In that study we observed a shift of $f_2 = f_{\textup{peak}}$ due to the magnetic field of $\Delta f_2 \approx 160$Hz for a SLy $M=2.6M_{\odot}$ merger with an initial pulsar-like purely poloidal magnetic field of strength $\vert B \vert_{\textup{max},\textup{insert}} \sim 6 \times 10^{16}$G compared to the hydro-only case (see Table IV in \cite{Ruiz:2021qmm}, SLyM2.6H and SLyM2.6P cases). This is comparable to the $\Delta f_2 \approx 170$Hz shift we observe in this work for the SLy $M=2.57M_{\odot}$ pulsar-like $\vert B \vert_{\textup{max},\textup{insert}} = 5.5 \times 10^{16}$G case with a higher resolution of $\dd x_{\textup{finest}}=90$m. Along with our (crude) convergence study we take this as evidence that our main results regarding a magnetic-field induced shift in the postmerger frequencies are robust with respect to numerical resolution, although some resolution dependence will be unavoidable due to the difficulty of achieving asymptotic convergence in resolving the KHI, as discussed in Sec. \ref{sec:magn_fields}.

\section{Additional tables: GW postmerger frequency data and initial values for $I_0$ and $A_b$.}
\label{app:I0}

The numerical values of the characteristic frequencies of the postmerger gravitational wave signals of the different cases are shown in Table \ref{tab:GW_freqs}. 

\begin{table*}
\begin{tabular}{lll|lllll}
\toprule \Tstrut
 EOS & M & $\vert B \vert_{\textup{max},\textup{insert}}$ & $f_2$ & $f_{2-0}$ & $f_{2+0}$ & $f_{\mathrm{spiral}}$ & $f_{0}$(from $\alpha_{\textup{min}}$) \Tstrut \\
 \hline
SLy & 2.57 &   0 & 3.224(7) & 2.1(2) & 4.36(10) & 2.244(14) & 1.194(18) \Tstrut \\
SLy & 2.57 & \textcolor{C0}{PU $5.5\times 10^{15}$} & 3.225(4) & 2.0(2) & 4.36(5) & 2.4(2) & 1.194(18) \\
SLy & 2.57 & \textcolor{C1}{PU $1.4\times 10^{16}$} & 3.229(6) & 2.11(12) & 4.39(8) & 2.4(3) & 1.20(4) \\
SLy & 2.57 & \textcolor{C2}{PU $5.5\times 10^{16}$} & 3.39(2) & 2.15(3) & 4.46(13) & 2.4(5) & 1.243(13) \\
SLy & 2.57 & \textcolor{C3}{PU $2.2\times 10^{17}$} & 3.434(17) & 2.20(3) & 4.51(15) & 2.4(2) & 1.19(4) \\
SLy & 2.57 & \textcolor{C4}{IT $5.5\times 10^{16}$} & 3.321(5) & 2.11(6) & 4.43(9) & 2.557(11) & 1.23(3) \\
SLy & 2.57 & \textcolor{C5}{IP $5.5\times 10^{16}$} & 3.37(3) & 2.15(6) & 4.43(11) & 2.5(2) & 1.218(10) \\
\hline
ALF2 & 2.57 &   0 & 2.647(13) & 1.39(11) & 3.72(10) & 2.14(9) & 1.22(10) \Tstrut \\
ALF2 & 2.57 & \textcolor{C0}{PU $5.5\times 10^{15}$} & 2.646(7) & 1.4(2) & 3.76(2) & 2.19(11) & 1.201(10) \\
ALF2 & 2.57 & \textcolor{C1}{PU $1.4\times 10^{16}$} & 2.685(8) & 1.48(8) & 3.80(11) & 2.15(11) & 1.199(10) \\
ALF2 & 2.57 & \textcolor{C2}{PU $5.5\times 10^{16}$} & 2.775(11) & 1.44(6) & 3.85(13) & 2.21(16) & 1.244(18) \\
ALF2 & 2.57 & \textcolor{C3}{PU $2.2\times 10^{17}$} & 2.80(3) & 1.52(8) & 3.88(15) & 2.23(11) & 1.23(2) \\
ALF2 & 2.57 & \textcolor{C4}{IT $5.5\times 10^{16}$} & 2.70(2) & 1.36(7) & 3.73(15) & 2.21(14) & 1.195(7) \\
ALF2 & 2.57 & \textcolor{C5}{IP $5.5\times 10^{16}$} & 2.76(4) & 1.40(6) & 3.78(16) & 2.20(18) & 1.207(18) \\
\hline
ALF2 & 2.70 &   0 & 2.84(3) & 1.702(19) & 3.86(4) & 2.26(15) & 1.01(2) \Tstrut \\
ALF2 & 2.70 & \textcolor{C0}{PU $5.5\times 10^{15}$} & 2.822(8) & 1.73(8) & 3.83(4) & 2.18(18) & 0.98(4) \\
ALF2 & 2.70 & \textcolor{C1}{PU $1.4\times 10^{16}$} & 2.90(5) & 1.8(2) & 3.86(4) & 2.29(19) & 1.00(3) \\
ALF2 & 2.70 & \textcolor{C2}{PU $5.5\times 10^{16}$} & 2.93(6) & 1.8(3) & 3.895(14) & 2.17(10) & 1.00(3) \\
ALF2 & 2.70 & \textcolor{C3}{PU $2.2\times 10^{17}$} & 2.853(15) & 1.70(6) & 3.96(2) & 2.14(9) & 1.179(8) \\
ALF2 & 2.70 & \textcolor{C4}{IT $5.5\times 10^{16}$} & 2.90(5) & 1.8(3) & 3.90(7) & 2.2(2) & 1.00(3) \\
ALF2 & 2.70 & \textcolor{C5}{IP $5.5\times 10^{16}$} & 2.92(6) & 1.74(3) & 3.96(13) & 2.3(2) & 1.125(5) \\
\botrule
 \end{tabular}
 \caption{Summary of the key characteristic frequencies of the postmerger GW signals, as shown in Fig. \ref{fig:GW_f_vs_B}. Frequencies are given in kHz, and uncertainties on the last digits are shown in parantheses.}
 \label{tab:GW_freqs}
 \end{table*}

\begin{table}
\begin{tabular}{lll|ll}
\toprule \Tstrut
 EOS & M & $\vert B \vert_{\textup{max},\textup{insert}}$ & $I_0$ & $A_b$ \Tstrut \\
 \hline
SLy & 2.57 & 0 & 0 & $\dots$ \Tstrut \\
SLy & 2.57 & \textcolor{C0}{PU $5.5\times 10^{15}$} & $3.0709\times 10^{-4}$ & $\dots$ \\
SLy & 2.57 & \textcolor{C1}{PU $1.4\times 10^{16}$} & $7.6771\times 10^{-4}$ & $\dots$ \\
SLy & 2.57 & \textcolor{C2}{PU $5.5\times 10^{16}$} & $3.0709\times 10^{-3}$ & $\dots$ \\
SLy & 2.57 & \textcolor{C3}{PU $2.2\times 10^{17}$} & $1.2283\times 10^{-2}$ & $\dots$ \\
SLy & 2.57 & \textcolor{C4}{IT $5.5\times 10^{16}$} & $\dots$ & $8.9031\times 10^{-5}$ \\
SLy & 2.57 & \textcolor{C5}{IP $5.5\times 10^{16}$} & $\dots$ & $4.3348\times 10^{-3}$ \\
\hline
ALF2 & 2.57 & 0                                      & 0 & $\dots$ \Tstrut \\
ALF2 & 2.57 & \textcolor{C0}{PU $5.5\times 10^{15}$} & $2.9129\times 10^{-4}$ & $\dots$  \\
ALF2 & 2.57 & \textcolor{C1}{PU $1.4\times 10^{16}$} & $7.3086\times 10^{-4}$ & $\dots$  \\
ALF2 & 2.57 & \textcolor{C2}{PU $5.5\times 10^{16}$} & $2.9129\times 10^{-3}$ & $\dots$  \\
ALF2 & 2.57 & \textcolor{C3}{PU $2.2\times 10^{17}$} & $1.1651\times 10^{-2}$ & $\dots$  \\
ALF2 & 2.57 & \textcolor{C4}{IT $5.5\times 10^{16}$} & $\dots$ & $6.3986\times 10^{-5}$ \\
ALF2 & 2.57 & \textcolor{C5}{IP $5.5\times 10^{16}$} & $\dots$ & $3.7583\times 10^{-3}$ \\
\hline
ALF2 & 2.70 &   0                                    & 0 & $\dots$  \Tstrut \\
ALF2 & 2.70 & \textcolor{C0}{PU $5.5\times 10^{15}$} & $1.9647\times 10^{-4}$ & $\dots$  \\
ALF2 & 2.70 & \textcolor{C1}{PU $1.4\times 10^{16}$} & $4.9119\times 10^{-4}$ & $\dots$  \\
ALF2 & 2.70 & \textcolor{C2}{PU $5.5\times 10^{16}$} & $1.9647\times 10^{-3}$ & $\dots$  \\
ALF2 & 2.70 & \textcolor{C3}{PU $2.2\times 10^{17}$} & $7.8590\times 10^{-3}$ & $\dots$ \\
ALF2 & 2.70 & \textcolor{C4}{IT $5.5\times 10^{16}$} & $\dots$ & $9.9803\times 10^{-4}$ \\
ALF2 & 2.70 & \textcolor{C5}{IP $5.5\times 10^{16}$} & $\dots$ & $3.9802\times 10^{-3}$ \\
\botrule
 \end{tabular}
 \caption{Summary of the initial values for $I_0$ and $A_b$ for the initial magnetic fields for the different cases. Note that only the pulsar-like topologies have $I_0$ values, and only the interior-only topologies have $A_b$ values. The current loop radius $r_0$ is set to $2R_x/3$.
 }
 \label{tab:I0_Ab}
 \end{table}

%\tableofcontents

\bibliography{apssamp}% Produces the bibliography via BibTeX.

\end{document}